\documentclass[fleqn, usenatbib]{mnras}
\usepackage{newtxtext, newtxmath}
\usepackage[T1]{fontenc}

\DeclareRobustCommand{\VAN}[3]{#2}
\let\VANthebibliography\thebibliography
\def\thebibliography{\DeclareRobustCommand{\VAN}[3]{##3}\VANthebibliography}

\usepackage{graphicx}   
\usepackage{amsmath}    
\usepackage{xfrac}
\usepackage{braket}
\usepackage[export]{adjustbox}
\usepackage{commath}
\usepackage{hyperref}
\usepackage{float}
\usepackage{subfigure}
\usepackage{wasysym}
\usepackage{xcolor}
\usepackage{todonotes}
\usepackage{booktabs}
\usepackage{pdfcomment}
\usepackage{xspace}
\usepackage{orcidlink}
\usepackage{booktabs}
\usepackage{etoolbox}

\newcommand{\solM}{$\mathrm{M}_{\odot}$} 
\newcommand{\rmin}{r_{\mathrm{min}}} 
\newcommand{\rcirc}{r_{\mathrm{circ}}} 
\newcommand{\vthermal}{v_{\mathrm{thermal}}}

\newcommand{\fsync}{f_{\mathrm{sync}}}
\newcommand{\qacc}{q_{\mathrm{acc}}}
\newcommand{\selfintersect}{f_{\mathrm{self\ intersect}}}
\newcommand{\otherintersect}{f_{\mathrm{other\ intersect}}}
\newcommand{\thetaintersect}{\theta_{\mathrm{intersect}}}
\newcommand{\thetathreshold}{\theta_{\mathrm{threshold}}}
\newcommand{\thetathresholdvalue}{45^{\circ}}
\newcommand{\streamdiameter}{D_{\mathrm{stream}}}

\DeclareRobustCommand{\Eqref}[1]{equation~\ref{#1}}
\DeclareRobustCommand{\Figref}[1]{Fig.~\ref{#1}}
\DeclareRobustCommand{\Tabref}[1]{Table~\ref{#1}}
\DeclareRobustCommand{\Secref}[1]{Section~\ref{#1}}
\DeclareRobustCommand{\Appref}[1]{Appendix~\ref{#1}}

\newcommand{\binaryc}{\textsc{binary\_c}}
\newcommand{\MESA}{\textsc{mesa}}
\newcommand{\COMPAS}{\textsc{compas}}

\renewcommand{\doi}[1]{\textsc{doi}: \href{http://dx.doi.org/#1}{\nolinkurl{#1}}}
\newcommand{\rom}[1]{%
  \textup{\uppercase\expandafter{\romannumeral#1}}%
}

\newcommand{\mytitle}{Mass-stream trajectories with non-synchronously
  rotating donors}
\newcommand{\myshorttitle}{Stream trajectories with
  non-synchronous donors}
\title[\myshorttitle]{\mytitle}

\hypersetup{
    colorlinks=true,
    linkcolor=blue,
    filecolor=magenta,
    urlcolor=cyan,
    pdftitle={\mytitle},
    bookmarksdepth=3
}

\author[D.D. Hendriks \& R.G. Izzard.]{
  D. D. Hendriks$^{1}$\thanks{E-mail: \href{mailto:dh00601@surrey.ac.uk}{dh00601@surrey.ac.uk}\ (DDH)} \orcidlink{0000-0002-8717-6046},
  R. G. Izzard$^{1}$ \orcidlink{0000-0003-0378-4843}
  \\
  $^{1}$Department of Physics, University of Surrey, Guildford, GU2 7XH, Surrey, UK
}

\date{Accepted 2023 July 6. Received 2023 July 5; in original form 2023 June 5}
\pubyear{2023}

\begin{document}
\label{firstpage}
\pagerange{\pageref{firstpage}--\pageref{lastpage}}
\maketitle

\begin{abstract} 
  Mass-transfer interactions in binary stars can lead to accretion
  disk formation, mass loss from the system and spin-up of the
  accretor. To determine the trajectory of the mass-transfer stream,
  and whether it directly impacts the accretor, or forms an accretion
  disk, requires numerical simulations. The mass-transfer stream is
  approximately ballistic, and analytic approximations based on such
  trajectories are used in many binary population synthesis codes as
  well as in detailed stellar evolution codes.
  We use binary population synthesis to explore the conditions under
  which mass transfer takes place. We then solve the reduced
  three-body equations to compute the trajectory of a particle in the
  stream for systems with varying system mass ratio, donor
  synchronicity and initial stream velocity.
  Our results show that on average both more mass and more time is
  spent during mass transfer from a sub-synchronous donor than from a
  synchronous donor.
  Moreover, we find that at low initial stream velocity the
  asynchronous rotation of the donor leads to self-accretion over a
  large range of mass ratios, especially for super-synchronous
  donors. The stream (self-)intersects in a narrow region of parameter
  space where it transitions between accreting onto the donor or the
  accretor.
  Increasing the initial stream velocity leads to larger areas of the
  parameter space where the stream accretes onto the accretor, but
  also more (self-)intersection. The radii of closest approach
  generally increase, but the range of specific angular momenta that
  these trajectories carry at the radius of closest approach gets
  broader.
  Our results are made publicly available.
\end{abstract}




\section{Introduction}
\label{sec:intro}
Binary stellar systems are ubiquitous and the proximity of a star to a
companion introduces a variety of interactions. These interactions
lead to a range of phenomena like the stripping of the outer envelope
of a star and the transfer of mass and angular momentum
\citep{lubowGasDynamicsSemidetached1975,
  ulrichAccretingComponentMassexchange1976}, tidal interactions
\citep{zahnReprint1977AAmp1977, zahnTidalDissipationBinary2008,
  ogilvieTidalDissipationStars2014,
  mirouhDetailedEquilibriumDynamical2023}, the formation and evolution
of accretion disks \citep{pringleAccretionDiscModels1972,
  shakuraBlackHolesBinary1973, papaloizouTidalTorquesAccretion1977,
  osakiTidalEffectsAccretion1993, hameuryReviewDiscInstability2020},
accretion induced supernovae \citep{nomotoFateAccretingWhite1986,
  ruiterTypeIaSupernovae2010, claeysTheoreticalUncertaintiesType2014},
the (high velocity) ejection of companions
\citep{blaauwOriginBtypeStars1961, taurisRunawayVelocitiesStellar1998,
  renzoMassiveRunawayWalkaway2019}, quasi chemically-homogeneous
evolution \citep{ghodlaEvaluatingChemicallyHomogeneous2023}, Be stars
\citep{shaoFormationBeStars2014, postnovSpinupSpindownNeutron2015},
and circumbinary disk formation \citep{kashiCircumbinaryDiscFinal2011,
  pejchaBinaryStellarMergers2016, izzardCircumbinaryDiscsStellar2022}.

A comprehensive review of these binary interactions is given in
\citet{demarcoImpactCompanionsStellar2017}, but the most relevant
interactions to the current study are the transfer of mass and tidal
interactions between the stars. Both these interchange orbital and
rotational angular momentum of the system and the stars. Tidal
interactions circularise the orbit, i.e. reduce the eccentricity, and
synchronise the stars, i.e. force the stellar rotation rate to equal
the orbital rotation rate. Mass transfer, among other effects,
de-synchronises the stars by angular momentum transfer from the donor
to the accretor.

In a semi-detached system, where the accretor significantly underfills
its Roche-Lobe, the mass transfer process can be split into three main
parts: The ejection from the donor, the flight of the particles in the
potential between the stars, and the accretion onto the accretor
\cite{kruszewskiExchangeMatterClose1964a}. During the flight stage,
the gravitational interaction between the binary system and the
particle leads to a torque and subsequent exchange of angular momentum
between the binary system and the particle. It is during this stage
that the final outcome of the trajectory is determined, i.e. accretion
onto the companion star, accretion back onto the donor star or loss
from the system entirely. The flight stage is approximately ballistic,
and it is the stage that we focus on in this study.

\begin{figure}
  \centering
  \includegraphics[width=\columnwidth]{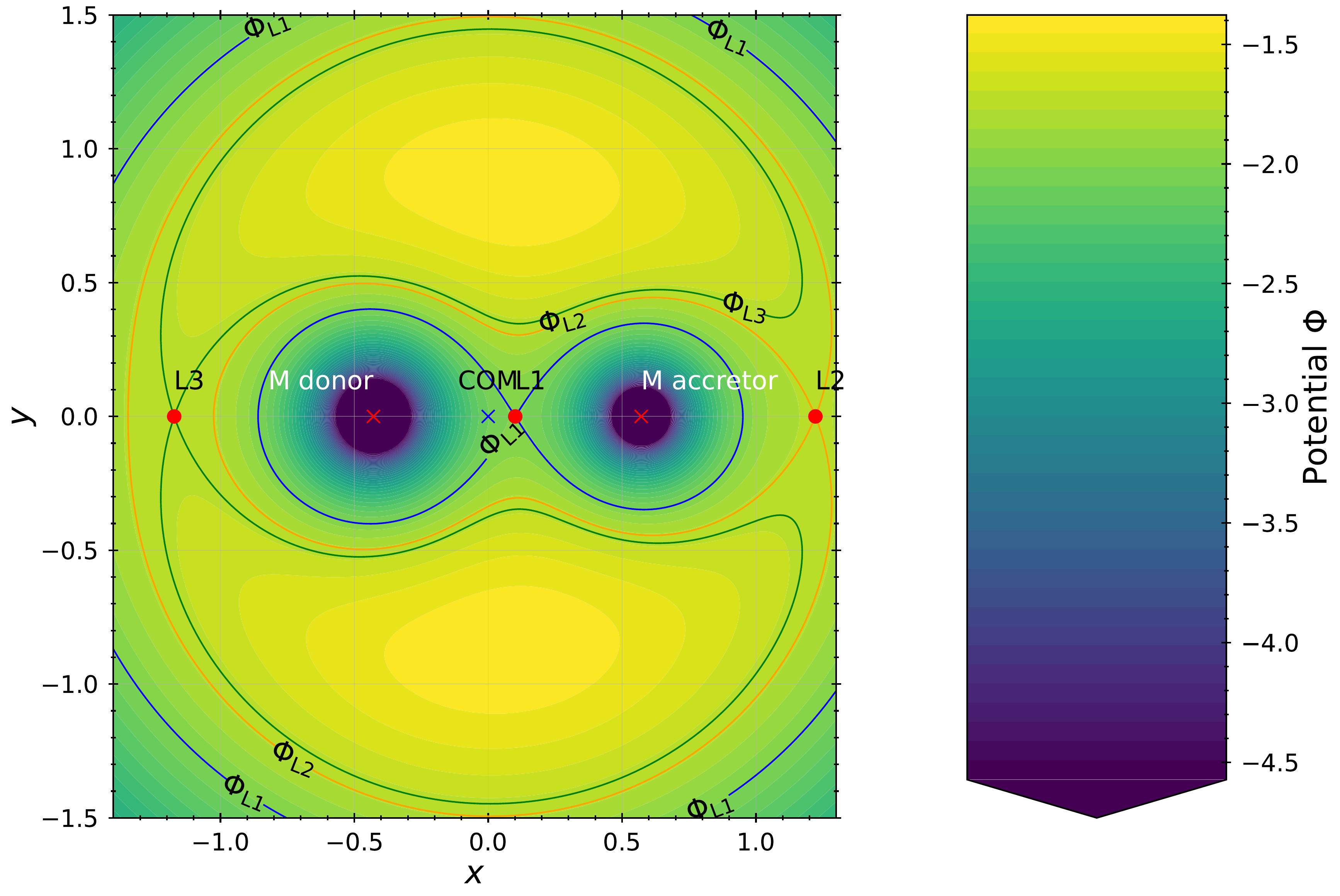}
  \caption{The Roche potential, $\Phi$, in the co-rotating frame of
    reference, with the positions of the two stars and the centre of
    mass denoted with red and blue crosses respectively. We indicate
    several equilibrium points within the co-rotating frame of
    reference, the L1, L2 and L3 Lagrange points, by red dots. The
    lines of constant potential energy that belong to each of these
    points are indicated with the solid blue, orange and green lines
    and labelled $\Phi_{L1}$, $\Phi_{L2}$, and $\Phi_{L3}$
    respectively. The colour-scale indicates the potential. The system
    is characterised by two masses of which the ratio of the accretor
    to the donor mass is $\qacc = 0.75$. Moreover, the ratio of the
    rotation rate of the donor to the orbital rotation rate is
    $\fsync = 1$}
  \label{fig:schematic_overview_frame}
\end{figure}

The potential that is used to calculate when and how mass is
transferred from one star to the other is often calculated under the
assumption that the orbit of the binary system is circular and that
the donor rotates synchronously with the orbit
\citep{lubowGasDynamicsSemidetached1975}. Together with the
approximation that the stars are point particles this setup is often
called the Roche potential (\Figref{fig:schematic_overview_frame}).

The points in this potential where accelerations vanish are called
Lagrange points. The first Lagrange point lies on the critical
equipotential surface and is located between the two stars. While
generalisations of the equipotential surface and the inclusion of
additional physical effects have been studied, binary
stellar-evolution codes often still use simplified analytical formulae
for the mass stream properties based on circular and synchronous
systems.
Some examples of extensions to this simple Roche model that relax some
assumptions, or add additional physics, are those that allow the
asynchronous rotation of the donor with respect to the orbital
rotation \citep{plavec49DynamicalInstability1958,
  limberSurfaceFormsMass1963, kruszewskiExchangeMatterClose1963},
eccentric orbits \citep{avniEclipseDurationXray1976,
  sepinskyEquipotentialSurfacesLagrangian2007}, spin-orbit
misalignment \citep{limberSurfaceFormsMass1963,
  avniGeneralizedRochePotential1982}, effects of external radiation
\citep{podsiadlowskiSupermassiveBlackHoles1994,
  drechselRadiationPressureEffects1995,
  phillipsIrradiationPressureEffects2002,
  tsantilasInfluenceRadiationPressure2006,
  dermineRadiationPressurePulsation2009} or combinations of these
\citep{vanbeverenInfluenceCriticalSurface1977}. These extensions
change the shape of the critical surface and the location of the
Lagrange points, most notably the first.

Asynchronous rotation of the donor induces time-dependent tides,
exerted by the companion, which then affect the potential. The
above-mentioned extensions that take the asynchronous rotation of the
donor into account \citep[e.g.][]{plavec49DynamicalInstability1958,
  limberSurfaceFormsMass1963, kruszewskiExchangeMatterClose1963} rest
on several assumptions. First, the shape of the donor is assumed to
conform instantaneously to the shape dictated by the
potential. Secondly, the motion of mass in the donor is assumed to
move primarily along the axis of rotation (i.e. primarily zonal,
instead of meridional). These two assumptions are called the first
approximation \citep{limberSurfaceFormsMass1963,
  savonijeRochelobeOverflowXray1978,
  sepinskyEquipotentialSurfacesLagrangian2007}. Asynchronous rotation
of the donor can occur due to, e.g., rapid expansion of the donor star
leading to sub-synchronous rotation when it fills its Roche Lobe.

Given L1 it is possible to calculate the initial conditions and
subsequent trajectory of the mass flow away from the donor star.
\citet{lubowGasDynamicsSemidetached1975} analyse the behaviour of
donor material at L1 and the trajectory of the stream of matter
flowing from L1 to the accretor. Their perturbative analysis provides
mass-transfer stream properties over a range of orbital configurations
of the binary based on ballistic trajectories of particles in the
Roche potential. Critical to the study of
\citet{lubowGasDynamicsSemidetached1975} are the assumptions that the
donor rotates synchronously with the orbit, that the stream at L1 has
a low thermal-velocity (cold) of compared to the orbital velocity,
that the gas remains isothermal throughout the flow, and that the mass
contained in the stream is negligible compared to the total mass of
the system. \citet{ulrichAccretingComponentMassexchange1976} provide
analytical fits to this data and study the response of the accretor
when the mass-transfer stream either directly impacts the accretor or
misses the accretor and forms an accretion
disk. \citet{kruszewskiExchangeMatterClose1963,
  kruszewskiExchangeMatterClose1964,
  kruszewskiExchangeMatterClose1964a,
  kruszewskiExchangeMatterClose1967} calculates properties of the mass
transfer in non-synchronous rotating donors, including the effects of
kinematic acceleration due to the bulging motion of the donor star as
a result of its non-synchronicity. \citet{warnerLocationSizeHot1972}
and \citet{flanneryLocationHotSpot1975} study the effect of initial
thermal-velocity of the stream particles on the location of hotspots
in cataclysmic variable
systems. \citet{sepinskyInteractingBinariesEccentric2010} and
\citet{davisMassTransferEccentric2013, davisBinaryEvolutionUsing2014}
calculate the ballistic trajectories to include in their osculating
orbit calculations and consider asynchronous rotating donors. They do
not make the results of these calculations public, however.

The aim of our paper is to publicly release interpolation tables that
contain the results of our ballistic-stream trajectories calculations
over a wide range of mass ratios and degrees of asynchronicity of the
donor, as well as mass-stream surface areas and initial thermal
velocities at L1. These can be used in combination with osculating
orbit calculations \citep{davisMassTransferEccentric2013,
  davisBinaryEvolutionUsing2014,
  dosopoulouORBITALEVOLUTIONMASSTRANSFERRING2016,
  dosopoulouRochelobeOverflowEccentric2017}, and as tables in stellar
evolution codes like {\MESA}
\citep{jermynModulesExperimentsStellar2023} and population synthesis
codes like \binaryc~\citep{izzardCircumbinaryDiscsStellar2022} or
\COMPAS~\citep{rileyRapidStellarBinary2022}.

Our paper is structured as follows. In \Secref{sec:theory} we explain
the theoretical basis of our project, and in \Secref{sec:Method} we
lay out the methods used to calculate our ballistic trajectories and
our approach to dataset interpolation. In \Secref{sec:results} we show
the results of our ballistic trajectory calculations for several
initial properties of the mass transfer stream. We discuss and
conclude in Sections \ref{sec:discussion}
and~\ref{sec:conclusions}. \Appref{sec:fiduc-source-distr} provides a
description of our interpolation datasets, and
\Appref{sec:lagrange-point-plot} contains a visual overview of the
first three Lagrange point locations in two different frames of
reference.


\section{Theory}
\label{sec:theory}
In this section we lay out the theoretical basis of the calculations
of the trajectory of a particle flowing through L1. We first determine
the potential that the particle experiences when attached to the donor
star and when moving freely through the system, and we then determine
the cross-sectional surface area of the stream and the initial
velocity of the particles L1.

\subsection{Generalised Roche potential and Lagrange points}
\label{sec:roche-potent-reduc}
To calculate the particle trajectory through the potential of the
binary system, we consider the reduced three-body problem in a
Cartesian coordinate system $Oxyz$ in the co-rotating frame of the
binary, which rotates with angular frequency $\omega$, with the origin
$O$ of the frame of reference located on the centre of mass of the
system \citep{hubovaKinematicsMasslossOuter2019}. The $x$-coordinate
is defined parallel to the line connecting the centres of the stars,
the $y$-coordinate defined perpendicular to the $x$-coordinate and in
the plane of the orbit and the $z$-coordinate perpendicular to the
orbital plane. Throughout our calculations we consider particle motion
only in the plane of the orbit, i.e. $z = 0$.

The donor and accretor are regarded as point masses,
$M_{\mathrm{don}}$ and $M_{\mathrm{acc}}$, with their positions fixed
at $\boldsymbol{x}_{\mathrm{don}} = [-\mu_{\mathrm{acc}},\ 0]$ and
$\boldsymbol{x}_{\mathrm{acc}} = [1-\mu_{\mathrm{acc}},\ 0]$
respectively, where
$\mu_{\mathrm{acc}} = M_{\mathrm{acc}}/(M_{\mathrm{don}} +
M_{\mathrm{acc}})$, and $\qacc =
M_{\mathrm{acc}}/M_{\mathrm{don}}$. Our units of length, time,
velocity, and potential are the semi-major axis $a$, the inverse
orbital frequency $\omega^{-1}$, the orbital velocity $a\omega$, and
$a^{2}\omega^{2}$ respectively, unless otherwise indicated.

A particle freely moving in a binary star system in a co-rotating
frame experiences the gravitational potential of both stars, and a
centrifugal potential due to the co-rotation, and a Coriolis force due
to movement relative to the co-rotating frame.  When we assume that
both stars are centrally condensed, i.e. the Roche model, the
potential is,
\begin{equation}
  \label{eq:roche_potential_COM_particle}
  \begin{aligned}
    \Phi(x,y) =
    -\frac{\mu_{\mathrm{acc}}}{\left[(x-1+\mu_{\mathrm{acc}})^2 +
        y^2\right]^{1/2}} -&
    \frac{1-\mu_{\mathrm{acc}}}{\left[\left(x+\mu_{\mathrm{acc}}\right)^2
        + y^2\right]^{1/2}} \\- \frac{1}{2}\left(x^{2}+y^{2}\right).
  \end{aligned}
\end{equation}
This is valid for a freely moving particle, i.e. not inside either
star, because there is no other force acting on the particle. This
potential is also valid to calculate the critical surface beyond which
mass starts flowing away from the donor, in the case the donor rotates
synchronously with the orbit and its rotation is along an axis
parallel to the orbital rotation. We show an example of the Roche
potential in \Figref{fig:schematic_overview_frame}.

To calculate the location at which mass starts flowing from the donor
we need to find the critical surface of the donor, i.e. the last
surface at which the net inward force of the potential is balanced by
the pressure of the star. We assume the rotation of the donor is in
the same direction as the orbit of the binary system, the dynamic
timescale is shorter than the tidal timescale, and that the orbit is
circular, in the rest of our study. The potential felt by a
non-synchronously rotating donor is
\begin{equation}
  \label{eq:roche_potential_COM_don}
  \begin{aligned}
    \Phi_{\mathrm{don}}(x,y, \fsync) = -&\frac{\mu_{\mathrm{acc}}}{\left[(x-1+\mu_{\mathrm{acc}})^2 + y^2\right]^{1/2}}\\
    -\frac{1-\mu_{\mathrm{acc}}}{\left[\left(x+\mu_{\mathrm{acc}}\right)^2
        + y^2\right]^{1/2}}
    &-\frac{1}{2}\fsync^{2}\left(x^{2}+y^{2}\right) -
    \left(\fsync^{2}-1\right)\mu\,x.
  \end{aligned}
\end{equation}
Here the potential acting on the donor depends on the synchronicity
factor,
\begin{equation}
  \label{eq:synchronicity_factor}
  \fsync = \Omega_{\mathrm{don}} / \omega,
\end{equation}
where $\Omega_{\mathrm{don}}$ is the rotation rate of the donor.

We calculate the location of the first three Lagrange points of the
donor, determining the critical equipotential surface, by taking the
derivative of the potential in \Eqref{eq:roche_potential_COM_don} with
respect to $x$ and setting $y = 0$,
\begin{equation}
  \label{eq:critical_surface}
  \begin{aligned}
    \frac{\mathrm{d}\Phi_{\mathrm{don}}(y=0)}{\mathrm{d}x} &= \frac{\left(1 - \mu_{\mathrm{acc}}\right)}{\left(\mu_{\mathrm{acc}} + x\right)^{2}} + \frac{\mu_{\mathrm{acc}}}{\left(\mu_{\mathrm{acc}} + x - 1\right)^{2}}\\
    &- \fsync^{2} x - \mu_{\mathrm{acc}} \left(\fsync^{2} - 1\right) =
    0.
  \end{aligned}
\end{equation}
We solve this equation for $x$ which gives the first three Lagrange
points. In \Secref{sec:lagrange-point-plot} we show these points for a
selection of $\fsync$.

In the potential acting on particles in the donor
(\Eqref{eq:roche_potential_COM_don}) we assume that the dynamical
timescale of the donor is much shorter than the timescale of the tides
induced by the secondary star and the non-synchronous rotation of the
donor \citep{limberSurfaceFormsMass1963,
  sepinskyEquipotentialSurfacesLagrangian2007}, and thus the potential
is approximately static. We express the validity of this approximation
as
\begin{equation}
  \label{eq:validity_equation}
  \eta_{\mathrm{static}} = \frac{P_{\mathrm{orb}}}{\tau_{\mathrm{dyn,\ don}} \alpha(e, f, \nu)} \gg 1,
\end{equation}
where $P_{\mathrm{orb}}$ is the orbital period of the system,
$\tau_{\mathrm{dyn,\ don}} = \sqrt{R^{3}/2GM_{\mathrm{don}}}$ is the
dynamical timescale of the donor where $R$ is its radius and
$M_{\mathrm{don}}$ is its mass, $G$ is the gravitational constant, and
\begin{equation}
  \label{eq:alpha_sepinsky_circular}
  \alpha(f, e=0, \nu=0) = \left|1-\fsync\right|
\end{equation}
is generally a function of synchronicity $\fsync$, eccentricity $e$
and mean anomaly $\nu$, but here we focus on circular systems
(i.e. $e=0$, $\nu$ is irrelevant)
\citep{sepinskyEquipotentialSurfacesLagrangian2007}.
$\alpha = \tau_{\mathrm{tide}} \omega/ 2\pi$ captures the timescale,
$\tau_{\mathrm{tide}}$, on which tides induced by asynchronous
rotation operate. If $\eta_{\mathrm{static}} \gg 1$, the response of
the donor to a change in the potential is much faster than the
timescale of the tides induced by the asynchronous rotation of the
donor. The potential can then be regarded as static.

\subsection{Mass-stream particle properties}
\label{sec:mass-stream-particle}
In this section we describe the relevant properties of the particles
in the mass stream at and around the first Lagrange point, L1.

\subsubsection{Thermal velocity of stream particles at L1}
\label{sec:therm-veloc-stre-1}
The initial velocity with which material flows through L1 is set by
the thermal velocity of the material at L1
\citep{warnerLocationSizeHot1972, lubowGasDynamicsSemidetached1975,
  flanneryLocationHotSpot1975}. The thermal velocity, $\vthermal$,
depends on the properties of the photosphere of the donor,
\begin{equation}
  \label{eq:sigma_thermal}
  \vthermal = \tilde{v}_{\mathrm{thermal}}/a\omega = \sqrt{\frac{3kT_{\mathrm{eff,\ don}}}{m}} \frac{1}{a\omega}= \sqrt{\frac{3kT_{\mathrm{eff,\ don}}}{\mu_{\mathrm{phot,\ don}}\,m_{\mathrm{a}}}}\frac{1}{a\omega},
\end{equation}
where $\tilde{v}_{\mathrm{thermal}}$ is the dimensionful thermal
velocity, $k$ is the Boltzmann constant, $T_{\mathrm{eff,\ don}}$ is
the effective temperature of the donor, $m$ and
$\mu_{\mathrm{phot,\ don}}$ are the average mass and the mean
molecular weight of the particles in the photosphere respectively,
$m_{\mathrm{a}}$ is the atomic mass unit, $a$ is the semi-major axis
of the system and $\omega$ is the orbital frequency of the
system. Here we have assumed the equation of state behaves like an
ideal gas.

\subsubsection{Stream surface area at L1}
\label{sec:stream-surface-area}
\begin{figure}
    \centering \includegraphics[width=\columnwidth]{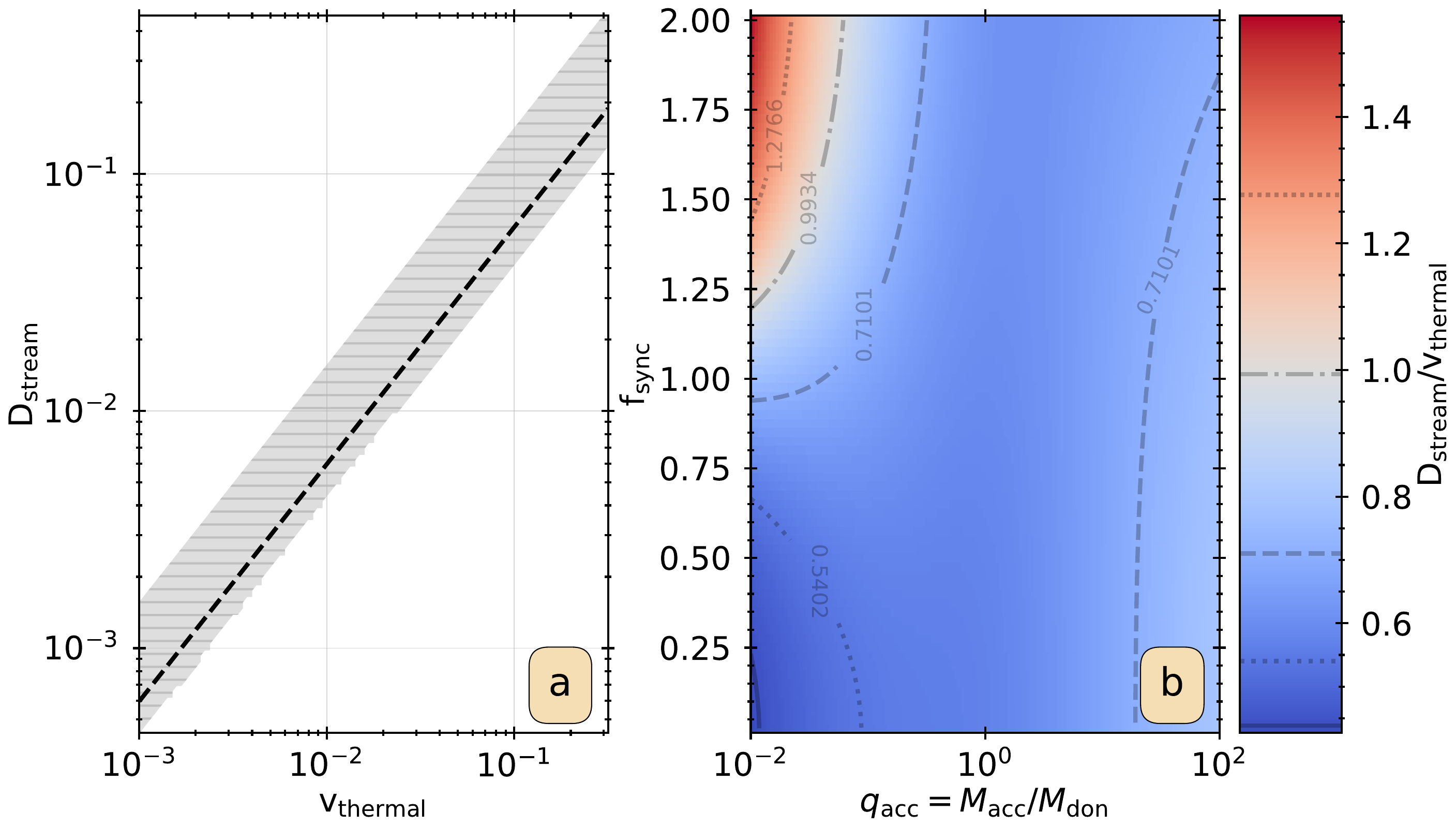}
    \caption{(a): The diameter of the mass-stream, $\streamdiameter$
      (ordinate), as a function of the thermal-velocity, $\vthermal$
      (abscissa). The black dashed line indicates the mass-stream
      diameter for $\qacc = 1$ and $\fsync = 1$, and the grey
      horizontal-dashed region indicates the range of
      $\streamdiameter$ spanned by the ranges of $\qacc$ and $\fsync$
      adopted in our study. (b): Ratio of $\streamdiameter$ to
      $\vthermal$ as a function of mass ratio $\qacc$ (abscissa) and
      synchronicity factor $\fsync$ (ordinate).}
  \label{fig:combined_area_plot}
\end{figure}

The mass-transfer stream at L1 has a non-zero surface area such that
particles are distributed around L1. We calculate the surface area of
the stream, $A_{\mathrm{stream}}$
\citep{meyerModelStandstillCamelopardalis1983,
  ritterTurningMassTransfer1988, davisMassTransferEccentric2013,
  davisBinaryEvolutionUsing2014}, assuming a circular cross-section,
as,
\begin{equation}
  \label{eq:stream_surface_area}
  \begin{aligned}
    A_{\mathrm{stream}} = \tilde{A}_{\mathrm{stream}}/a^{2} &=
    \frac{2\mathrm{\pi}\,k\,T_{\mathrm{eff,\
          don}}}{\mu_{\mathrm{phot,\
          don}}\,m_{a}}\frac{\qacc}{\mu_{\mathrm{acc}} \omega^{2}}\\
    &\times\left\{g(\qacc)\left[g(\qacc)-\left(1+\qacc\right)\fsync^{2}\right]\right\}^{-1/2}\frac{1}{a^{2}},
  \end{aligned}
\end{equation}
where $\tilde{A}_{\mathrm{stream}}$ is the dimensionful stream area,
$\fsync$ is the synchronicity factor
(\Eqref{eq:synchronicity_factor}). The geometric factor, $g(\qacc)$,
is,
\begin{equation}
  \label{eq:geometry_at_L1}
  g(\qacc) = \frac{1}{d_{\mathrm{L1,\ don}}^{3}} + \frac{\qacc}{\left(1-d_{\mathrm{L1,\ don}}\right)^{3}},
\end{equation}
where $d_{\mathrm{L1,\ don}}$ is the distance from the centre of the
donor to L1 in terms of the separation of the binary system.
We reformulate the mass stream area in terms of the thermal-velocity
of the particle $\vthermal$ at L1, as,
\begin{equation}
  \label{eq:normalized_surface_area}
  \begin{aligned}
    A_{\mathrm{stream}} =
    \frac{2\mathrm{\pi}}{3}&\vthermal^{2}\left(1+\qacc\right)\\
    &\times\left\{g(\qacc)\left[g(\qacc)-\left(1+\qacc\right)\fsync^{2}\right]\right\}^{-1/2}.
\end{aligned}
\end{equation}
\Figref{fig:combined_area_plot} (a) shows the stream diameter,
$\streamdiameter$, as a function of the thermal-velocity, $\vthermal$,
mass ratio and synchronicity factor. The solid line indicates
$\qacc = 1$ and $\fsync = 1$, and the grey transparent area indicates
the extent of diameters spanned by the ranges of $\qacc$ and
$\fsync$. At fixed thermal-velocity, the extent of stream diameters
spans about a factor of $3-4$, and from $\vthermal \gtrapprox 0.06$
the diameter of the stream reaches a significant fraction
($\streamdiameter \gtrapprox 0.1$) of the separation of the
system. \Figref{fig:combined_area_plot} (b) shows the ratio of the
stream diameter and the thermal-velocity as a function of mass ratio
$\qacc$ and $\fsync$. Overall, in most of the parameter space this
ratio does not exceed $\approx\,0.7$, except for $\fsync > 1.1$ and
$\qacc < 10^{-1}$. Only in the extreme case of $\fsync \gtrapprox 1.5$
and $\qacc \approx 10^{-2}$ does the ratio exceed unity, indicating
that for most of the parameter space, the stream diameter is close to
that of the case of a synchronous and equal mass-ratio system.

The density distribution in the stream at L1 is approximately Gaussian
\citep{lubowGasDynamicsSemidetached1975,
  raymerThreedimensionalHydrodynamicSimulations2012},
\begin{equation}
  \label{eq:stream_density_distribution}
  \xi\left(\tilde{l}\right) = \eta e^{-\tilde{l}^{\,2}/2\sigma^{2}},
\end{equation}
where reduced position offset $|\tilde{l}| < 1$ and position offset
$l = \tilde{l}\,\sqrt{A/\mathrm{\pi}}$ $\sigma = 0.4$ such that at
$l = \pm 1$ the density equals that of the photosphere of the donor
\citep{davisBinaryEvolutionUsing2014}, and,
\begin{equation}
  \label{eq:eta_normalisation}
  \eta = \frac{1}{\int_{-1}^{1}\xi\left(\tilde{l}\right)\,\mathrm{d}\tilde{l}}.
\end{equation}

In a given system with $\qacc$, $\fsync$ and $\vthermal$, we calculate
trajectories with $N_{A\, \mathrm{stream}}$ equally-spaced initial
positions relative to L1, sampled in the range [$-\streamdiameter/2$,
$\streamdiameter/2$], and weigh each according to
\Eqref{eq:stream_density_distribution}. We use these trajectories to
calculate averaged quantities.


\section{Method}
\label{sec:Method}
In this section we explain how we calculate the trajectory of a
particle and how we classify its trajectory in the potential of
\Secref{sec:roche-potent-reduc}, as well how we calculate the relevant
properties of mass transfer in a binary population.

\subsection{Particle trajectories in the Roche potential}
In the following sections we explain our method of calculating the
trajectory of particles in the Roche potential.

\subsubsection{Reduced three-body equations and ballistic integration}
\label{sec:reduced-three-body}
The trajectory of a particle is found by integrating the equations of
motion of the particle in the rotating frame,
\begin{equation}
  \label{eq:equations_of_motion_x}
  \ddot{x} = -\frac{\partial \Phi}{\partial x} + 2 \dot{y}
\end{equation}
and
\begin{equation}
  \label{eq:equations_of_motion_y}
  \ddot{y} = -\frac{\partial \Phi}{\partial y} - 2 \dot{x},
\end{equation}
where $x$ and $y$ are the position components of the particle with
respect to the centre of mass of the binary system, $\dot{x}$ and
$\dot{y}$ the velocity components of the particle, $\ddot{x}$ and
$\ddot{y}$ the acceleration components of the particle, and $\Phi$ is
the potential experienced by the particle
(\Eqref{eq:roche_potential_COM_particle}). The first terms in
equations \ref{eq:equations_of_motion_x}
and~\ref{eq:equations_of_motion_y} are the gradient of the potential
and the second terms are the Coriolis force in each direction.

We calculate the specific energy and angular momentum of the particle
in the inertial frame, with respect to the centre of mass, using
quantities defined in the co-rotating frame,
\begin{equation}
  \label{eq:energy_particle}
  \varepsilon = \Phi + \frac{1}{2}\left(\dot{x}^{2} + \dot{y}^{2}\right) + x^{2} + y^{2} + x\dot{y}-\dot{x}y
\end{equation}
and
\begin{equation}
  \label{eq:angular_momentum_particle}
  h = x^{2} + y^{2} + x\dot{y}-\dot{x}y,
\end{equation}
respectively, in units $a^{2}\omega^{2}$ and $a^{2}\omega$.

In the circular reduced three-body problem, the only first integral of
motion is the Jacobi constant \citep{moulton1914,
  ovendenUseJacobiIntegral1961},
\begin{equation}
  \label{eq:jacobi_constant}
  C = \Phi + \frac{1}{2}\left(\dot{x}^{2} + \dot{y}^{2}\right) = \varepsilon-h,
\end{equation}
which is the difference between the energy and the angular momentum of
the particle with respect to the observer frame. We use the Jacobi
constant to determine the accuracy of our calculations.

\subsubsection{Initial position and velocity}
\label{sec:init-posit-veloc}
We integrate trajectories from a given initial position,
$\boldsymbol{x}_{i}$, relative to L1 and initial velocity,
$\boldsymbol{v}_{i}$, relative to the co-rotating frame.

The initial position is,
\begin{equation}
  \label{eq:initial_position}
  \boldsymbol{x}_{i} = \boldsymbol{x}_{\mathrm{minor\ offset}} + \boldsymbol{x}_{\mathrm{stream\ area\ offset}}.
\end{equation}
Here $\boldsymbol{x}_{\mathrm{minor\ offset}} = [\delta x,\ 0]$ is a
minor offset to prevent the particle starting exactly on L1, where
$\delta x = |x_{\mathrm{L}_{1}}-x_{\mathrm{acc}}|/100$,
$x_{\mathrm{acc}}$ is the position of the accretor, and
$x_{\mathrm{L}_{1}}$ is the $x$-coordinate of L1, and
$\boldsymbol{x}_{\mathrm{stream\ area\ offset}} = [x_{\mathrm{stream\
    area\ offset}},\ 0]$ is an offset to sample the surface area of
the stream at L1 (\Secref{sec:stream-surface-area}).

\begin{figure}
  \centering
  \includegraphics[width=\columnwidth]{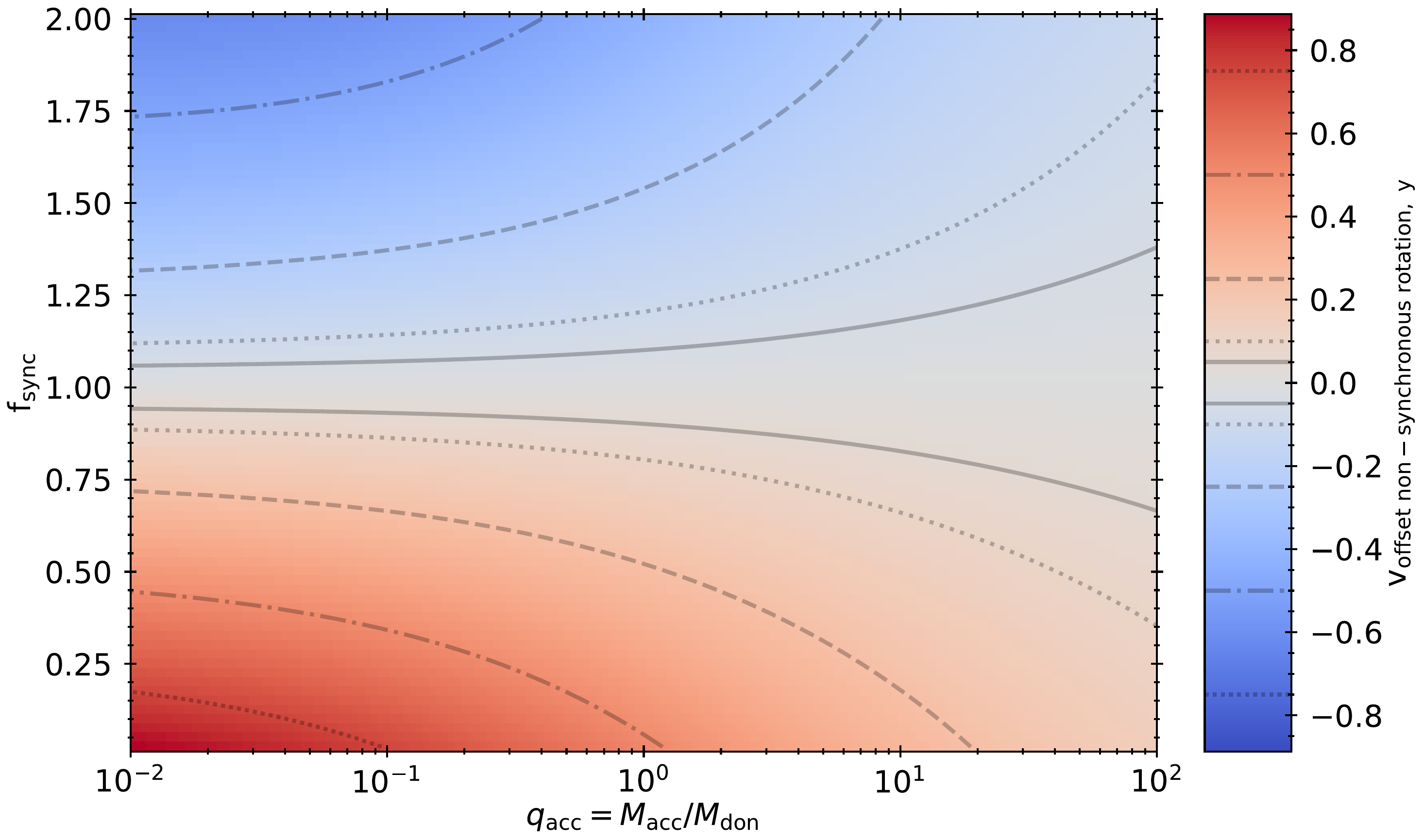}
  \caption{The $y$-component of the velocity offset due to
    non-synchronous rotation of the donor
    $\boldsymbol{v}_{\mathrm{non-synchronous\ offset}}$ as a function
    of $\qacc$ (abscissa) and $\fsync$ (ordinate). Lines of equal and
    opposite velocities are indicated with black transparent lines,
    where solid indicates 0.05 and -0.05, dotted indicates 0.1 and
    -0.1, dashed indicates 0.25 and -0.25, dashed-dotted indicates 0.5
    and -0.5, and fine-dotted indicates 0.75 and -0.75. This figure
    illustrates the magnitude of the velocity due to asynchronous
    rotation in units of the orbital velocity. Generally, with higher
    mass-ratio, the lower the velocity offset due to asynchronous
    rotation is. This is due to the increasingly smaller size of the
    donor relative to the system. At low mass-ratio there is a clear
    asymmetry, with at low ($\fsync \sim 0.2$) synchronicity the
    velocity offset is larger in absolute terms than at high
    ($\fsync \sim 1.8$) synchronicity. This is due to that the L1
    point moves outward for lower synchronicity which increases the
    velocity offset.}
\label{fig:non_synchronous_rotation_schematic}
\end{figure}

The initial velocity is,
\begin{equation}
  \label{eq:initial_velocity}
  \boldsymbol{v}_{i} = \boldsymbol{v}_{\mathrm{non-synchronous\ offset}} + \boldsymbol{v}_{\mathrm{thermal}},
\end{equation}
where $\boldsymbol{v}_{\mathrm{thermal}} = [\vthermal,\ 0]$ is the
thermal velocity of the particle in the stream
(\Secref{sec:therm-veloc-stre-1}).
$\boldsymbol{v}_{\mathrm{asynchronous\ offset}} = [0, \ (\fsync-1)\
d_{\mathrm{don,\ L1}}]$ is the velocity relative to the co-rotating
frame due to the non-synchronous rotation of the donor, and
$d_{\mathrm{don,\ L1}}$ is the normalised distance from the centre of
the donor to L1 \citep{kruszewskiExchangeMatterClose1964}. The
synchronicity changes the tangential velocity offset in two ways. It
determines the angular velocity offset, $\fsync-1$, and it affects the
distance, $d_{\mathrm{don,\ L1}}$ (\Eqref{eq:critical_surface} and
\Figref{fig:lagrange_point_plot}). We show the $y$-component of
$\boldsymbol{v}_{\mathrm{non-synchronous\ offset}}$ as a function of
$\fsync$ and $\qacc$ in
\Figref{fig:non_synchronous_rotation_schematic}. Generally, with
higher mass-ratio, the lower the velocity offset due to asynchronous
rotation is. This is due to the increasingly smaller size of the donor
relative to the system. At low mass-ratio this effect is reversed, and
there is a clear asymmetry, with at low ($\fsync \sim 0.2$)
synchronicity the velocity offset is larger in absolute terms than at
high ($\fsync \sim 1.8$) synchronicity. This is due to that the L1
point moves outward for lower synchronicity which increases the
velocity offset.

\begin{figure}
  \centering
  \includegraphics[width=\columnwidth]{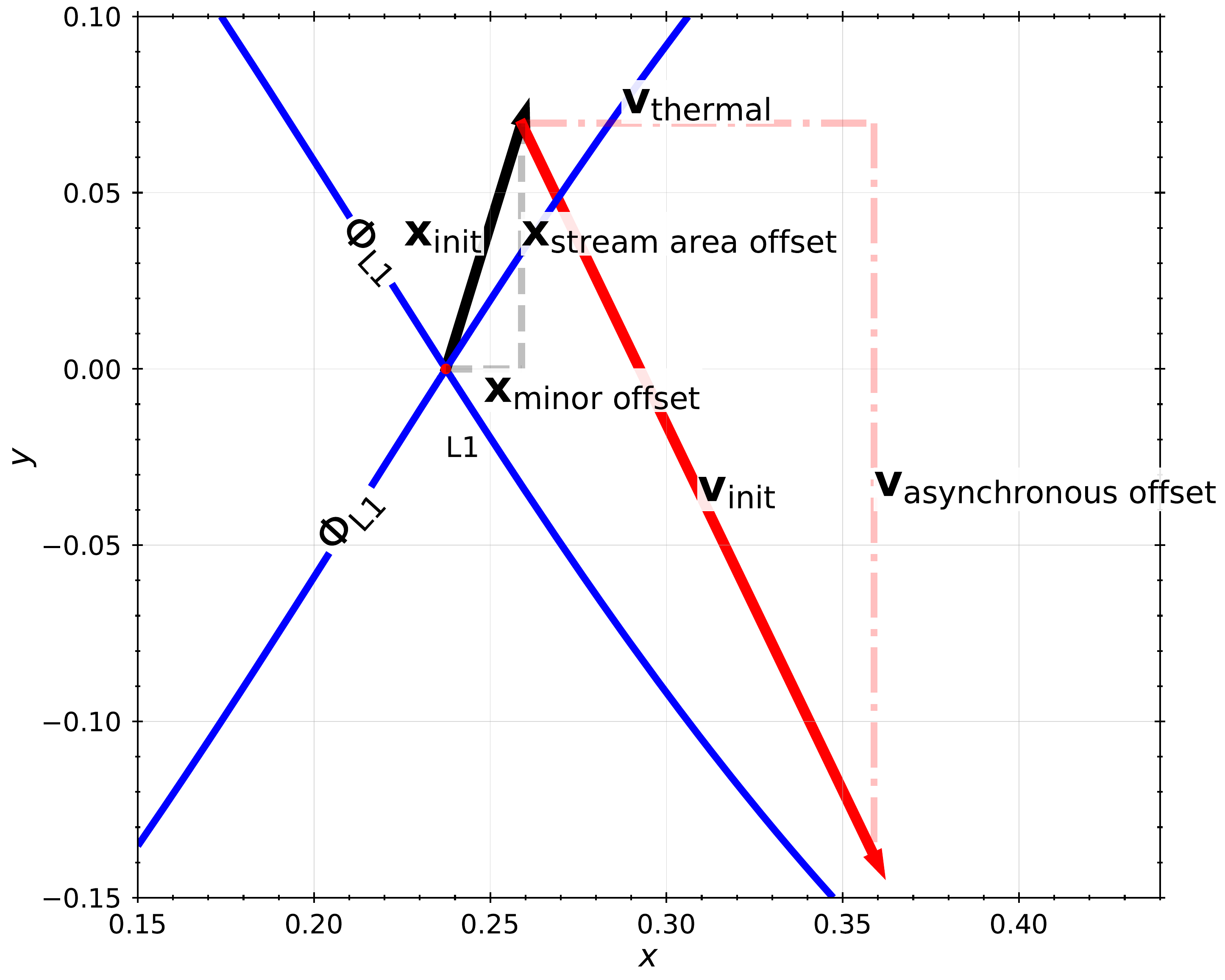}
  \caption{
    The initial position and initial
    velocity of the particle in our ballistic integrations.
    The solid blue line indicates the equipotential corresponding to the Roche lobe.
    The grey-dashed lines indicate the components of the initial position
    due to the mass stream sampling and the minor offset. The solid black
  line indicates the initial position vector relative to L1.
  The red-dashed lines indicate the components of the initial
  velocity due to thermal velocity and asynchronous rotation. The
  solid red line indicates the initial velocity vector relative to the
  rotation rate of the system.
  The length of the vectors is not to scale and only serves for
  illustration purposes, but the L1 equipotential line corresponds to
  a sub-synchronously rotating donor ($\fsync = 0.6$) in an equal mass
  binary ($\qacc = 1$) with a hot stream ($\vthermal = 0.1$). The
  position offset is enhanced for illustrative purpose.  }
\label{fig:initial_position_velocity_schematic}
\end{figure}

We show the initial position and velocity components for an equal mass
binary ($\qacc = 1$) with a sub-synchronously rotating donor
($\fsync = 0.6$) and a hot stream ($\vthermal = 0.1$) in
\Figref{fig:initial_position_velocity_schematic}, where the thick
black and red arrows indicate the position and momentum vectors
respectively, and the thin dashed lines indicate their component
vectors.

\subsection{Integration method}
\label{sec:integration-method}
We calculate ballistic trajectories by solving the equations of motion
(equations \ref{eq:equations_of_motion_x}
and~\ref{eq:equations_of_motion_y}) with an explicit 4th order
Runge-Kutta method using the \textsc{dopri5} ODE solver
\citep{hairerSolvingOrdinaryDifferential2008} from the \textsc{Python}
\textsc{SciPy} package
\citep{virtanenSciPyFundamentalAlgorithms2020}. We use an adaptive
method that rejects the model and halves the time step if the relative
error on the Jacobi constant exceeds $10^{-6}$. We either terminate
the integration based on a classification of the trajectory
(\Secref{sec:classifying-and-averaging}) or when the integrator fails
to conserve the Jacobi constant and the time step is shorter than
$10^{-20} \omega^{-1}$.

\subsection{Classifying and averaging trajectories}
\label{sec:classifying-and-averaging}
For each set of parameters $[\vthermal$, $\fsync$, $\qacc]$ we
integrate $N_{A,\ \mathrm{stream}}$ trajectories, each with a position
offset $\boldsymbol{x}_{\mathrm{stream\ area\ offset,\ i}}$ and a
weighting $w_{A_{\mathrm{stream}}}$
(\Secref{sec:mass-stream-particle},
\Eqref{eq:stream_density_distribution}).

\begin{figure}
  \centering
  \includegraphics[width=\columnwidth]{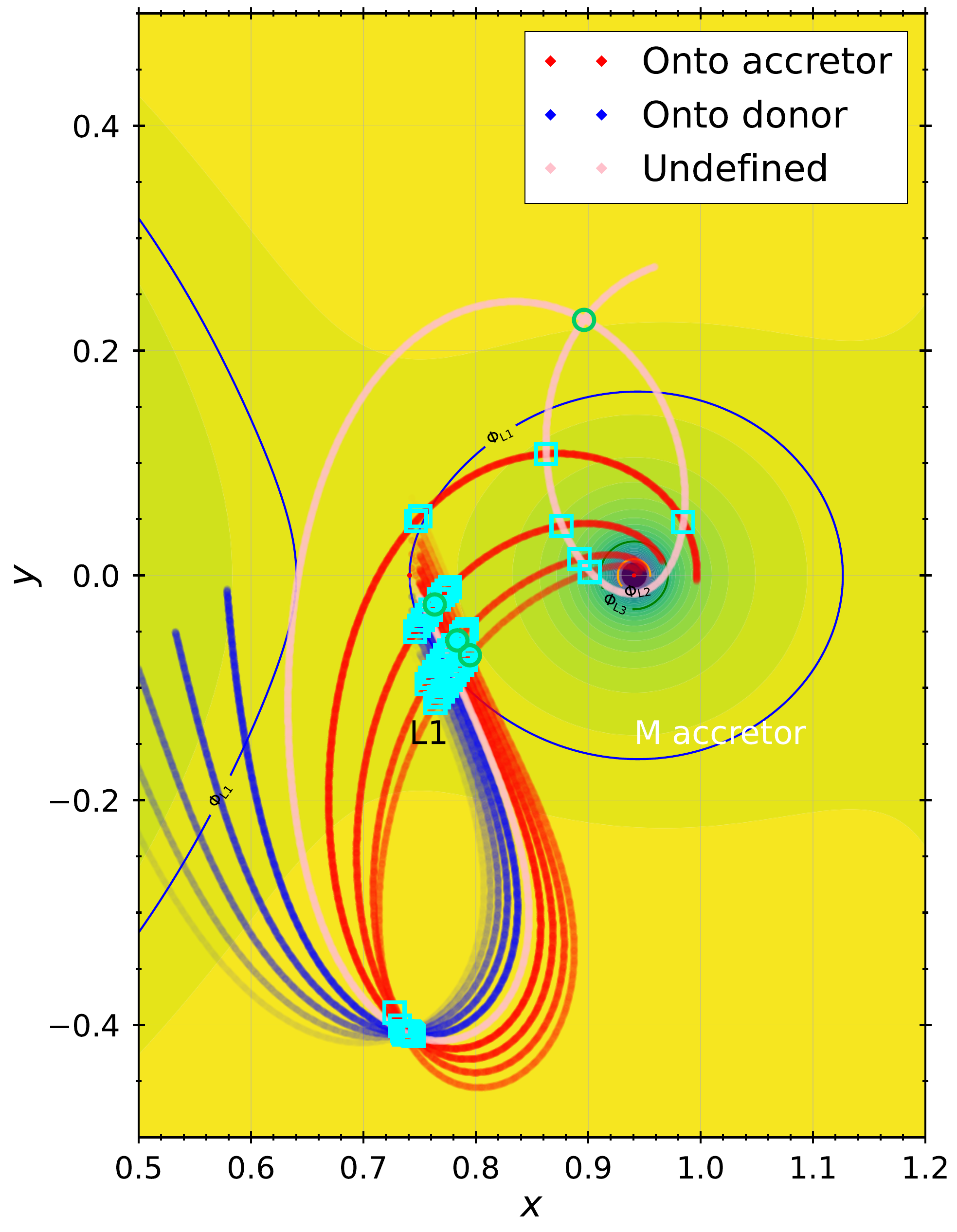}
  \caption{Trajectories for $\qacc = 10^{-1.2}$, $\fsync = 0.22$, and
    $\vthermal = 10^{-0.5}$, and $N_{A,\ \mathrm{stream}} = 12$
    equally spaced sampled trajectories around the mass transfer
    stream centre. The red trajectories are classified as accreting
    onto the accretor. The dark-blue trajectories are classified as
    accreting back onto the donor. The green circles are locations of
    self-intersection and the light-blue squares are locations of
    intersection with other trajectories. This illustrates the
    different trajectory classes as well as the intersections that
    occur between the trajectories.}
  \label{fig:trajectory_classification_overview}
\end{figure}

The trajectories are classified by their behaviour and
outcome. Particles accrete onto either the accretor or the donor, or
are lost from the system. Classification happens during integration,
and changes how the calculation is terminated.
\begin{enumerate}
\item \textbf{Accretion onto accretor}: Classified by motion towards
  the accretor, away from the donor, away from L1, into a deeper
  potential than L1, and within the Roche lobe of the
  accretor. Terminated at the moment the particle starts moving away
  from the accretor.
\item \textbf{Accretion onto donor}: Classified by motion towards the
  donor, away from the accretor, away from L1, into a deeper potential
  than L1 and within the Roche lobe of the donor. Terminated at the
  moment of classification.
\item \textbf{Lost from system}: Classified by distance from centre of
  mass $>3$. Terminated on classification.
\end{enumerate}
We show an example of different classifications in
\Figref{fig:trajectory_classification_overview}

Of the trajectories that are not terminated for numerical reasons, we
calculate weighted averages of their properties.
We determine the fraction, $\beta_{\mathrm{acc}}$, of our trajectories
that accrete onto the accretor,
\begin{equation}
  \label{eq:fraction_accretion_accretor}
  \beta_{\mathrm{acc}} = \frac{\sum_{i \in \mathcal{C}} \delta_{i} w_{A\, \mathrm{stream},\ i}}{\sum_{i \in \mathcal{C}} w_{A\, \mathrm{stream},\ i}},
\end{equation}
where $w_{A\, \mathrm{stream},\ i}$ is the weight of the sampled
position offset along the mass stream cross-section, $\mathcal{C}$ is
the set of classified trajectories, and
\begin{equation}
  \label{eq:delta_accretion_accretor}
  \delta_{i} = \begin{cases}
    1 \quad &\mathrm{if\ trajectory_{\it{i}}\ classification\ is\ accretion\ onto\ accretor,\ and}\\
    0 \quad &\mathrm{otherwise}.\\
     \end{cases}
\end{equation}
We calculate the fraction that accretes back onto the donor,
$\beta_{\mathrm{don}}$, in the same way as $\beta_{\mathrm{acc}}$
(equations \ref{eq:fraction_accretion_accretor}
and~\ref{eq:delta_accretion_accretor}).
We calculate the fraction of trajectories that is lost from the system
or classified as
$\beta_{\mathrm{lost}} = 1 - \beta_{\mathrm{acc}} -
\beta_{\mathrm{don}}$.

We denote the total weight of all trajectories that are successfully
categorised with,
\begin{equation}
  \label{eq:all_succesfull}
  w_{\mathrm{successful}} = \sum_{i \in \mathcal{C}} w_{A\, \mathrm{stream},\ i},
\end{equation}
and the total weight of all those that fail or are rejected with,
\begin{equation}
  \label{eq:all_fail}
  w_{\mathrm{fail}} = 1-w_{\mathrm{successful}},
\end{equation}
which can occur when our integrator is not able to conserve the Jacobi
constant within the minimum time step threshold
(\Secref{sec:integration-method}).

With these weights and fractions we can quickly identify how
successful our calculations are for a given set of parameters
$[\vthermal$, $\fsync$, $\qacc]$, and how the trajectories are
classified.

\subsection{Intersecting orbits}
\label{sec:intersecting-orbits}
At each coordinate in our parameter space we evolve a set of
trajectories sampled along the stream diameter
(\Secref{sec:stream-surface-area}). We treat each of these
trajectories independently, even though these trajectories can cross
either themselves or each-other.
To find intersecting trajectories we use the
\textit{SweepIntersectorLib}\footnote{\url{https://github.com/prochitecture/sweep_intersector}},
which is a Python implementation of the \textit{Sweep line algorithm}
of \citet{mehlhornImplementationSweepLine1994}.
Orbits that self-intersect are always flagged as such, but only
trajectories with an angle of intersection, $\thetaintersect$, with
another trajectory larger than
$\thetathreshold = \thetathresholdvalue$ get flagged as intersecting
with others. While the exact threshold angle is not strongly
motivated, we argue that low-angle intersecting trajectories would
merge and be well approximated by their weighted average, high angles
of intersection could significantly change the outcome of both
trajectories.
\Figref{fig:trajectory_classification_overview} shows the different
types of intersection for a system with $\qacc = 10^{-1.2}$,
$\fsync = 0.22$, and $\vthermal = 10^{-0.5}$, and
$N_{A,\ \mathrm{stream}} = 12$ equally spaced sampled trajectories.
At each coordinate we record the weighted fraction of trajectories
that self-intersect, $\selfintersect$, as well as those that intersect
with other trajectories, $\otherintersect$, if their intersection
angle exceeds the threshold.

\subsection{Radii, specific angular momenta and torques}
\label{sec:onto-accretor-stream-properties}
When the mass stream misses the accretor it loops back around and form
an accretion disk. This disk forms at the circularisation radius,
defined as the radius where the specific angular momentum,
$h_{\mathrm{stream,\ min,\ acc}}$, with respect to the accretor at the
moment of closest approach, $\rmin$, equals that of a circular
Keplerian orbit around the accretor with radius
\begin{equation}
  \label{eq:circularisation radius}
  \rcirc = \frac{h_{\mathrm{stream,\ min,\ acc}}^{2}}{\mu_{\mathrm{acc}}}.
\end{equation}
The specific angular momentum of a particle with respect to the
accretor is,
\begin{equation}
  \label{eq:angmom_wrt_acc}
  \begin{aligned}
    h_{\mathrm{acc}} &= (x-x_{\mathrm{acc}})^{2} + (y-y_{\mathrm{acc}})^{2} + (x-x_{\mathrm{don}})\dot{y}-\dot{x}(y-y_{\mathrm{don}}),\\
    &= (x-1+\mu_{\mathrm{acc}})^{2} + y^{2} +
    (x-1+\mu_{\mathrm{acc}})\dot{y}-\dot{x}y.
  \end{aligned}
\end{equation}
We calculate $h_{\mathrm{stream,\ min,\ acc}}$ by evaluating
\Eqref{eq:angmom_wrt_acc} at the radius of closest approach.

\begin{figure}  \includegraphics[width=\columnwidth]{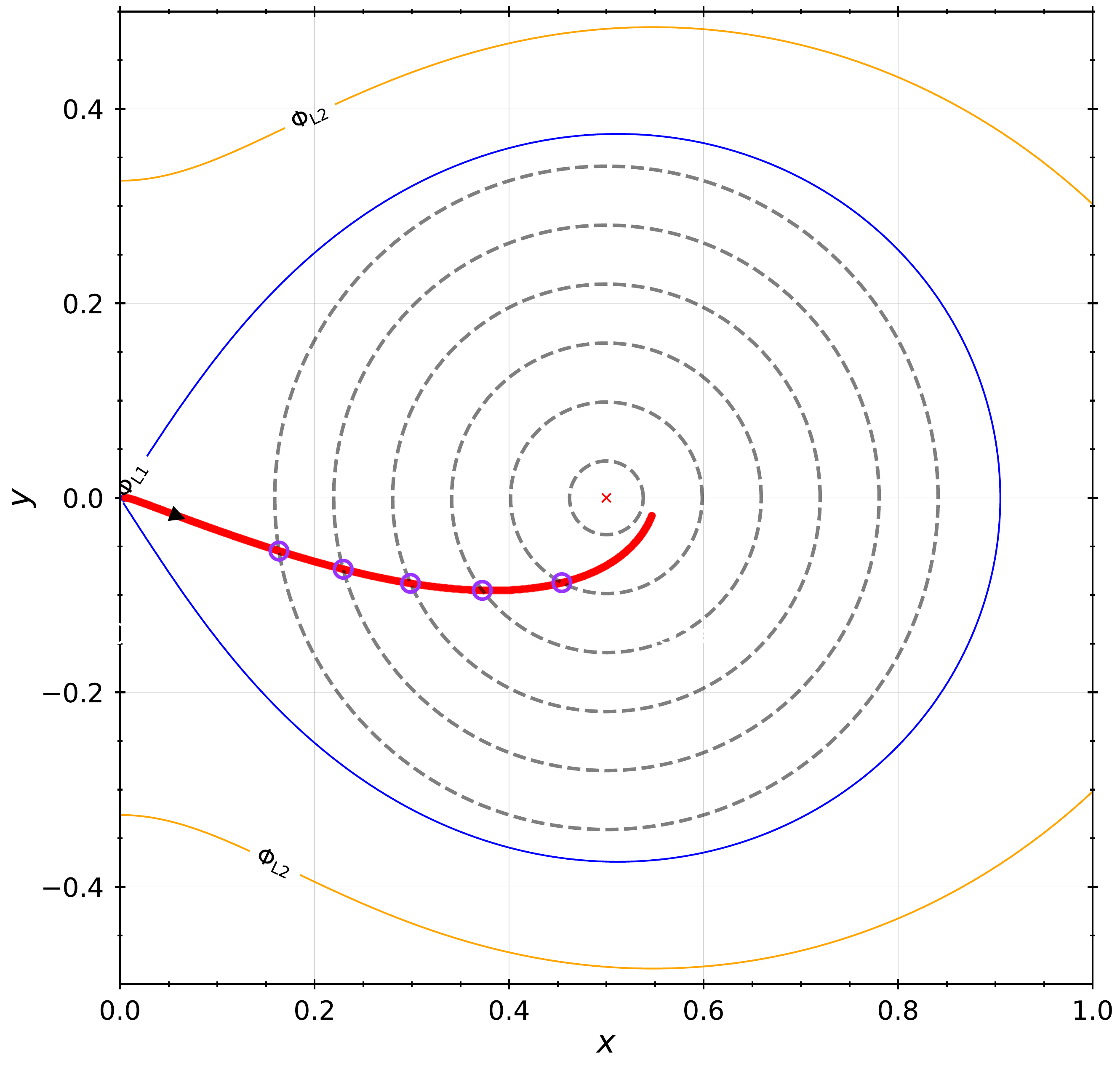}
  \caption{Locations at which we record the specific angular momentum
    of the stream along the stream trajectory. The solid blue line
    indicates the equipotential surface coinciding with L1 (i.e.
    Roche-Lobe). The solid orange line indicates the equipotential
    surface coinciding with L2. The solid red line indicates the
    stream trajectory. The dashed grey lines indicates equidistant
    lines from the centre of mass of the accretor. The purple circles
    indicate the intersection point of the stream with the
    equidistant. The black arrow indicates the direction of travel of
    the stream. The number and radii of the equidistances are chosen
    to illustrate the situation, but may differ in our datasets. We
    record the specific angular momentum along the stream to determine
    the torque on the accretor in the case of direct-impact
    accretion.}
  \label{fig:stream_interpolation_schematic}
\end{figure}

While in our ballistic trajectory calculations we implicitly assume
that the stream will miss the accretor and will form an accretion disk
around the star, many interacting binaries actually transfer mass
through direct-impact accretion. When the stream collides with the
accretor, i.e. direct-impact accretion
$r_{\mathrm{stream}} < r_{\mathrm{accretor}}$, the specific angular
momentum of the stream (\Eqref{eq:angmom_wrt_acc}) at that point is
different than at the point of closest approach during disk formation.
We calculate the specific angular momentum of the stream with respect
to the accretor as a function of the distance to the centre of the
accretor. This allows a more accurate determination of the specific
angular momentum accretion rate when the stream directly impacts the
accretor.
We record the (averaged) specific angular momentum of the stream at
fixed distances from the accretor, with a minimum distance of
$d_{\mathrm{stream\, min}}$, a maximum distance of
$d_{\mathrm{stream\, max}}$ at $N_{\mathrm{radii}}$ equally spaced
radii, located at,
\begin{equation}
  \label{eq:distance_for_stream_interpolation}
  d_{\mathrm{stream\, i}} = d_{\mathrm{stream\, min}} + i \times \frac{(d_{\mathrm{stream\, max}} - d_{\mathrm{stream\,  min}})}{N_{\mathrm{radii}}}.
\end{equation}
Here $d_{\mathrm{stream\, i}}$ indicates the i-th radius from the
centre of the accretor in units of the Roche-lobe radius of the
accretor, at which we record the i-th specific angular momentum along
the stream $h_{\mathrm{stream\, i}}$. We show a schematic example of
the locations at which we record the specific angular momentum of the
stream in \Figref{fig:stream_interpolation_schematic}.

\subsubsection{Self-accretion torque}
\label{sec:self-accr-torq}
Accretion of (part of) the mass transfer stream back onto the donor
exerts a torque on the donor star. We calculate the specific angular
momentum of a particle at the moment of impact on the donor, with
respect to the donor,
\begin{equation}
  \label{eq:angmom_wrt_don}
  \begin{aligned}
    h_{\mathrm{don}} &= (x-x_{\mathrm{don}})^{2} + (y-y_{\mathrm{don}})^{2} + (x-x_{\mathrm{don}})\dot{y}-\dot{x}(y-y_{\mathrm{don}}),\\
    &= (x+\mu_{\mathrm{acc}})^{2} + y^{2} +
    (x+\mu_{\mathrm{acc}})\dot{y}-\dot{x}y.
  \end{aligned}
\end{equation}
We calculate the initial $h_{\mathrm{i,\ don}}$ and final
$h_{\mathrm{f,\ don}}$ specific angular momentum of a particle
accreting back onto the donor by evaluating \Eqref{eq:angmom_wrt_don}
with the initial and final positions and velocities respectively, and
we use these specific angular momenta to calculate the total torque on
the donor due to self-accretion.

\subsection{Properties of mass transfer in binary populations}
\label{sec:expl-param-rang}
To inform us of the ranges of $\qacc$, $\fsync$ and $\vthermal$ we
should cover, we evolve a binary population with the rapid binary
population synthesis framework \binaryc\
\citep{izzardNewSyntheticModel2004,
  izzardPopulationNucleosynthesisSingle2006,
  izzardPopulationSynthesisBinary2009, izzardBinaryStarsGalactic2018,
  izzardCircumbinaryDiscsStellar2022,
  hendriksBinaryCpythonPythonbased2023}, which is based on the
algorithm from \citet{hurleyComprehensiveAnalyticFormulae2000,
  hurleyEvolutionBinaryStars2002}, and makes use of the single star
models of \citet{polsStellarEvolutionModels1998} and provides
analytical fits to their evolution as in
\citet{toutRapidBinaryStar1997}.
Specifically relevant to this study are the tidal interactions between
binary stars. These are implemented as in
\citet{hurleyEvolutionBinaryStars2002}, in which dynamical tides are
based on \citep{zahnDynamicalTideClose1975, zahnTidalFrictionClose1977} and equilibrium tides are based
on \citet{hutTidalEvolutionClose1981}.

Our population contains binary systems with an initial primary mass
$M_{1}$, secondary mass $M_{2}$ and orbital period $P$, and we assign
weights to each system according to the distribution functions of
their birth properties of \citet{moeMindYourPs2017}.
$M_{1}$ is sampled logarithmically in the range 0.8 to 120 \solM.
$M_{2}$ is sampled from a flat mass-ratio distribution between
$0.1 \mathrm{M}_{\odot}/M_{1}$ and $1$.
$P$ is sampled from a logarithmically-spaced distribution of periods
between $1$ day and $10^{8}$ days.
We evolve
$N_{M_{1}} \times N_{M_{2}} \times N_{P} = 80 \times 80 \times 80$
binary systems sampled with the distributions described above at
near-solar metallicity ($Z = 0.02$).

During Roche-lobe overflow we record the mass transfer quantities
$\vthermal$, $\fsync$ and $\qacc$, and we weigh them by the
time spent transferring mass and the mass transferred,
\begin{equation}
  \label{eq:weightings}
  \begin{aligned}
    W_{\mathrm{time},\ i}\,&= p_{i} * dt\,\left[\mathrm{yr}\right]\\
    W_{\mathrm{mass},\ i}\,&= p_{i} * dt\,\dot{M}_{\mathrm{don}}\,\left[\mathrm{M}_{\odot}\right],\\
  \end{aligned}
\end{equation}
where $W_{\mathrm{time},\ i}$ is the time-weighted probability,
$W_{\mathrm{mass},\ i}$ is the mass-weighted probability, $p_{i}$ is
the probability of the $i$-th system according to the distribution
functions of \citet{moeMindYourPs2017} $dt$ is the time-step taken in
{\binaryc} and $\dot{M}_{\mathrm{don}}$ is the mass-transfer rate of
the donor.

Based on our results of our binary population, we determine the
parameter ranges for our ballistic interpolation calculations
(\Tabref{tab:interpolation_table_properties}). We use these ranges to
span a hypercube of initial parameters for our ballistic calculations.


\section{Results}
\label{sec:results}
We present our results in the following sections. First, we show our
binary population which contain data on the properties of the mass
transfer in many systems. We then take these results and use them to
determine the ranges of the parameters in our trajectory
calculations. We then show our ballistic trajectory results for
``cold'' (narrow and slow) and ``hot'' (wide and fast) streams.

\subsection{Mass transfer in binary populations}
\label{sec:mass-transfer-binary}
With the results of our stellar population generated in
\Secref{sec:expl-param-rang}, we calculate the ranges of the
parameters of interest in a population of interacting binary
systems. Our results include the average time spent, and the average
mass transferred, of each system configuration.

\begin{figure}
  \centering
  \includegraphics[width=\columnwidth]{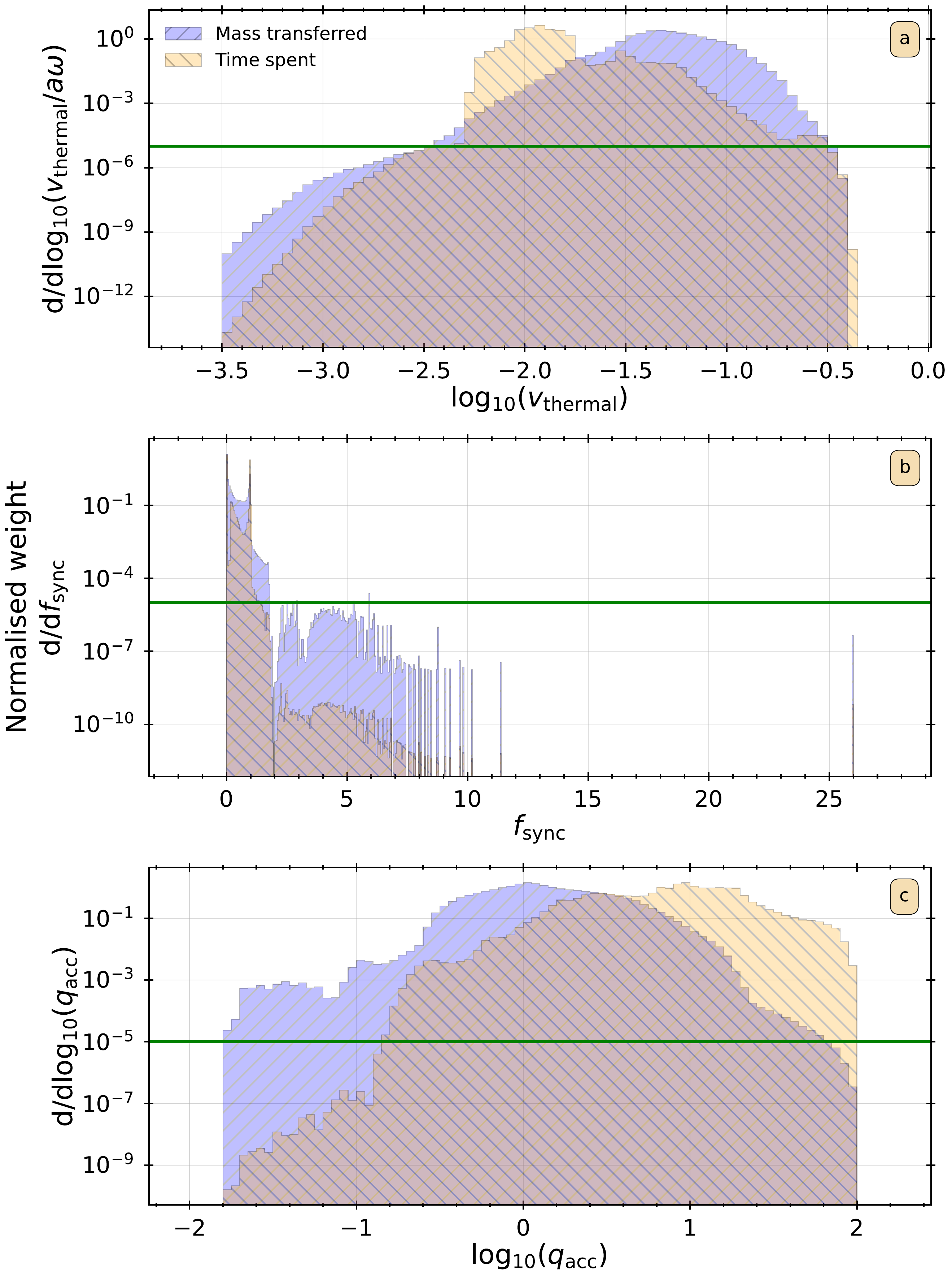}
  \caption{Parameter distributions obtained from our binary stellar
    population simulations (\Secref{sec:expl-param-rang}). (a) shows
    the thermal-velocity $\mathrm{log}_{10}\left(\vthermal\right)$,
    (b) shows the synchronicity factor $\fsync$ and (c) shows the mass
    ratio $\mathrm{log}_{10}\left(\qacc\right)$. The blue hatched
    areas show the mass-weighted distribution of each parameter, and
    the orange hatched areas show the time-weighted distribution. The
    green horizontal line indicates the weighted probability density
    of $10^{-5}$.}
  \label{fig:exploration_results_parameters}
\end{figure}

\Figref{fig:exploration_results_parameters} shows the distributions of
the parameters of interest, weighted either by time spent transferring
mass or mass transferred. We normalise the area under each of the
curves to unity, and we define values $<10^{-5}$ as rare and indicate
them by a green horizontal line.

\Figref{fig:exploration_results_parameters} (a) shows the logarithmic
thermal-velocity, $\mathrm{\log}_{10}\left(\vthermal\right)$,
distributions. All systems have thermal-velocities between $10^{-3.5}$
and $10^{-0.5}$.
\Figref{fig:exploration_results_parameters} (b) shows the
synchronicity fraction, $\fsync$, distributions. These are
mostly between $0$ and $2$, with a peak around both $0$ and $1$ for
both the time spent transferring mass and mass-transferred
weights. While the time-spent distribution peaks at synchronous
rotation rates ($\fsync = 1$), the mass-transferred
distribution peaks at very sub-synchronous rotation rates
($\fsync \sim 0$). There is a large tail of synchronicity
fractions from $\fsync = 2$ to
$\fsync \simeq 10$, but their probability is low.
\Figref{fig:exploration_results_parameters} (c) shows the mass ratio,
$\mathrm{\log}_{10}(\qacc)$, distribution. We see a single main range
between $\mathrm{\log}_{10}(\qacc) = -2$ and $2$ for both the time
spent and mass transferred weights. The data show that at small mass
ratios ($\qacc < 1)$ hardly any time is spent transferring mass
(probabilities up to $10^{-4}$), while at larger mass ratios
($\qacc > 1$) the opposite is true. This is understood by the mass
ratio reversal during mass transfer and the transition from thermal
timescale mass transfer (high mass-transfer rate, short time) to
nuclear timescale mass transfer (low mass-transfer rate, long time).

\begin{figure}
  \centering
  \includegraphics[width=\columnwidth]{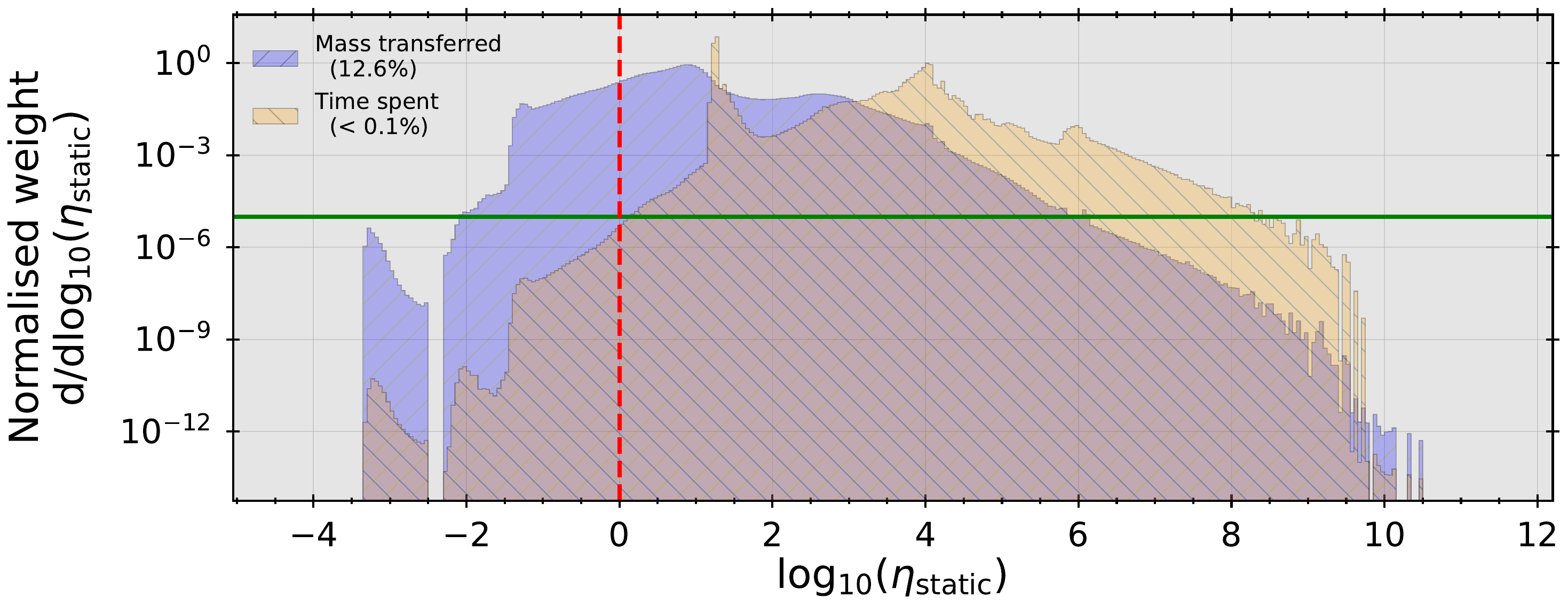}
  \caption{Normalised weighted distribution of
    $\mathrm{log}_{10}\left(\eta_{\mathrm{static}}\right)$, the
    quantity that captures the validity of the static tides
    approximation (equations \ref{eq:validity_equation} and
    \ref{eq:alpha_sepinsky_circular}). Blue indicates weighted by
    mass-transferred, orange indicates weighted by time-spent
    transferring mass (\Eqref{eq:weightings}). The regions where both
    histograms overlap are coloured darker red. The green horizontal
    line indicates the normalised weight of $10^{-5}$ and the
    red-dashed vertical line indicates
    $\mathrm{log}_{10}\left(\eta_{\mathrm{static}}\right) = 0$, above
    which the tides induced by the non-synchronous rotation are
    approximately static. The legend includes the fraction of the
    total weighted mass-transferred or time-spent with
    $\mathrm{log}_{10}\left(\eta_{\mathrm{static}}\right) < 0$.}
  \label{fig:exploration_results_alpha}
\end{figure}

We show the distributions of the logarithm of the ratio of the
dynamical timescale of the donor to the tidal timescale,
$\mathrm{log}_{10}\left(\eta_{\mathrm{static}}\right)$ in
\Figref{fig:exploration_results_alpha}. We indicate equal-valued
timescales, $\mathrm{log}_{10}\left(\eta_{\mathrm{static}}\right) = 0$
with a red-dashed vertical line. The area on the right of this line
indicates that the static-tide approximation is justified, and vice
versa. The numbers in the legend indicate the total fraction for
either weights with
$\mathrm{log}_{10}\left(\eta_{\mathrm{static}}\right) < 0$. The data
show a broad range of
$\mathrm{log}_{10}\left(\eta_{\mathrm{static}}\right)$, and clearly
show that in terms of time-spent transferring mass, the static
approximation is overall valid (less than $0.1$ per cent below
$\mathrm{log}_{10}\left(\eta_{\mathrm{static}}\right) = 0$). This is
not always the case for the mass-transferred, because a significant
fraction ($13$ per cent) of all mass transferred occurs when the
static-tide approximation is invalid.

\begin{figure*}
  \centering
  \includegraphics[width=\textwidth]{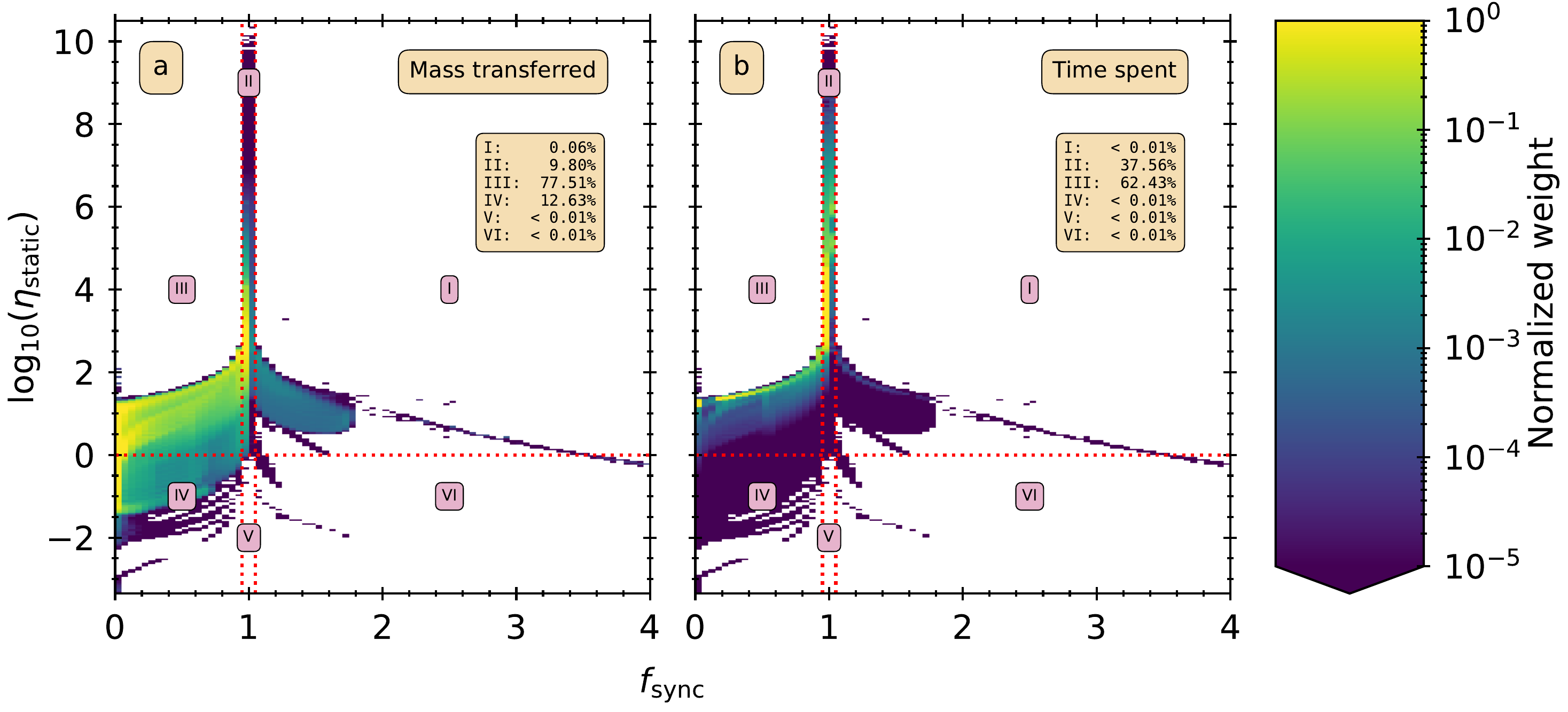}
  \caption{$\mathrm{log}_{10}\left(\eta_{\mathrm{static}}\right)$
    (ordinate) as a function of $\fsync$ (abscissa) in binary
    population (\Secref{sec:expl-param-rang}). (a) shows the
    distribution of
    $\mathrm{\log}_{10}\left(\eta_{\mathrm{static}}\right)$ vs.
    $\fsync$ in a population of binaries in terms of mass transferred,
    and (b) shows the distribution in terms of time spent transferring
    mass.  We indicate $6$ sections, separated by red-dotted
    lines. Section \rom{1} indicates super-synchronous
    ($\fsync > 1.025$) systems where the potential is approximately
    static
    ($\mathrm{log}_{10}\left(\eta_{\mathrm{static}}\right) >= 0$),
    section \rom{2} indicates near-synchronous systems
    ($0.975 <= \fsync <= 1.025$) with a static potential
    ($\mathrm{log}_{10}\left(\eta_{\mathrm{static}}\right) >= 0$) and
    section \rom{3} indicates sub-synchronous systems
    ($0.975 < \fsync$) with a static potential
    ($\mathrm{log}_{10}\left(\eta_{\mathrm{static}}\right) >=
    0$). Section \rom{4} indicates sub-synchronous systems
    ($0.975 < \fsync$) where the static approximation is not valid
    ($\mathrm{log}_{10}\left(\eta_{\mathrm{static}}\right) < 0$,
    i.e. with a dynamic potential), section \rom{5} indicates
    near-synchronous systems ($0.975 <= \fsync <= 1.025$) with a
    dynamic potential
    ($\mathrm{log}_{10}\left(\eta_{\mathrm{static}}\right) < 0$) and
    section \rom{6} indicates super-synchronous systems
    ($\fsync > 1.025$) with a dynamic potential
    ($\mathrm{log}_{10}\left(\eta_{\mathrm{static}}\right) < 0$). The
    range of $\fsync$ in the near-synchronous regions is determined by
    the bin-width in our simulations.}
  \label{fig:exploration_results_alpha_vs_synchronicity}
\end{figure*}

We show the normalised distribution of
$\mathrm{log}_{10}\left(\eta_{\mathrm{static}}\right)$ as a function
of $\fsync$ in
\Figref{fig:exploration_results_alpha_vs_synchronicity}, where in
\Figref{fig:exploration_results_alpha_vs_synchronicity} (a) we show
the distribution weighted by mass-transferred, and in
\Figref{fig:exploration_results_alpha_vs_synchronicity} (b) we show
the data in terms of time-spent transferring mass. We indicate $6$
sections, separated by red-dotted lines. Section \rom{1} indicates
super-synchronous ($\fsync > 1.025$) systems where the potential is
approximately static
($\mathrm{log}_{10}\left(\eta_{\mathrm{static}}\right) >= 0$), section
\rom{2} indicates near-synchronous systems
($0.975 <= \fsync <= 1.025$) with a static potential
($\mathrm{log}_{10}\left(\eta_{\mathrm{static}}\right) >= 0$) and
section \rom{3} indicates sub-synchronous systems ($0.975 < \fsync$)
with a static potential
($\mathrm{log}_{10}\left(\eta_{\mathrm{static}}\right) >= 0$). Section
\rom{4} indicates sub-synchronous systems ($0.975 < \fsync$) where the
static approximation is not valid
($\mathrm{log}_{10}\left(\eta_{\mathrm{static}}\right) < 0$, i.e. with
a dynamic potential), section \rom{5} indicates near-synchronous
systems ($0.975 <= \fsync <= 1.025$) with a dynamic potential
($\mathrm{log}_{10}\left(\eta_{\mathrm{static}}\right) < 0$) and
section \rom{6} indicates super-synchronous systems ($\fsync > 1.025$)
with a dynamic potential
($\mathrm{log}_{10}\left(\eta_{\mathrm{static}}\right) < 0$). The
range of $\fsync$ in the near-synchronous regions is determined by the
bin-width in our simulations.
\Figref{fig:exploration_results_alpha_vs_synchronicity} (a) shows that
the transferred mass is mostly transferred in three sections. Only
$9.8$ per cent of the systems normalised by mass-transferred are
synchronous and are well approximated by the static potential (section
\rom{2}). The large majority ($77.5$ per cent) of transferred mass
takes place in systems with a sub-synchronous donor that still
responds rapidly enough to regard the potential as static (section
\rom{3}). Most of the remaining systems ($12.6$ per cent) have donors
that rotate sub-synchronously for which the static potential
approximation does not hold (section \rom{4}). The rest of the
sections cover less than $0.08$ per cent of all transferred mass,
which indicates that super-synchronous rotation does not occur much in
field binaries ($<0.07$ per cent), and especially not in cases where
the static potential approximation breaks down.
\Figref{fig:exploration_results_alpha_vs_synchronicity} (b) shows that
the time-spent transferring mass mostly is spent in just two sections,
section \rom{2} and \rom{3}. Between these two sections, the
synchronous case where the static potential approximation holds
(section \rom{2}) covers $37.6$ per cent of all time-spent
transferring mass. The majority, thus, is spent where systems have
donors that rotate sub-synchronously but effectively experience a
static potential. The contribution of the other regions is negligible
($<0.04$), indicating that like in the mass-transferred case
super-synchronous rotation is not common in field binaries, but also
that not much time is spent in the case where the donor effectively
experiences dynamical tides.

With our results shown in \Figref{fig:exploration_results_parameters}
and \Figref{fig:exploration_results_alpha_vs_synchronicity} we
determine the parameter ranges for the trajectory simulations.
\begin{enumerate}
\item For the thermal velocity, $\mathrm{log}_{10}(\vthermal)$, we
  consider the range between $-3.5$ and $-0.5$ for our trajectory
  calculations.
\item For the synchronicity factor, $\fsync$, we consider the range
  between $0$ and $2$ for our trajectory calculations. A small
  fraction of systems has $\fsync > 2$, they are however clearly less
  frequent.
\item For the mass ratio, $\qacc$, we use the range between $-2$ and
  $2$ for our trajectory calculations.
\end{enumerate}
These values are listed in
\Tabref{tab:interpolation_table_properties}.

\begin{table*}
  \begin{center}
  \begin{tabular}{lll}
    Parameter &  [lower bound, upper bound, step size] & comment \\
    \midrule
    $\mathrm{log}_{10}\left(\vthermal\right)$ & [-3.5, -0.5, 0.5] & thermal velocity (\Secref{sec:therm-veloc-stre-1} and \Eqref{eq:sigma_thermal}).\\
    $\fsync$ & [0.1, 2.0, 0.1] & Synchronicity factor (\Secref{sec:roche-potent-reduc} and \Eqref{eq:synchronicity_factor}).\\
    $\mathrm{log}_{10}\left(\qacc\right)$ & [-2, 2, 0.1] & Mass ratio $\qacc = M_{\mathrm{acc}}/M_{\mathrm{don}}$.
  \end{tabular}
\end{center}
\caption{Parameters and ranges used to generate our interpolation
  dataset. The first column describes the parameter, the second column
  describes the parameter range and step size, and the third column
  contains extra descriptions and references to the appropriate
  formulae.}
  \label{tab:interpolation_table_properties}
\end{table*}

The above results indicate that sub-synchronous mass-transfer is
common, both for the time-spent ($>60$ per cent) and for the
mass-transferred ($90$ per cent). This further motivates the remainder
of this study.

\subsection{Ballistic trajectory properties}
\label{sec:ball-traj-prop}
In this section we show our results of the ballistic trajectory
calculations. While our results span a large parameter space, we
choose to highlight the two extreme cases, with the results for cold
and narrow streams ($\vthermal = 10^{-3}$ and
$\streamdiameter \approx 10^{-4}-10^{-3}$) in \Secref{sec:cold-narrow}
and hot and wide streams ($\vthermal = 10^{-0.5}$ and
$\streamdiameter \approx 0.1-0.4$) in \Secref{sec:traj-prop-hot}.
Before looking at the results let us highlight several effects that
are relevant to the evolution of the trajectories.

In sub-synchronous systems ($\fsync < 1$) L1 moves outward relative to
the synchronous case, the velocity offset due to asynchronous rotation
at L1 is downward ($v_{\mathrm{non-synchronous\ offset}}$ is
negative), the Coriolis force for downward motion leads to a rightward
acceleration ($a_{\mathrm{Coriolis,\ y}}$ is positive), and at the
moment of release the particle is located within the Roche lobe of the
accretor.
In super-synchronous systems ($\fsync > 1$) L1 moves inward relative
to the synchronous case, the velocity offset due to asynchronous
rotation at L1 is upward ($v_{\mathrm{non-synchronous\ offset}}$ is
positive), the Coriolis force for upward motion leads to a leftward
acceleration ($a_{\mathrm{Coriolis,\ y}}$ is negative) and at the
moment of release the particle is located within the Roche lobe of the
donor.
In low mass ratio systems ($\qacc < 1$) the velocity offset due to
asynchronous rotation is larger relative to equal mass-ratio systems
due to the large size of the Roche-lobe of the donor and the velocity
is even higher for sub-synchronous rotation as L1 moves outward.
In high mass ratio systems ($\qacc > 1$) the velocity offset due to
asynchronous rotation is smaller relative to the equal mass-ratio
systems due to the small size of the Roche-lobe of the donor.
These effects are visualised and quantified in Figures
\ref{fig:combined_area_plot},
\ref{fig:non_synchronous_rotation_schematic} and
\ref{fig:lagrange_point_plot}, and \Eqref{eq:equations_of_motion_x}.

\subsubsection{Cold and narrow streams}
\label{sec:cold-narrow}
We show our cold and narrow ballistic integrations,
$\vthermal = 10^{-3}$, in the ranges of mass ratio, $\qacc$, and
synchronicity factor, $\fsync$, described in
\Tabref{tab:interpolation_table_properties}. From
\Figref{fig:combined_area_plot} we know that the stream diameter is
small, $\streamdiameter \approx 10^{-4}-10^{-3}$, so all the
trajectories sampled along the stream effectively have the same
initial position. From \Figref{fig:non_synchronous_rotation_schematic}
we know that for asynchronous systems ($\fsync \neq 1$) at low mass
ratios $\qacc < 1$ the initial radial velocity is low compared to the
tangential asynchronous velocity offset,
$v_{\mathrm{asynchronous\ offset}}$, which indicates that results in
that part of the parameter space will deviate most from the
synchronous case explored by \citep{lubowGasDynamicsSemidetached1975}.

\begin{figure}
  \centering
  \includegraphics[width=\columnwidth]{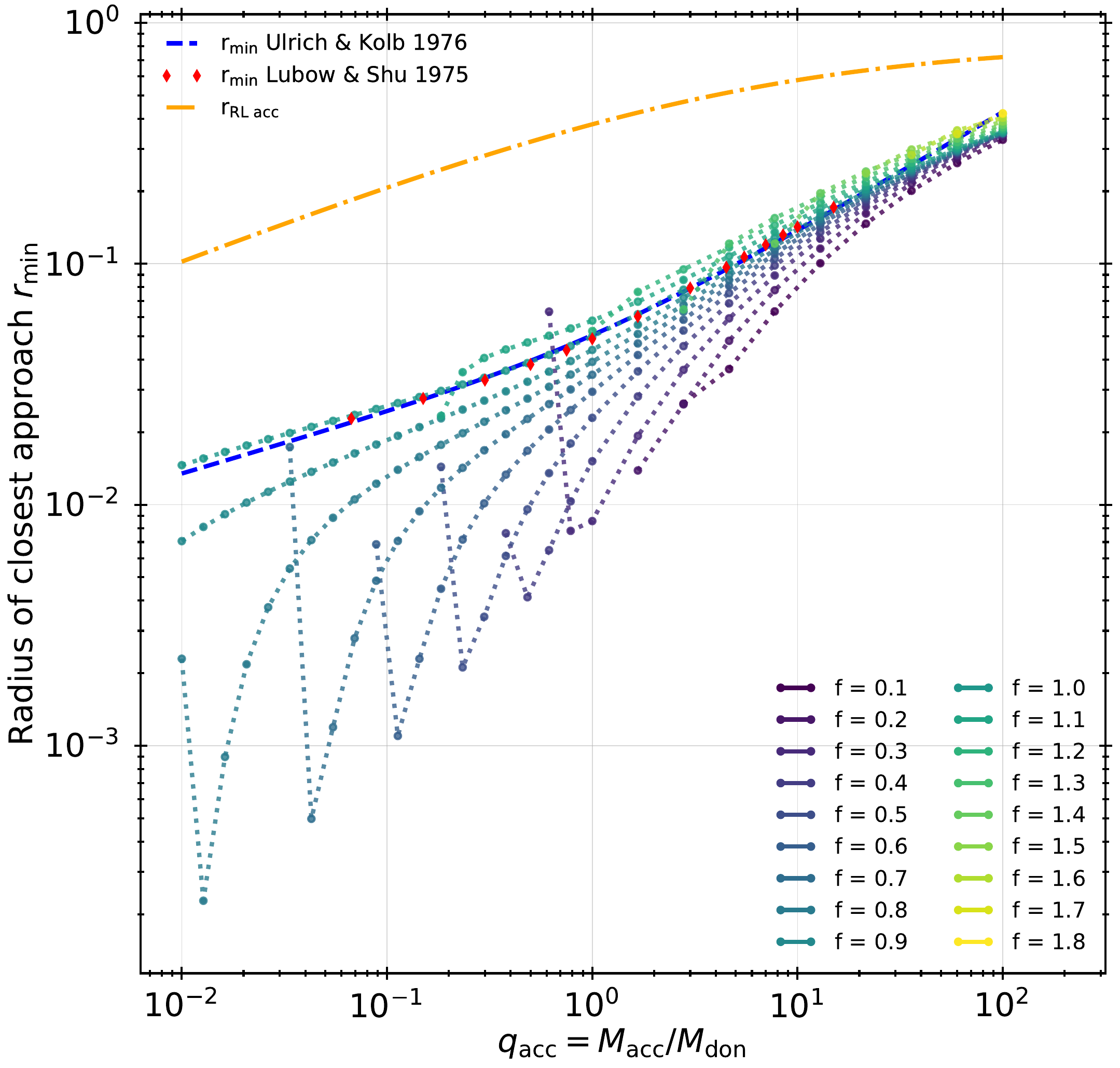}
  \caption{Radius of closest approach $\rmin$ (ordinate) from our
    cold, $\vthermal = 10^{-3}$, ballistic calculations and
    prescriptions in the literature as a function of mass-ratio,
    $\qacc$ (abscissa), and synchronicity factor, $\fsync$
    (colour-scale).  The red diamonds indicate the original data
    points from \citet{lubowGasDynamicsSemidetached1975}, the blue
    dashed line indicates the prescription from
    \citet{ulrichAccretingComponentMassexchange1976} and the orange
    dashed-dotted line indicates the Roche-lobe radius of the
    accretor.}
  \label{fig:rmin_low_v}
\end{figure}

In \Figref{fig:rmin_low_v} we show the radii of closest approach,
$\rmin$, of particles that accrete onto the accretor as a function of
mass ratio, $\qacc$ (abscissa), and donor synchronicity, $\fsync$
(colour scale). The triangles indicate the orientation of the
particle, where the upward triangle indicates prograde (same direction
as the binary orbit) orientation and the downward triangle indicates
retrograde (opposite direction).  The red diamonds are from
\citet{lubowGasDynamicsSemidetached1975}, and the blue dashed line
indicates the prescription of
\citet{ulrichAccretingComponentMassexchange1976}.
The radii of closest approach in synchronously-rotating
$\fsync = 1$ donor systems match closely to the results of
\citet{lubowGasDynamicsSemidetached1975}.

Overall, in the range that covers the parameters of
\citet[][$8\times10^{-2} < \qacc < 2\times10^{1}$ and
$\fsync = 1$]{lubowGasDynamicsSemidetached1975} we find a good match,
confirming that our method works as it should, given the assumptions
and approach.
Our data show the super-synchronous donors only accrete onto the
accretor at high mass ratios ($\qacc > 10$ and $\fsync > 1.5$), with a
decrease in minimum $\qacc$ required for accretion onto the donor with
a decrease in $\fsync$. At high mass-ratio the donor is not able to
exert enough force to turn the stream back onto itself even though the
particle is released within its Roche-Lobe, due to its low mass.
With sub-synchronous donors we find an increase in the minimum
mass-ratio that accretes onto the accretor, with decreasing
$\fsync$. Moreover, given a synchronicity factor, $\fsync$, the radius
of closest approach decreases with decreasing mass ratio, $\qacc$.
Systems with a low mass ratio and a low synchronicity factor
($\qacc < 1$ and $\fsync < 0.4$) experience a high negative velocity
offset due to asynchronous rotation and, even though they initially
start in the Roche lobe of the accretor, they experience an
acceleration towards the donor because of the Coriolis force, which is
strong enough to steer the trajectory onto the donor.
Generally, in high mass-ratio systems, the effect of asynchronous
rotation on the radius of closest approach is small, with a spread of
only a factor of 2 at $\qacc \sim 30$. This is because the velocity
offset due to the asynchronous rotation for systems with mass-ratio
$\qacc \geq 30$ is generally low
($|v_{\mathrm{asynchronous\ offset}}| < 0.2$,
\Figref{fig:non_synchronous_rotation_schematic}), so the trajectories
do not differ much from the synchronous case.

\begin{figure}
  \centering
\includegraphics[width=\columnwidth]{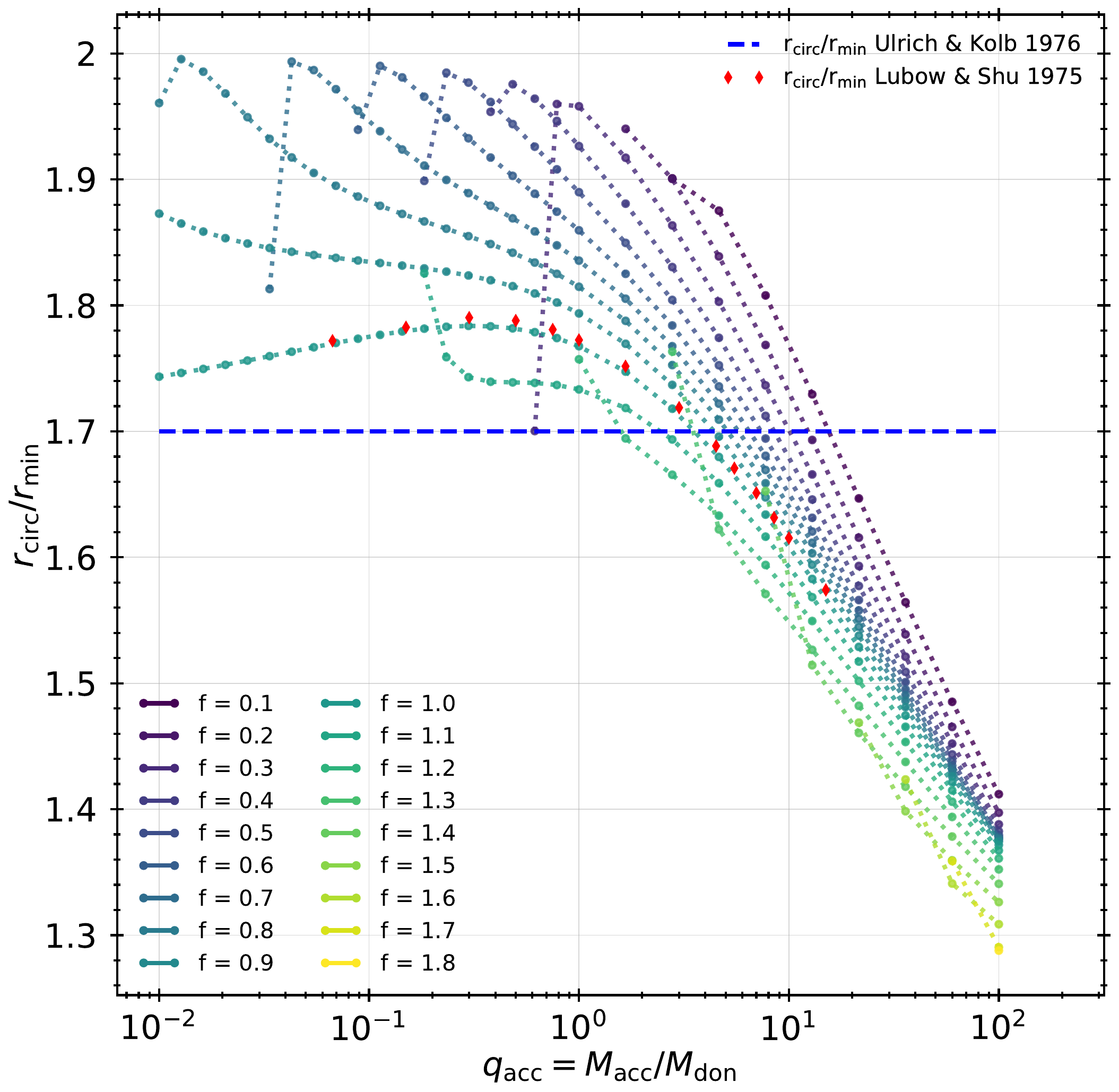}
\caption{Ratio of circularisation radius to radius of closest approach
  ${r}_{\mathrm{circ}}/{r}_{\mathrm{min}}$ (ordinate) as a function of
  mass ratio $\qacc$ (abscissa) and synchronicity factor $\fsync$
  (colour-scale). The red-dashed horizontal line indicates the ratio
  ${r}_{\mathrm{circ}}/{r}_{\mathrm{min}}$ from
  \citet{ulrichAccretingComponentMassexchange1976} }
  \label{fig:rcirc_over_rmin}
\end{figure}

In \Figref{fig:rcirc_over_rmin} we show the ratio of circularisation
radius to radius of closest approach, $\rcirc/\rmin$, as a function of
mass ratio, $\qacc$, and synchronicity factor, $\fsync$. This data is
a measure of the specific angular momentum at the radius of closest
approach and how much it differs from that of a circular orbit at the
radius of closest approach. Moreover, this data is used to calculate
the radius at which an accretion disk forms. The red diamonds are from
\citet{lubowGasDynamicsSemidetached1975} and the blue-dashed
horizontal line is from
\citet{ulrichAccretingComponentMassexchange1976}.
At high mass-ratios ($\qacc > 10$), we see a general decrease of the
${r}_{\mathrm{circ}}/{r}_{\mathrm{min}}$ with increasing mass ratio
$\qacc$, regardless of the synchronicity factor, with a spread of at
most 0.2. This indicates that the specific angular momentum at the
radius of closest approach tends to that of a circular orbit at the
radius of closest approach, and that the asynchronous rotation of the
donor does not affect this quantity strongly either.

At mass-ratios $\qacc < 1$ the trajectories of most asynchronous
donors accrete onto the accretor. All the trajectories have a ratio of
radii between $1.7$ and $2.0$, indicating that the stream carries much
more specific angular momentum at the radius of closest approach than
a circular orbit would. Because of its low mass, the torque exerted by
the accretor is insufficient to circularise the stream.

Overall, the ratio is between $1.3$ and $2$, indicating that the
stream always carries more specific angular momentum than a circular
orbit at the radius of closest approach would. Moreover, the commonly
used constant ratio $1.7$ used by
\citet{ulrichAccretingComponentMassexchange1976} is up to $30$ per
cent off.

\begin{figure}
  \centering \includegraphics[width=\columnwidth]{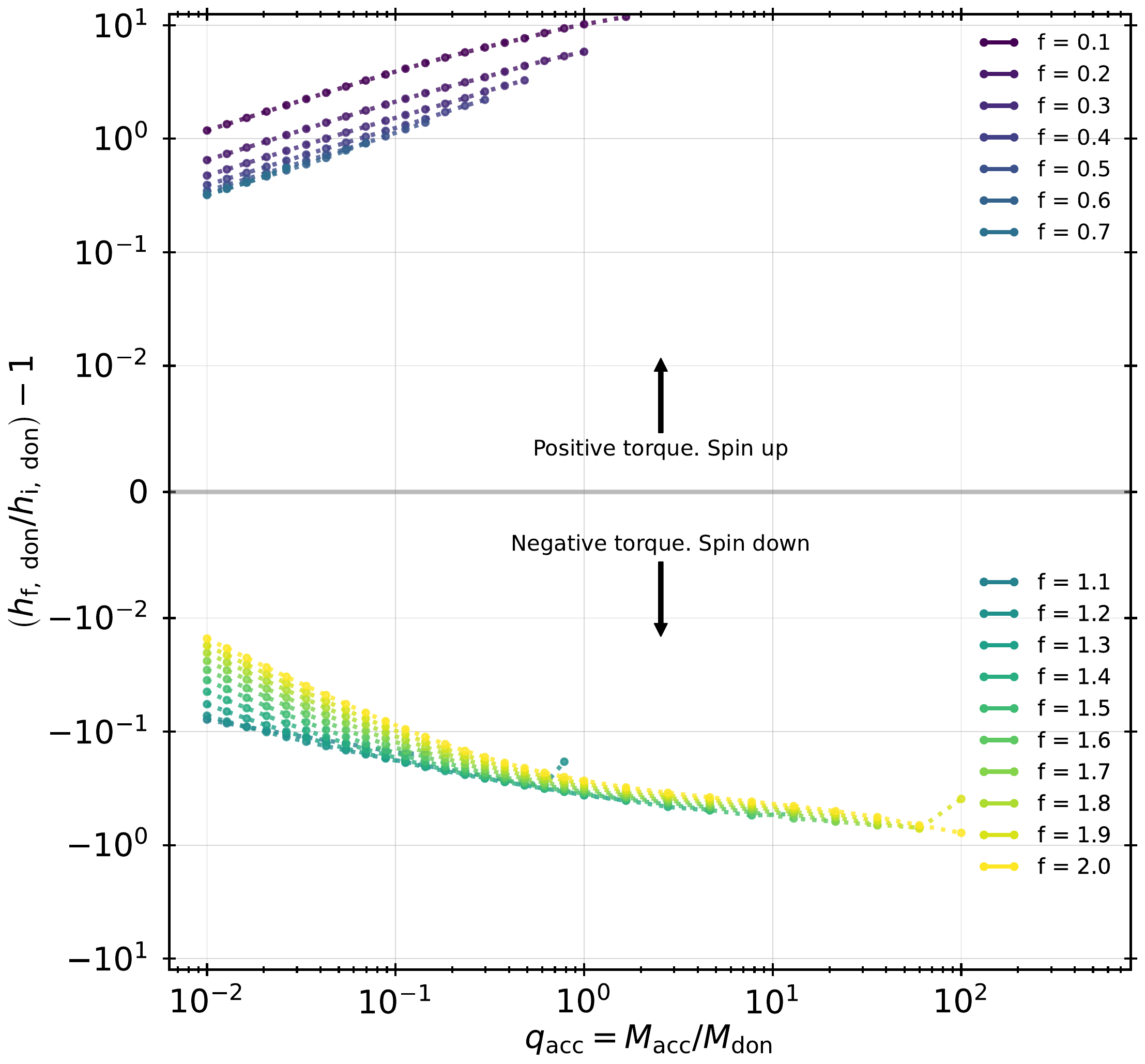}
  \caption{Fractional difference between the final
    ($h_{\mathrm{f,\ don}})$ and initial ($h_{\mathrm{i, don}}$)
    specific angular momenta (ordinate) of particles that accrete back
    onto the donor as a function of mass ratio $\qacc$ (abscissa) and
    synchronicity fraction $\fsync$ (colour scale). The set of values
    at the top originates from sub-synchronously rotating donors. The
    set of values at the bottom originates from super-synchronously
    rotating donors.}
  \label{fig:self_accretion_specific_angular_momentum_factor}
\end{figure}

In \Figref{fig:self_accretion_specific_angular_momentum_factor} we
show the fractional difference between the final
($h_{\mathrm{f,\ don}}$) and initial ($h_{\mathrm{i, don}}$) specific
angular momenta (ordinate) of particles that accrete back onto the
donor as a function of mass ratio, $\qacc$ (abscissa), and
synchronicity fraction, $\fsync$. The data show two distinct regions.

The trajectories from sub-synchronous donors ($\fsync \leq 0.7$) show
an increasingly larger final specific angular momentum
$h_{\mathrm{f,\ don}}$ compared to the initial specific angular
momentum, $h_{\mathrm{i, don}}$, of the stream for decreasing
synchronicity factor, $\fsync$. Moreover, the lower $\fsync$, the
larger the range in mass-ratios, $\qacc$, for which the stream
accretes onto the donor. This is because a larger deviation from
synchronism introduces a larger velocity offset, which requires an
increasingly massive accretor to completely turn the stream towards
itself. These trajectories all exert a positive torque on the donor
that leads to the donor becoming more synchronous.

Trajectories from super-synchronous donors show a decrease in specific
angular momentum relative to their initial specific angular momentum,
with a decrease in specific angular momentum relative to the initial
angular momentum as $\fsync$ increases. This is because of a decrease
in angle of incidence with the donor with increasing asynchronicity
for super-synchronous donors, caused by a combination of a lower
velocity offset and an acceleration towards the donor, and vice versa
for sub-synchronous donors. Trajectories that accrete onto
super-synchronous donors all exert a negative torque that again leads
to the donor becoming more synchronous.

For both the super-synchronous (negative torque) and the
sub-synchronous (positive torque) torque self-accretion, the magnitude
of the difference between the initial and final specific angular
momenta increases, for a given synchronicity factor, with increasing
mass ratio $\qacc$. At higher mass-ratios the trajectory is affected
more, due to the stronger gravitational effect of the accretor. This
increasingly affects the final angular momentum of the stream, which
leads to the increasing difference. In the low mass-ratio systems, the
stream angular momentum is hardly affected, and thus the difference
remains small (e.g. at $\qacc = 0.01$,
$h_{\mathrm{f,\ don}}/h_{\mathrm{i,\ don}}-1 > -10^{-1}$ for
sub-synchronous donors, and
$h_{\mathrm{f,\ don}}/h_{\mathrm{i,\ don}}-1 < 5\times10^{-1}$ for
super-synchronous donors).

\begin{figure*}
  \centering
  \includegraphics[width=\textwidth]{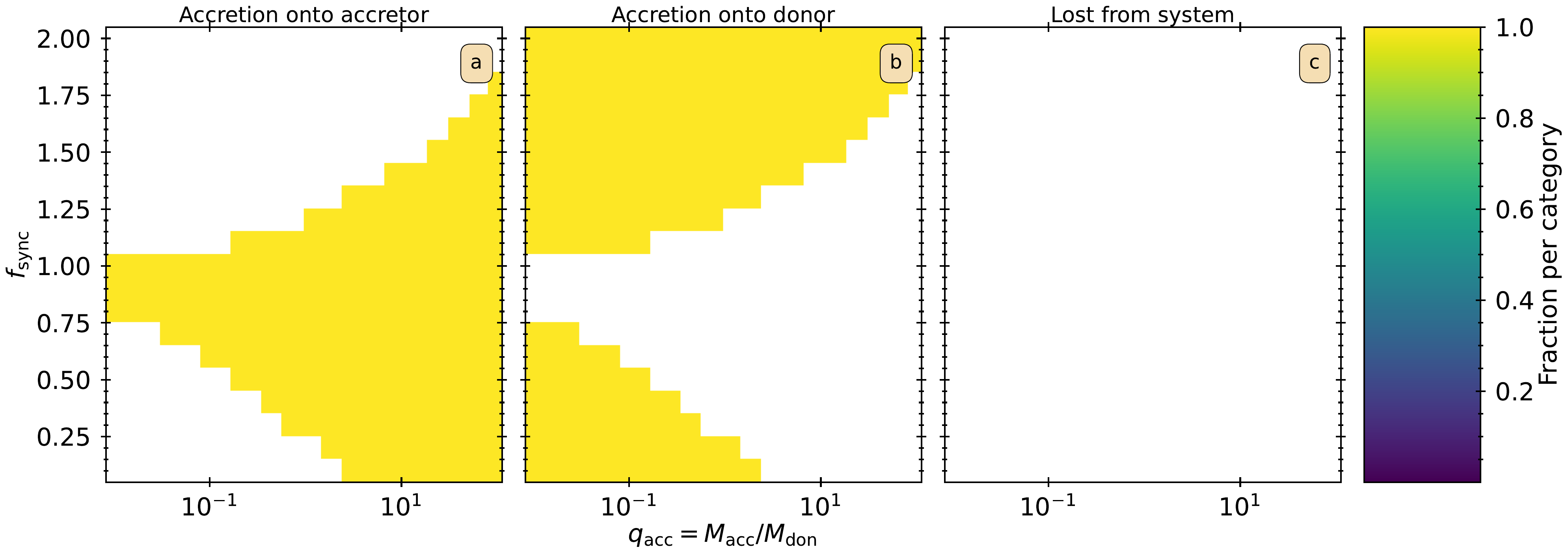}
  \caption{Fractions of trajectory classifications as a function of
    mass ratio $\qacc$ (abscissa) and synchronicity factor $\fsync$
    (ordinate, \Secref{sec:classifying-and-averaging},
    \Eqref{eq:fraction_accretion_accretor}). (a) shows the fraction of
    trajectories accreting onto the accretor, (b) shows the fraction
    of trajectories accreting back onto the donor and (c) shows the
    trajectories that escape the system. Colour indicates the fraction
    of each classification to all the classified trajectories. White
    indicates a fraction of zero. The red lines indicate the fraction
    of all trajectories that failed to evolve correctly
    (\Secref{sec:integration-method} and \Eqref{eq:all_fail}).}
  \label{fig:classification_fractions}
\end{figure*}

We show the fractions of each classification as a function of mass
ratio, $\qacc$ (abscissa), and synchronicity factor, $\fsync$
(ordinate, \Secref{sec:classifying-and-averaging},
\Eqref{eq:fraction_accretion_accretor}), in
\Figref{fig:classification_fractions}. \Figref{fig:classification_fractions}
(a) shows the fraction of all trajectories accreting onto the
accretor, \Figref{fig:classification_fractions} (b) shows those
accreting onto the donor and \Figref{fig:classification_fractions} (c)
shows those that are lost from the system. The colour-scale indicates
a non-zero fraction, where a white indicates a fraction of zero. The
red lines indicate the fraction of all trajectories that failed to
evolve correctly (\Secref{sec:integration-method} and
\Eqref{eq:all_fail}).

The data in \Figref{fig:classification_fractions} (a) show that, at
low mass-ratio, ($\qacc < 0.1$), only the near-synchronous donors
accrete onto the accretor. The region of synchronicity factor,
$\fsync$, that corresponds to accretion onto the accretor increases
both to sub- and super-synchronous donors with increasing mass ratio
$\qacc$. This is due to the decrease in velocity offset due to
asynchronous rotation with increasing $\qacc$
(\Figref{fig:non_synchronous_rotation_schematic}). This reduces the
effect of asynchronicity is reduced and the makes the trajectories
behave like ones from synchronous systems. The asymmetry in the shape
of the fraction accreted onto the accretor is caused by the Coriolis
force, which accelerates the particle towards the accretor for
sub-synchronous donors and away for super-synchronous donors.
The data in \Figref{fig:classification_fractions} (b) show an exact
inversion of the data in \Figref{fig:classification_fractions} (a),
and \Figref{fig:classification_fractions} (c) shows that for the low
thermal-velocity (cold) there are no trajectories that escape the
system.

The transition between each region is sharp, caused by the narrow
stream associated with the low thermal-velocity (cold), which
indicates that for every classification at a given coordinate either
none (white) of the trajectories or all (yellow) of the trajectories
are classified as such. Moreover, we find no failing systems for our
cold and narrow trajectories.

\begin{figure} \includegraphics[width=\columnwidth]{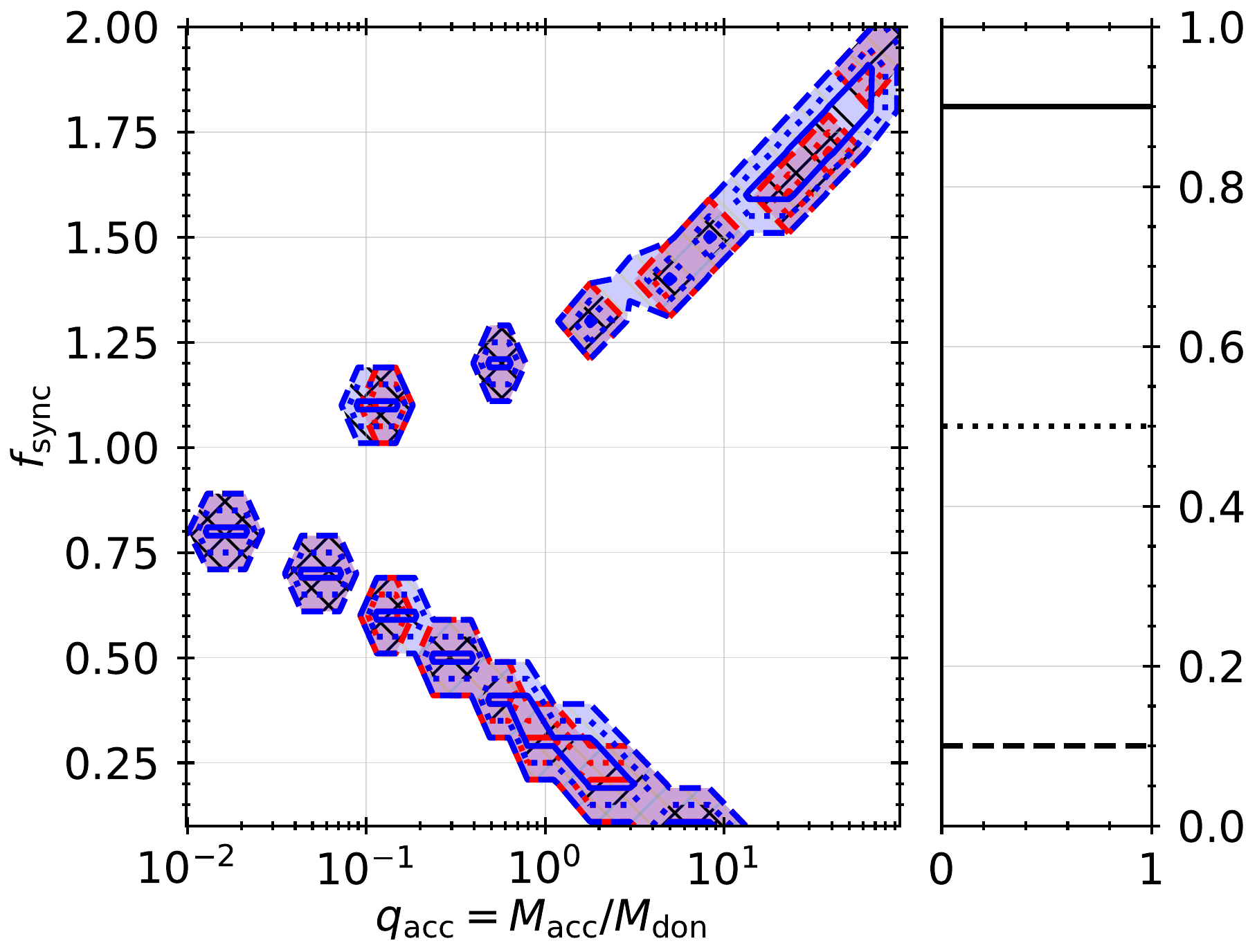}
  \caption{Fractions of trajectory intersections
    (\Secref{sec:intersecting-orbits}) as a function of $\qacc$
    (abscissa) and synchronicity factor $\fsync$ (ordinate). The red
    contours show the fraction of self-intersecting trajectories,
    $\selfintersect$ , the blue contours show the fraction of
    intersection with other trajectories, $\otherintersect$. The
    dashed line indicates a weighted fraction of at least 0.1 of all
    the trajectories, a dotted line indicates a weighted fraction of
    at least 0.5 and the solid line indicates a weighted fraction of
    at least 0.9 of all trajectories.}
  \label{fig:intersection_low_v}
\end{figure}

\Figref{fig:intersection_low_v} shows the fractions of trajectory
intersections (\Secref{sec:intersecting-orbits}) as a function of
$\qacc$ and $\fsync$, for low thermal-velocity (cold,
$\vthermal = 10^{-3}$) streams. The red contours show the fraction of
self-intersecting trajectories, the blue contours show the fraction of
intersection with other trajectories. The dashed line indicates a
weighted fraction of at least 0.1 of all the trajectories, a dotted
line indicates a weighted fraction of at least 0.5 and the solid line
indicates a weighted fraction of at least 0.9 of all trajectories.

We find that self-intersecting orbits occur at the edges of the
transition regions between accretion onto the accretor and accretion
onto the donor (\Figref{fig:classification_fractions}). The fraction
is always high, since the stream is itself so narrow that the
trajectories stay bundled and follow approximately the same path.
Intersection with other trajectories, with angles of incidence above
the threshold $\thetathreshold$, occurs in the same narrow region of
($\qacc$, $\fsync$) parameter space as the self-intersecting
orbits. This is because the stream is so narrow that the trajectories
effectively follow the same path as each other.

\begin{figure}
  \includegraphics[width=\columnwidth]{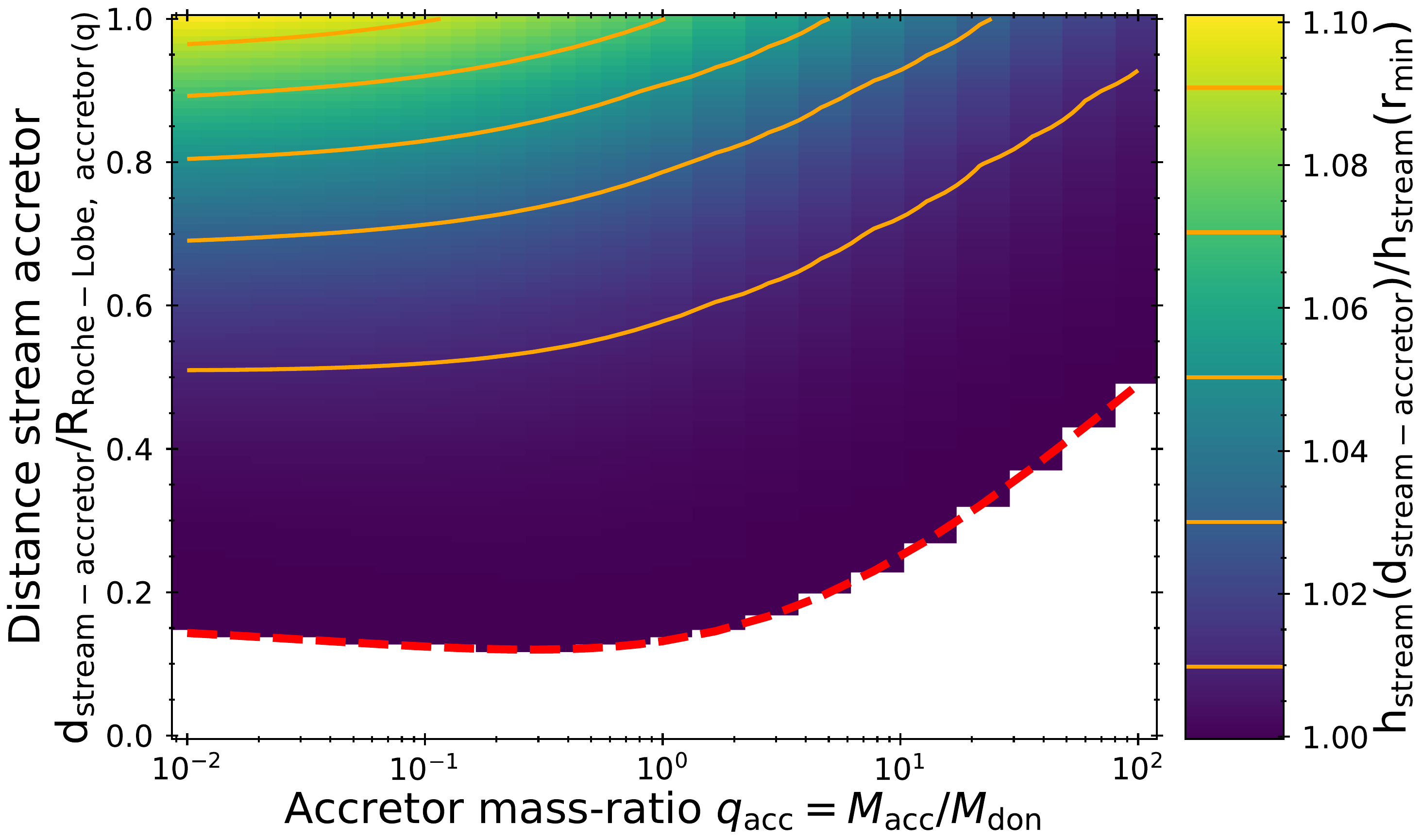}
  \caption{Specific angular momentum of the mass-transfer stream in
    terms of the specific angular momentum at the radius of closest
    approach (colour-scale) as a function of mass ratio $\qacc$
    (abscissa) and distance of the stream to the accretor in units of
    the Roche-lobe radius of the accretor (ordinate), when
    $\fsync = 1$. The red-dashed line indicates the radius of closest
    approach of the stream in terms of the Roche-lobe radius of the
    accretor. The orange lines are contour lines of equal spacing
    between 1.0 and 1.1. For a system with these properties, the
    specific angular momentum of the stream is generally higher than
    at the radius of closest approach, but the change in specific
    angular momentum of the stream is at maximum $\sim 10$ per cent,
    specifically at low mass-ratios ($\qacc > 10$) and furthest away
    from the accretor
    ($d_{\mathrm{stream-accretor}}=1.0 R_{\mathrm{Roche-Lobe,
        accretor}}$). At high mass-ratios ($\qacc > 10$), the maximum
    change over the length of the stream is less ($ < 5$ per
    cent). This indicates that direct-impact accretion is more
    efficient the larger the accretor is, but the increase in torque
    is marginal (up to $\sim 10$ per cent).}
  \label{fig:stream_interpolation_low_v}
\end{figure}

In \Figref{fig:rmin_low_v} and \Figref{fig:rcirc_over_rmin} we focus
on properties of the stream at its radius of closest approach to the
accretor. In many situations, though, the radius of the accretor
exceeds the radius of closest approach and the stream directly impacts
the accretor. In that case, the stream has less travel time through
the potential and experiences less torque by the binary system, which
affects the specific angular momentum of the stream upon impact with
the accretor.
We show the evolution of the specific angular momentum of the
mass-transfer stream as a function of its distance to the accretor and
the mass ratio $\qacc$ for systems with synchronously rotating donors
($\fsync = 1$) in \Figref{fig:stream_interpolation_low_v}. The color
indicates the specific angular momentum of the stream in units of that
of the specific angular momentum at the radius of closest
approach. The orange lines show $5$ equally spaced lines where this
specific angular momentum is constant.
For all mass ratios the specific angular momentum of the stream starts
out higher than what the specific angular momentum at the radius of
closest approach is. For systems with high mass ratios the difference
between the initial specific angular momentum and that at $\rmin$ is
minor (a few percent), but this difference increases with decreasing
mass ratio (up to ten percent).
We note that the qualitative behaviour of the stream systems with a
different synchronicity factor, $\fsync$, and thermal velocity,
$\vthermal$, is not necessarily the same as described above.


\subsubsection{Hot and wide streams}
\label{sec:traj-prop-hot}
In this section we show the trajectory properties of systems with a
hot and wide stream ($\vthermal = 10^{-0.5}$ and
$\streamdiameter \approx 0.1-0.4$). Whereas in the low
thermal-velocity (cold) regime the stream area is negligible, here the
stream area is sufficiently large as to cause a relevant offset
between the initial positions of the particles. Moreover, the high
thermal-velocity (hot) provides a large initial radial velocity
towards the accretor, and the Coriolis force subsequently provides a
large downward (negative y-direction) acceleration on the particles.

\begin{figure}
  \centering  \includegraphics[width=\columnwidth]{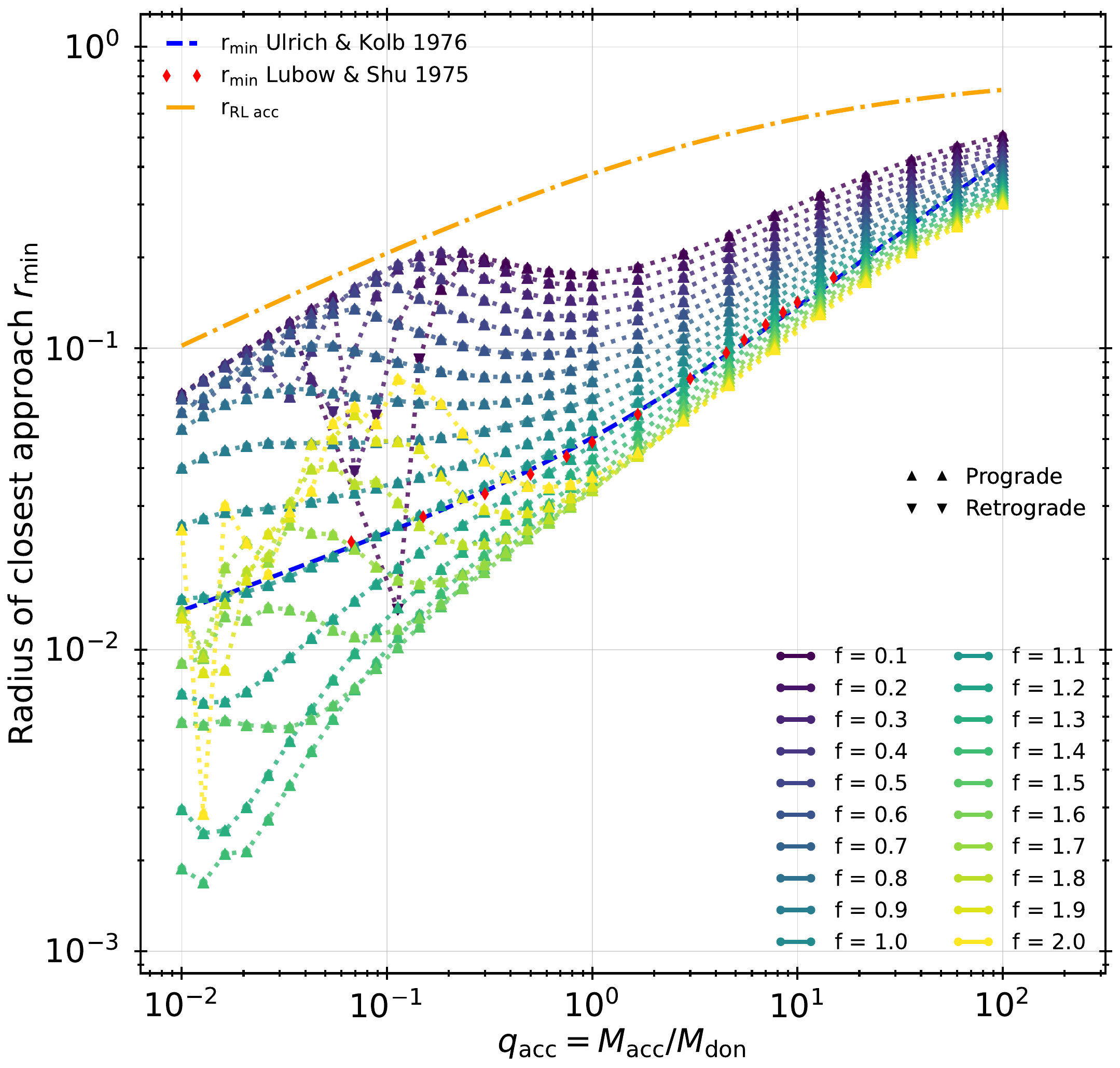}
  \caption{As \Figref{fig:rmin_low_v}, but with
    $\vthermal = 10^{-0.5}$.}
  \label{fig:rmin_high_v}
\end{figure}

We show the radii of closest approach in our high thermal-velocity
(hot) calculations in \Figref{fig:rmin_high_v}.
Overall, we again find a small spread of radii ($\rmin \sim 0.3-0.5$)
at large mass-ratios $\qacc = 100$, but we now see a much larger
spread ($\rmin \sim 0.01-0.8$) at low mass ratios ($\qacc \sim
0.1$). Notably, a wider range ($\fsync = 0.1-2.0$) of initial
asynchronicities lead to the accretion onto the accretor. This is due
to the larger ($\vthermal \approx 0.32$) initial radial velocity that
makes it harder to deflect the stream.

\begin{figure*}
  \centering
  \includegraphics[width=\textwidth]{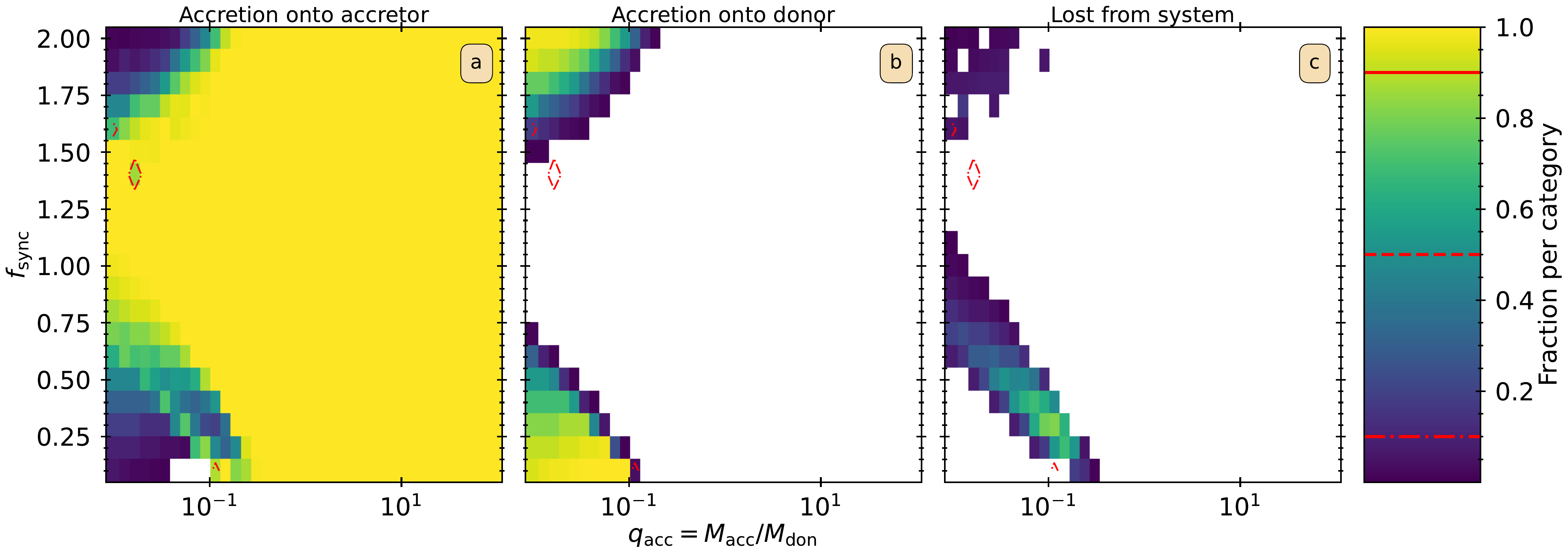}
  \caption{As \Figref{fig:classification_fractions}, but with
    $\vthermal = 10^{-0.5}$.}
  \label{fig:classification_fractions_high_v}
\end{figure*}

Accretion onto the donor now only occurs for systems with a low
mass-ratio ($\qacc < 0.1$, significantly lower than in the cold-stream
case), either with $\fsync < 0.4$ or with $\fsync > 1.5$
(\Figref{fig:classification_fractions_high_v}).

Sub-synchronous donors show a general increase of $\rmin$ with
decreasing synchronicity factor. This is due to the initially negative
transversal velocity from the sub-synchronous rotation directing the
stream further away from the accretor. This eventually leads to a
fraction of the stream escaping from the system, but for very
sub-synchronous rotating donors many trajectories self-intersect.

\begin{figure} \includegraphics[width=\columnwidth]{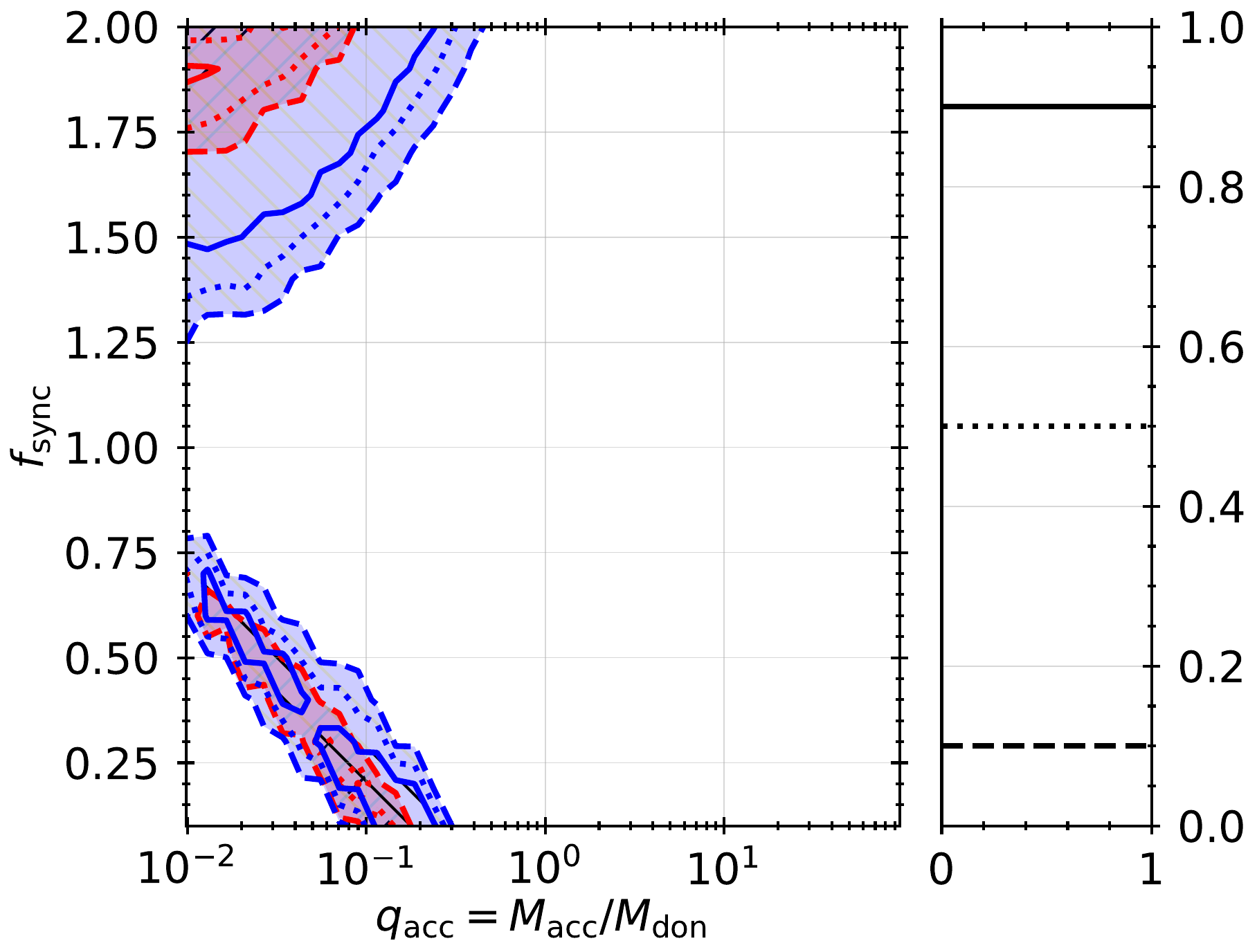}
  \caption{As \Figref{fig:intersection_low_v}, but with
    $\vthermal = 10^{-0.5}$.}
  \label{fig:intersection_high_v}
\end{figure}

At low mass-ratios, super-synchronous donors show a general decrease
of $\rmin$ with increasing synchronicity factor, but this behaviour
turns around for highly super-synchronous donors ($\fsync > 1.7$) at
low mass ratios ($\qacc < 0.1$). This is because part of the stream
for these systems starts accreting onto the donor, and the
trajectories that do still accrete onto the accretor on average have a
large radius of closest approach. This region of parameter space
contains many (self-)intersecting trajectories
(\Figref{fig:intersection_high_v}), and since we do not treat
intersecting orbits differently, this indicates that this region
requires a more sophisticated approach than our current one.

\begin{figure}
  \centering
  \includegraphics[width=\columnwidth]{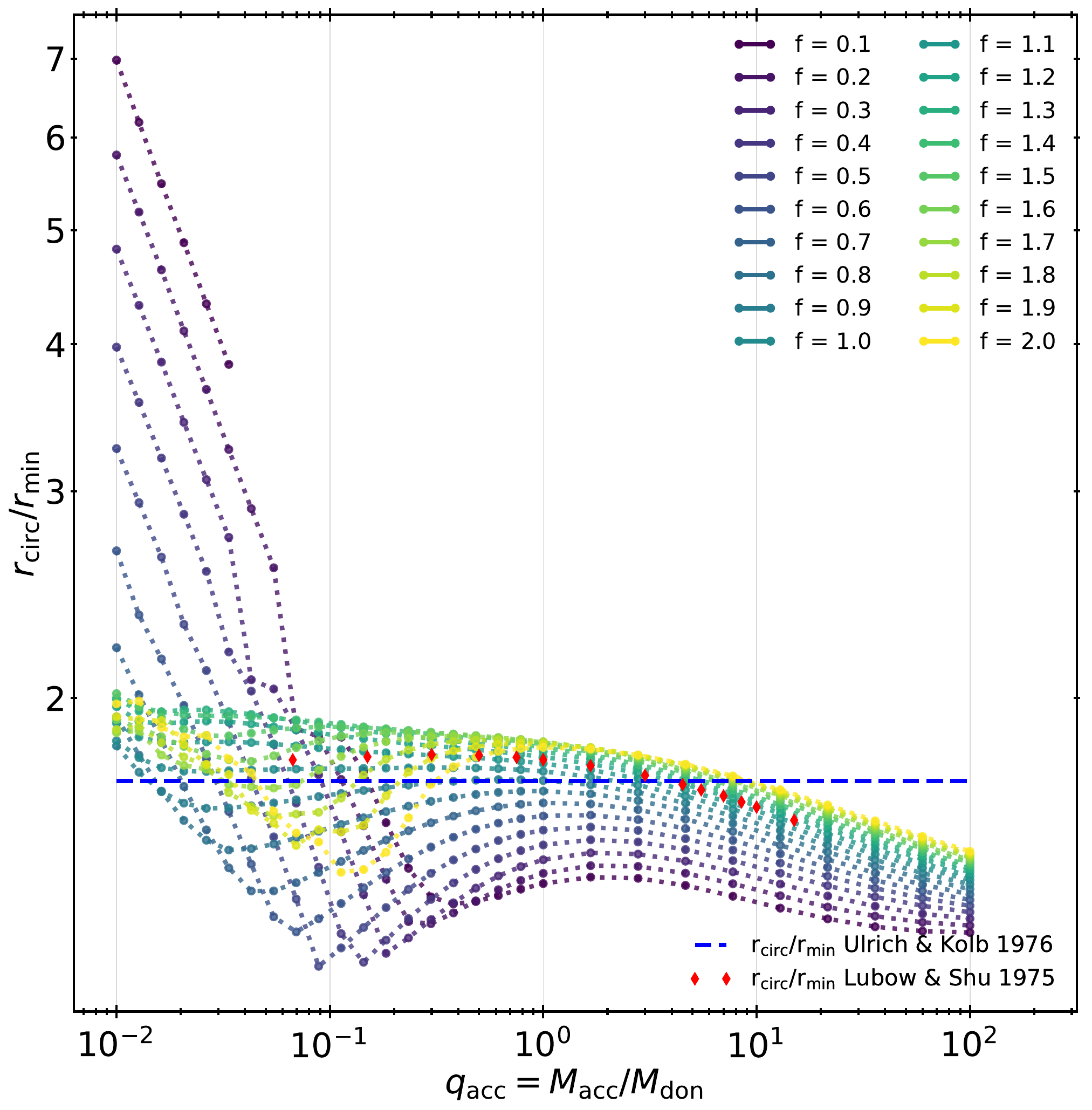}
  \caption{As \Figref{fig:rcirc_over_rmin}, but with
    $\vthermal = 10^{-0.5}$.}
  \label{fig:rcirc_over_rmin_high_v}
\end{figure}

We show the ratio of $\rcirc/\rmin$ in our high thermal-velocity (hot)
calculations in \Figref{fig:rcirc_over_rmin_high_v}.

At high mass-ratios ($\qacc > 1$), while as a function of mass ratio
the results are similar to the low thermal-velocity (cold) case,
i.e. with higher the mass ratio a lower $\rcirc/\rmin$, the behaviour
as a function of synchronicity is now reversed. The lower the
synchronicity, the lower the ratio $\rcirc/\rmin$, indicating that the
trajectories on averages at the radius of closest approach are similar
to circular orbits at that radius, and vice versa.

At small mass-ratios ($\qacc < 1$) the behaviour is similar as above,
but from $\qacc \leq 0.4$ the sub-synchronous systems get an
increasingly high ratio $\rcirc/\rmin$ with decreasing mass ratio
$\qacc$. This coincides with regions of the parameter space where part
of the stream either escapes from the system, or starts accreting onto
the donor. The remaining trajectories that barely do not escape often
fall back into the Roche-lobe of the accretor near radially, or they
find their radius of closest approach very early in their
trajectory. For these trajectories, the first radius of closest
approach is potentially not suitable to determine the angular momentum
of the ring that would form when the stream circles around the
accretor and hits itself.

\begin{figure}
  \centering
  \includegraphics[width=\columnwidth]{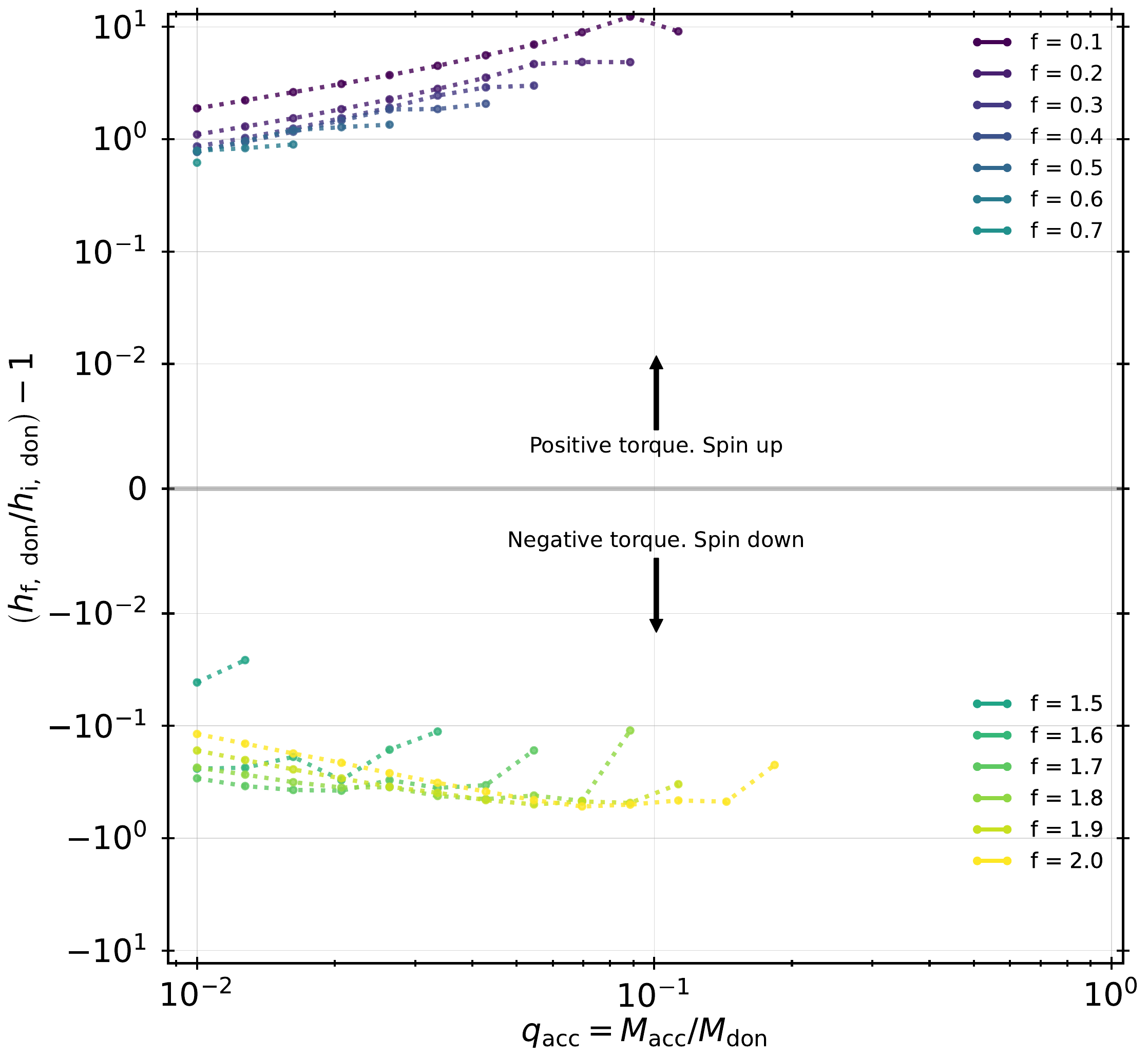}
  \caption{As
    \Figref{fig:self_accretion_specific_angular_momentum_factor}, but
    with $\vthermal = 10^{-0.5}$.}
  \label{fig:self_accretion_specific_angular_momentum_factor_high_v}
\end{figure}

We show the ratio of final and initial specific angular momenta of
self-accreting material in our high thermal-velocity (hot)
calculations in
\Figref{fig:self_accretion_specific_angular_momentum_factor_high_v}
While again there are two distinct regions of positive and negative
torque of self-accreting material, both regions are smaller and
require a higher degree of asynchronicity (i.e. $\fsync > 1.5$) and/or
a lower mass ratio ($\qacc < 0.1$) to self-accrete. Systems with
super-synchronous that self-accrete tend to experience a higher torque
for a given mass ratio, e.g. for $\qacc = 0.01$,
$h_{\mathrm{don,\ f}}/h_{\mathrm{don,\ i}}-1 > -10^{-1}$ for
$\vthermal = 10^{-0.5}$ compared to
$h_{\mathrm{don,\ f}}/h_{\mathrm{don,\ i}}-1 = [-10^{-2},
-10^{-1}]$. The angles of incidence of these trajectories with the
donor are much larger, nearing perpendicular to its surface.

In \Figref{fig:classification_fractions_high_v} we show the fractions
of trajectories in each classification for the hot stream
calculations.
\Figref{fig:classification_fractions_high_v} (a) shows that compared
to our low thermal-velocity (cold) results
(\Figref{fig:classification_fractions}), a larger fraction of mass
ratios and synchronicity factors accrete onto the accretor,
e.g. systems with $0.1 < \qacc < 10$ and $\fsync > 1.5$ or $< 0.5$ now
accrete onto the accretor instead of onto the donor in the low
thermal-velocity case ($\vthermal = 10^{-3.0}$). This is mainly
attributed to the larger initial radial velocity, which gives the
particles more momentum to start with and make it more difficult to
change the course of their trajectories. This, in turn, leads to a
smaller region of the parameter space in $\qacc$ and $\fsync$ that
accretes back onto the donor.
\Figref{fig:classification_fractions_high_v} (c) shows the fraction of
trajectories that escapes as a function of $\qacc$ and $\fsync$. While
at low ($\vthermal = 10^{-3.0}$) thermal-velocity (cold) there is no
trajectory that escapes, the high ($\vthermal = 10^{-0.5}$)
thermal-velocity allows trajectories to pass the accretor and escape
through the Lagrange point behind the accretor (at
$x > x_{\mathrm{acc}}$). This primarily occurs in sub-synchronous
systems, again due to the Coriolis force accelerating the particle
towards positive $x$.
Overall, the data in \Figref{fig:classification_fractions_high_v} (a)
and (b) show that instead of the sharp transition between accretion
onto accretor and self-accretion, there is a much more gradual
transition between the regions where the fractions transition from $0$
to $1$ over a larger range of parameters. This is because the high
thermal-velocity leads to a wide stream, i.e. a wider range of initial
positions around L1 for our trajectories for a given system. In
systems with e.g. $\qacc = 0.1$ and $\fsync = 1.9$, about half of the
trajectories that make up the stream accrete onto the donor, and half
accrete onto the accretor.
Moreover, some trajectories fail to stay accurate within the given
minimum time step, but the total fraction of the failing systems is
negligible, and they only occur in small regions.

\Figref{fig:intersection_high_v} shows the intersection fractions for
high thermal-velocity (hot, $\vthermal \approx 0.32$) mass
transfer. The structure of the figure is the same as in
\Figref{fig:intersection_low_v}.

We find that the self-intersecting orbits again occur on the edges of
the transition regions between accretion onto the donor and accretion
onto the accretor. The fraction itself it not always high, due to the
stream being wider and parts the transition region is more gradual
(i.e. for a wider range in $\fsync$, $\qacc$, parts of the stream can
accrete onto different regions). For sub-synchronous rotation
($\fsync < 0.75$) the region of parameter space where self-accretion
occurs is narrow and is more confined to the transition region than
self-accretion in super-synchronously ($\fsync > 1.75$) rotating
systems. Super-synchronous systems with high-thermal velocity streams
have very wide streams, but the asynchronous velocity offset is lower
than in the equivalent sub-synchronous configurations
($\fsync < 1.75$,
\Figref{fig:non_synchronous_rotation_schematic}). This leads to
trajectories in a larger region in the parameter space
self-intersecting.
Intersection with other trajectories again coincides with regions of
self-intersection, where for sub-synchronous systems the regions
overlap strongly but for super-synchronous the region where
trajectories intersect with others extends to a larger part of the
parameter space ($\qacc < 0.5$ and $\fsync > 1.25$). The increase in
stream diameter in this leads to the initial conditions of each
trajectory to be sufficiently different to cross at high angles of
incidence ($\thetaintersect > \thetathresholdvalue$).

Overall, the regions of self-intersection and other-intersection are
confined to regions of low mass-ratios ($\qacc < 0.5$), due to the
higher radial velocity that occurs at high-thermal velocity, which
gives the stream more momentum and makes it harder to deflect or
rotate. Only for low mass-ratios is the donor massive enough to turn
the trajectories and lead to (self-)intersections.

\section{Discussion}
\label{sec:discussion}


We use binary population synthesis to evolve populations of binary
systems and record their properties during mass transfer. We do this
to find the ranges of the mass ratios of the accretor, $\qacc$,
synchronicity factors of the donor, $\fsync$, and thermal velocities
$\vthermal$ of the stream, that we should cover in our ballistic
stream trajectory calculations. At the same time we use these results
as a motivation for this study. Most notably, we find that mass
transfer takes place with non-synchronous donors for a significant
fraction of either mass transferred ($\approx 90$ per cent), as well
as time spent transferring mass ($\approx 60$ per cent,
\Figref{fig:exploration_results_alpha_vs_synchronicity}).
We find that the approximation of static tides does not always hold,
especially the fraction of mass transferred while the static
approximation fails is significant ($\approx 10$ per cent). This
indicates that mass transfer in those systems occurs in a
time-dependent potential, the effects of which are not captured by our
modelling approach and these systems likely require detailed stellar
evolution models and time-averaging to model the mass transfer
correctly.
We note that, while the results shown in
\Secref{sec:mass-transfer-binary} indicate the extent of the
parameters relevant to this study, they should be used just for
that. We currently calculate the population statistics for a starburst
population at a specific metallicity, and we do not convolve with any
star formation rate. This means that our population results are not a
directly observable quantity, even if the assumption of a single
metallicity is not entirely wrong for populations of dwarfs in the
solar neighbourhood
\citep{haywoodRevisionSolarNeighbourhood2001}. Moreover, our results
depend on the details of the population synthesis
calculations. Changes in, e.g.,\ tidal interaction physics
\citep{mirouhDetailedEquilibriumDynamical2023,
  preeceEquilibriumTideUpdated2022} or birth distributions
\citep{moeMindYourPs2017, moeCloseBinaryFraction2019} of the binary
components will change our results, although the extent to which is
not clear.

Recently \citet{davisMassTransferEccentric2013,
  davisBinaryEvolutionUsing2014} performed calculations with a similar
approach to ours. While they don't supply a data-release, the
behaviour of their stream models is described in some cases. They find
that, in all cases of self-accretion, the donor experiences a positive
torque, effectively spinning up the donor and removing angular
momentum from the orbit. This agrees with the results of
\citet{sepinskyInteractingBinariesEccentric2010}, as well as with
those of \citet{belvedereRoleSecondaryRotation1993}. The focus of all
these studies is on sub-synchronous donors. We find that
self-accretion onto super-synchronous donors lead to a spin-down of
the donor.
Our results imply that if a donor rotates asynchronously and
self-accretes, this self-accretion always works to synchronise the
donor even if it rotates super-synchronously.

We capture the effects of a large mass transfer stream cross-section
by simulating a set of trajectories with initial position offsets
along the stream. We treat these trajectories as individual, and we do
not include any interaction between these trajectories. In some cases,
however, the trajectories along the mass stream intersect at large
angles with other trajectories
\citep{warnerLocationSizeHot1972}. Realistically, these would be swept
up by parts of the stream with a higher density and momentum
\citep{flanneryLocationHotSpot1975}. We track whether trajectories
intersect with either themselves or with others
(\Secref{sec:intersecting-orbits}) and we find self-intersection and
intersection with other trajectories (at angles larger than the
threshold) occurs primarily in the transition regions between
accretion onto the accretor and accretion onto the donor
(Figures~\ref{fig:intersection_low_v}
and~\ref{fig:intersection_high_v}). Especially in the high
thermal-velocity (hot) stream super-synchronous cases we find that
where a high fraction of intersection with other trajectories takes
place ($\fsync > 1.25$ and $\qacc < 0.5$) extends to a larger part of
the parameter space than the region where self-intersection occurs
($\fsync > 1.75$ and $\qacc < 0.25$,
\Figref{fig:intersection_high_v}). The very wide stream causes the
particles along it to have a large spread in initial conditions and to
follow significantly varying trajectories.
We currently do not post-process any of these trajectories to alter
their outcome or to reject them based on intersection. The regions
where a high degree of (self-)intersection occurs likely require an
approach that is more sophisticated than approximating the stream by a
series of non-interacting ballistic trajectories.

Our ballistic approach imposes some assumptions on the starting
conditions of the particle, especially in asynchronous rotating
donors.
We take the transversal velocity offset due to asynchronous rotation
$\boldsymbol{v}_{\mathrm{non-synchronous\ offset}}$ to scale linearly
with the synchronicity factor. \citet{lubowIssuesTheoryMass1993}
critiques this approach, and argues that this axisymmetric velocity
assumption is not valid \citep[also][]{limberSurfaceFormsMass1963},
and that the problem requires a hydrodynamical analysis.
This is based on two studies that look at the gas dynamics of material
at L1 in non-synchronous donors using polytropic models for the
radiative \citep{lubowEquilibriumStatesNonsynchronous1979} and
convective \citep{campbellPossibilityNonsynchronismConvective1983}
stars, specifically the shape of the flow field at L1.
They both find that in the linearised and low-asynchronicity case the
velocity field tends to zero as it approaches L1, and hence and flow
towards L1 slows down and tend to zero before flowing through L1 and
increasing. This is in contrast with our assumption of a transverse
velocity component linearly dependent on the non-synchronicity factor
$\fsync$. A lower velocity offset with the same asynchronous rotation
of the donor leads to stream properties that are more like the
synchronous case.
Because of the initial supersonic velocity relative to the L1 point
(\Figref{fig:non_synchronous_rotation_schematic}) the slow-down can be
accompanied by shocks \citep{lubowIssuesTheoryMass1993}. The heating
of the shock dissipation could change the initial properties of the
stream (e.g. increase the local temperature at L1), and could be
observable as an excess luminosity around L1. With observations of
mass-transferring systems it might be possible to discern whether this
slow-down to L1 actually occurs, and whether mass-stream trajectories
behave like those in synchronous systems even for asynchronous donors.

The aim of this paper was to include the effects of non-synchronous
rotation of the donor on the particles in the mass transfer stream in
the ballistic approach, where we treat the accretor as a point
particle with no physical size. Our method, however, is suitable for
extensions like treating direct impact accretion onto the accretor and
adding properties of the particle during its flight to the
interpolation dataset, and the inclusion of additional physical
effects like post-Newtonian potentials for the accretor
\citep{kramarevDynamicsDirectImpact2023}, the effects of kinematic
acceleration \citep{kruszewskiExchangeMatterClose1963} or those of
irradiation by the secondary
\citep{podsiadlowskiSupermassiveBlackHoles1994,
  drechselRadiationPressureEffects1995,
  phillipsIrradiationPressureEffects2002} on the critical surface of
the donor.

\section{Conclusions}
\label{sec:conclusions}
Motivated by the lack of publicly available data of stream properties
in systems with non-synchronously rotating donor stars, we hereby
present our results of ballistic trajectory calculations. We calculate
ballistic trajectories with varying mass ratio $\qacc$, synchronicity
factor $\fsync$ and initial thermal-velocity $\vthermal$, and we
assume the accretor radius is infinitely small. We make use of binary
population synthesis to inform us of the ranges of the initial
parameters of the ballistic calculations and to provide further
motivation for the importance of this study and the need for a
publicly accessible data set on ballistic trajectories for
non-synchronous donors.

The main results of our study are summarised below.
\begin{enumerate}
\item Our binary population calculations with metallicity $Z=0.02$
  indicate that a large fraction of binary systems transfer mass
  sub-synchronously, but they transfer more mass ($90.14$ per cent)
  sub-synchronously than spend time doing so ($62.44$). Only a very
  low fraction of systems transfers mass super synchronously ($<0.07$
  per cent mass transferred and $<0.02$ per cent time spent
  transferring mass). Moreover, while only a small fraction of time is
  spent during which the static tide approximation breaks down, a
  non-negligible fraction of mass ($12.64$ per cent) is transferred
  when the donor experiences a dynamic potential. This does, however,
  mean that the static potential approximation is valid for the
  majority ($87.36$ per cent) of mass transferred with
  sub-synchronously rotating donors.
\item Our ballistic trajectory calculations indicate that at low
  initial thermal-velocity (cold, $\vthermal = 10^{-3.0}$) there are
  clear distinctions between accretion onto the accretor and accretion
  onto the donor within the parameter space of $\qacc$ and $\fsync$,
  and no trajectories escape from the system. The minimum radius of
  approach can be as low as $10^{-3}$, indicating a near head-on
  stream. A larger region in the ($\qacc$, $\fsync$), parameter space
  leads to accretion onto the donor for super-synchronous donors
  ($\fsync > 1$ and $\qacc < 100$) than for sub-synchronous donors
  ($\fsync < 0.75$ and $\qacc < 5$), but the change in specific
  angular momentum of the self-accreting stream is overall lower for
  super-synchronous donors. Both for sub-synchronous as for
  super-synchronous donors the self-accretion always works
  synchronising. We find that intersecting trajectories only occur at
  the edge of the transition region between accretion onto the
  accretor and accretion onto the donor and that the self-intersection
  regions overlap with that of intersection with other trajectories.
\item High initial thermal-velocities (hot, $\vthermal = 10^{-0.5}$)
  correspond to a wider mass stream, and lead to a less sharp
  transition between the regions of accretion onto the donor and
  accretion onto the accretor. Fewer configurations of $\qacc$ and
  $\fsync$, i.e. $\fsync > 1.5$ and $\qacc < 0.2$ for
  super-synchronous donors and $\fsync < 0.75$ and $\qacc < 0.1$ for
  sub-synchronous donors, lead to accretion onto donor because of the
  larger initial radial velocity of the stream that makes it more
  difficult to deflect the stream. We find some trajectories can
  escape the system through the Lagrange point behind the accretor
  ($x > x_{\mathrm{acc}}$), especially in systems with a
  sub-synchronous donor. Intersecting trajectories again occur at the
  edge of the transition region, but for super-synchronous donors the
  intersection with other trajectories occurs for a larger part of the
  parameter space than self-intersecting orbits, i.e. $\fsync > 1.25$
  and $\qacc < 0.5$ for intersection with other trajectories and
  $\fsync > 1.7$ and $\qacc < 0.1$ for self-intersection.
\end{enumerate}

Our results are useful for orbital evolution and mass transfer
calculations, including determining the formation and properties of
accretion disks. They can be used in stellar evolution and population
synthesis code, and they are available online upon publication of the
paper.

\section*{Acknowledgements}
DDH thanks the UKRI and the University of Surrey for the funding grant
H120341A, and thanks Arman Aryaeipour, Dominika Hubov\`{a}, Giovanni
Mirouh, Ond\v{r}ej Pejcha, Natalie Rees and Mathieu Renzo for useful
discussions. RGI thanks STFC for funding grants
\href{https://gtr.ukri.org/projects?ref=ST%2FR000603%2F1}{ST/R000603/1}
and
\href{https://gtr.ukri.org/projects?ref=ST/L003910/2}{ST/L003910/2}.

\section*{Data Availability}
\label{sec:data_avail}
We make our ballistic trajectory integration code, as well as the
interpolation tables for the stream properties and the exploration
data generated through population synthesis available on
\href{https://doi.org/10.5281/zenodo.7007591}{https://doi.org/10.5281/zenodo.7007591}
upon publication.

\bibliographystyle{mnras}
\bibliography{ballistic_paper.bib} 

\begin{thebibliography}{}
\makeatletter
\relax
\def\mn@urlcharsother{\let\do\@makeother \do\$\do\&\do\#\do\^\do\_\do\%\do\~}
\def\mn@doi{\begingroup\mn@urlcharsother \@ifnextchar [ {\mn@doi@}
  {\mn@doi@[]}}
\def\mn@doi@[#1]#2{\def\@tempa{#1}\ifx\@tempa\@empty \href
  {http://dx.doi.org/#2} {doi:#2}\else \href {http://dx.doi.org/#2} {#1}\fi
  \endgroup}
\def\mn@eprint#1#2{\mn@eprint@#1:#2::\@nil}
\def\mn@eprint@arXiv#1{\href {http://arxiv.org/abs/#1} {{\tt arXiv:#1}}}
\def\mn@eprint@dblp#1{\href {http://dblp.uni-trier.de/rec/bibtex/#1.xml}
  {dblp:#1}}
\def\mn@eprint@#1:#2:#3:#4\@nil{\def\@tempa {#1}\def\@tempb {#2}\def\@tempc
  {#3}\ifx \@tempc \@empty \let \@tempc \@tempb \let \@tempb \@tempa \fi \ifx
  \@tempb \@empty \def\@tempb {arXiv}\fi \@ifundefined
  {mn@eprint@\@tempb}{\@tempb:\@tempc}{\expandafter \expandafter \csname
  mn@eprint@\@tempb\endcsname \expandafter{\@tempc}}}

\bibitem[\protect\citeauthoryear{Avni}{Avni}{1976}]{avniEclipseDurationXray1976}
Avni Y.,  1976, \mn@doi [The Astrophysical Journal] {10.1086/154752}, 209, 574

\bibitem[\protect\citeauthoryear{Avni \& Schiller}{Avni \&
  Schiller}{1982}]{avniGeneralizedRochePotential1982}
Avni Y.,  Schiller N.,  1982, \mn@doi [The Astrophysical Journal]
  {10.1086/160025}, 257, 703

\bibitem[\protect\citeauthoryear{Belvedere, Lanzafame  \& Molteni}{Belvedere
  et~al.}{1993}]{belvedereRoleSecondaryRotation1993}
Belvedere G.,  Lanzafame G.,   Molteni D.,  1993, \href
  {https://ui.adsabs.harvard.edu/abs/1993A\&A...280..525B} {280, 525}

\bibitem[\protect\citeauthoryear{Blaauw}{Blaauw}{1961}]{blaauwOriginBtypeStars1961}
Blaauw A.,  1961, Bulletin of the Astronomical Institutes of the Netherlands,
  \href {https://ui.adsabs.harvard.edu/abs/1961BAN....15..265B} {15, 265}

\bibitem[\protect\citeauthoryear{Campbell \& Papaloizou}{Campbell \&
  Papaloizou}{1983}]{campbellPossibilityNonsynchronismConvective1983}
Campbell C.~G.,  Papaloizou J.,  1983, \mn@doi [Monthly Notices of the Royal
  Astronomical Society] {10.1093/mnras/204.2.433}, 204, 433

\bibitem[\protect\citeauthoryear{Claeys, Pols, Izzard, Vink  \& Verbunt}{Claeys
  et~al.}{2014}]{claeysTheoreticalUncertaintiesType2014}
Claeys J. S.~W.,  Pols O.~R.,  Izzard R.~G.,  Vink J.,   Verbunt F. W.~M.,
  2014, \mn@doi [Astronomy \& Astrophysics] {10.1051/0004-6361/201322714}, 563,
  A83

\bibitem[\protect\citeauthoryear{Davis, Siess  \& Deschamps}{Davis
  et~al.}{2013}]{davisMassTransferEccentric2013}
Davis P.~J.,  Siess L.,   Deschamps R.,  2013, \mn@doi [Astronomy and
  Astrophysics] {10.1051/0004-6361/201220391}, 556, A4

\bibitem[\protect\citeauthoryear{Davis, Siess  \& Deschamps}{Davis
  et~al.}{2014}]{davisBinaryEvolutionUsing2014}
Davis P.~J.,  Siess L.,   Deschamps R.,  2014, \mn@doi [Astronomy and
  Astrophysics] {10.1051/0004-6361/201423730}, 570, A25

\bibitem[\protect\citeauthoryear{De~Marco \& Izzard}{De~Marco \&
  Izzard}{2017}]{demarcoImpactCompanionsStellar2017}
De~Marco O.,  Izzard R.~G.,  2017, \mn@doi [Publications of the Astronomical
  Society of Australia] {10.1017/pasa.2016.52}, 34

\bibitem[\protect\citeauthoryear{Dermine, Jorissen, Siess  \&
  Frankowski}{Dermine et~al.}{2009}]{dermineRadiationPressurePulsation2009}
Dermine T.,  Jorissen A.,  Siess L.,   Frankowski A.,  2009, \mn@doi [Astronomy
  \& Astrophysics] {10.1051/0004-6361/200912313}, 507, 891

\bibitem[\protect\citeauthoryear{Dosopoulou \& Kalogera}{Dosopoulou \&
  Kalogera}{2016}]{dosopoulouORBITALEVOLUTIONMASSTRANSFERRING2016}
Dosopoulou F.,  Kalogera V.,  2016, \mn@doi [The Astrophysical Journal]
  {10.3847/0004-637X/825/1/70}, 825, 70

\bibitem[\protect\citeauthoryear{Dosopoulou, Naoz  \& Kalogera}{Dosopoulou
  et~al.}{2017}]{dosopoulouRochelobeOverflowEccentric2017}
Dosopoulou F.,  Naoz S.,   Kalogera V.,  2017, \mn@doi [The Astrophysical
  Journal] {10.3847/1538-4357/aa7a05}, 844, 12

\bibitem[\protect\citeauthoryear{Drechsel, Haas, Lorenz  \& Gayler}{Drechsel
  et~al.}{1995}]{drechselRadiationPressureEffects1995}
Drechsel H.,  Haas S.,  Lorenz R.,   Gayler S.,  1995, \href
  {https://ui.adsabs.harvard.edu/abs/1995A\&A...294..723D} {294, 723}

\bibitem[\protect\citeauthoryear{Flannery \& Faulkner}{Flannery \&
  Faulkner}{1975}]{flanneryLocationHotSpot1975}
Flannery B.~P.,  Faulkner J.,  1975, \mn@doi [Monthly Notices of the Royal
  Astronomical Society] {10.1093/mnras/170.2.325}, 170, 325

\bibitem[\protect\citeauthoryear{Ghodla, Eldridge, Stanway  \& Stevance}{Ghodla
  et~al.}{2023}]{ghodlaEvaluatingChemicallyHomogeneous2023}
Ghodla S.,  Eldridge J.~J.,  Stanway E.~R.,   Stevance H.~F.,  2023, \mn@doi
  [Monthly Notices of the Royal Astronomical Society] {10.1093/mnras/stac3177},
  518, 860

\bibitem[\protect\citeauthoryear{Hairer, N{\o}rsett  \& Wanner}{Hairer
  et~al.}{2008}]{hairerSolvingOrdinaryDifferential2008}
Hairer E.,  N{\o}rsett S.~P.,   Wanner G.,  2008, Solving {{Ordinary
  Differential Equations I}}: {{Nonstiff Problems}}.
{Springer Science \& Business Media}

\bibitem[\protect\citeauthoryear{Hameury}{Hameury}{2020}]{hameuryReviewDiscInstability2020}
Hameury J.-M.,  2020, \mn@doi [Advances in Space Research]
  {10.1016/j.asr.2019.10.022}, 66, 1004

\bibitem[\protect\citeauthoryear{Haywood}{Haywood}{2001}]{haywoodRevisionSolarNeighbourhood2001}
Haywood M.,  2001, \mn@doi [Monthly Notices of the Royal Astronomical Society]
  {10.1046/j.1365-8711.2001.04510.x}, 325, 1365

\bibitem[\protect\citeauthoryear{Hendriks \& Izzard}{Hendriks \&
  Izzard}{2023}]{hendriksBinaryCpythonPythonbased2023}
Hendriks D.~D.,  Izzard R.~G.,  2023, \mn@doi [Journal of Open Source Software]
  {10.21105/joss.04642}, 8, 4642

\bibitem[\protect\citeauthoryear{Hubov{\'a} \& Pejcha}{Hubov{\'a} \&
  Pejcha}{2019}]{hubovaKinematicsMasslossOuter2019}
Hubov{\'a} D.,  Pejcha O.,  2019, \mn@doi [Monthly Notices of the Royal
  Astronomical Society] {10.1093/mnras/stz2208}, 489, 891

\bibitem[\protect\citeauthoryear{Hurley, Pols  \& Tout}{Hurley
  et~al.}{2000}]{hurleyComprehensiveAnalyticFormulae2000}
Hurley J.~R.,  Pols O.~R.,   Tout C.~A.,  2000, \mn@doi [Monthly Notices of the
  Royal Astronomical Society] {10.1046/j.1365-8711.2000.03426.x}, 315, 543

\bibitem[\protect\citeauthoryear{Hurley, Tout  \& Pols}{Hurley
  et~al.}{2002}]{hurleyEvolutionBinaryStars2002}
Hurley J.~R.,  Tout C.~A.,   Pols O.~R.,  2002, \mn@doi [Monthly Notices of the
  Royal Astronomical Society] {10.1046/j.1365-8711.2002.05038.x}, 329, 897

\bibitem[\protect\citeauthoryear{Hut}{Hut}{1981}]{hutTidalEvolutionClose1981}
Hut P.,  1981, Astronomy and Astrophysics, \href
  {https://ui.adsabs.harvard.edu/abs/1981A\&A....99..126H} {99, 126}

\bibitem[\protect\citeauthoryear{Izzard \& Jermyn}{Izzard \&
  Jermyn}{2022}]{izzardCircumbinaryDiscsStellar2022}
Izzard R.~G.,  Jermyn A.~S.,  2022, \mn@doi [Monthly Notices of the Royal
  Astronomical Society] {10.1093/mnras/stac2899}

\bibitem[\protect\citeauthoryear{Izzard, Tout, Karakas  \& Pols}{Izzard
  et~al.}{2004}]{izzardNewSyntheticModel2004}
Izzard R.~G.,  Tout C.~A.,  Karakas A.~I.,   Pols O.~R.,  2004, \mn@doi
  [Monthly Notices of the Royal Astronomical Society]
  {10.1111/j.1365-2966.2004.07446.x}, 350, 407

\bibitem[\protect\citeauthoryear{Izzard, Dray, Karakas, Lugaro  \& Tout}{Izzard
  et~al.}{2006}]{izzardPopulationNucleosynthesisSingle2006}
Izzard R.~G.,  Dray L.~M.,  Karakas A.~I.,  Lugaro M.,   Tout C.~A.,  2006,
  \mn@doi [Astronomy \& Astrophysics] {10.1051/0004-6361:20066129}, 460, 565

\bibitem[\protect\citeauthoryear{Izzard, Glebbeek, Stancliffe  \& Pols}{Izzard
  et~al.}{2009}]{izzardPopulationSynthesisBinary2009}
Izzard R.~G.,  Glebbeek E.,  Stancliffe R.~J.,   Pols O.~R.,  2009, \mn@doi
  [Astronomy and Astrophysics] {10.1051/0004-6361/200912827}, 508, 1359

\bibitem[\protect\citeauthoryear{Izzard, Preece, Jofre, Halabi, Masseron  \&
  Tout}{Izzard et~al.}{2018}]{izzardBinaryStarsGalactic2018}
Izzard R.~G.,  Preece H.,  Jofre P.,  Halabi G.~M.,  Masseron T.,   Tout C.~A.,
   2018, \mn@doi [Monthly Notices of the Royal Astronomical Society]
  {10.1093/mnras/stx2355}, 473, 2984

\bibitem[\protect\citeauthoryear{Jermyn et~al.,}{Jermyn
  et~al.}{2023}]{jermynModulesExperimentsStellar2023}
Jermyn A.~S.,  et~al., 2023, \mn@doi [The Astrophysical Journal Supplement
  Series] {10.3847/1538-4365/acae8d}, 265, 15

\bibitem[\protect\citeauthoryear{Kashi \& Soker}{Kashi \&
  Soker}{2011}]{kashiCircumbinaryDiscFinal2011}
Kashi A.,  Soker N.,  2011, \mn@doi [Monthly Notices of the Royal Astronomical
  Society] {10.1111/j.1365-2966.2011.19361.x}, 417, 1466

\bibitem[\protect\citeauthoryear{Kramarev \& Yudin}{Kramarev \&
  Yudin}{2023}]{kramarevDynamicsDirectImpact2023}
Kramarev N.,  Yudin A.,  2023, \mn@doi [Monthly Notices of the Royal
  Astronomical Society] {10.1093/mnras/stad1018}, 522, 626

\bibitem[\protect\citeauthoryear{Kruszewski}{Kruszewski}{1963}]{kruszewskiExchangeMatterClose1963}
Kruszewski A.,  1963, Acta Astronomica, \href
  {https://ui.adsabs.harvard.edu/abs/1963AcA....13..106K} {13, 106}

\bibitem[\protect\citeauthoryear{Kruszewski}{Kruszewski}{1964a}]{kruszewskiExchangeMatterClose1964}
Kruszewski A.,  1964a, Acta Astronomica, \href
  {https://ui.adsabs.harvard.edu/abs/1964AcA....14..231K} {14, 231}

\bibitem[\protect\citeauthoryear{Kruszewski}{Kruszewski}{1964b}]{kruszewskiExchangeMatterClose1964a}
Kruszewski A.,  1964b, Acta Astronomica, \href
  {https://ui.adsabs.harvard.edu/abs/1964AcA....14..241K} {14, 241}

\bibitem[\protect\citeauthoryear{Kruszewski}{Kruszewski}{1967}]{kruszewskiExchangeMatterClose1967}
Kruszewski A.,  1967, Acta Astronomica, \href
  {https://ui.adsabs.harvard.edu/abs/1967AcA....17..297K} {17, 297}

\bibitem[\protect\citeauthoryear{Limber}{Limber}{1963}]{limberSurfaceFormsMass1963}
Limber D.~N.,  1963, \mn@doi [The Astrophysical Journal] {10.1086/147711}, 138,
  1112

\bibitem[\protect\citeauthoryear{Lubow}{Lubow}{1979}]{lubowEquilibriumStatesNonsynchronous1979}
Lubow S.~H.,  1979, \mn@doi [The Astrophysical Journal] {10.1086/157036}, 229,
  1008

\bibitem[\protect\citeauthoryear{Lubow}{Lubow}{1993}]{lubowIssuesTheoryMass1993}
Lubow S.~H.,  1993, in Sahade J.,  McCluskey G.~E.,   Kondo Y.,  eds,
  Astrophysics and {{Space Science Library}}, The {{Realm}} of {{Interacting
  Binary Stars}}.
{Springer Netherlands}, {Dordrecht}, pp 25--37,
  \mn@doi{10.1007/978-94-011-2416-4\_5}, \url
  {https://doi.org/10.1007/978-94-011-2416-4_5}

\bibitem[\protect\citeauthoryear{Lubow \& Shu}{Lubow \&
  Shu}{1975}]{lubowGasDynamicsSemidetached1975}
Lubow S.~H.,  Shu F.~H.,  1975, \mn@doi [The Astrophysical Journal]
  {10.1086/153614}, 198, 383

\bibitem[\protect\citeauthoryear{Mehlhorn \& N{\"a}her}{Mehlhorn \&
  N{\"a}her}{1994}]{mehlhornImplementationSweepLine1994}
Mehlhorn K.,  N{\"a}her S.,  1994

\bibitem[\protect\citeauthoryear{Meyer \& {Meyer-Hofmeister}}{Meyer \&
  {Meyer-Hofmeister}}{1983}]{meyerModelStandstillCamelopardalis1983}
Meyer F.,  {Meyer-Hofmeister} E.,  1983, Astronomy and Astrophysics, \href
  {https://ui.adsabs.harvard.edu/abs/1983A\&A...121...29M} {121, 29}

\bibitem[\protect\citeauthoryear{Mirouh, Hendriks, Dykes, Moe  \&
  Izzard}{Mirouh et~al.}{2023}]{mirouhDetailedEquilibriumDynamical2023}
Mirouh G.~M.,  Hendriks D.~D.,  Dykes S.,  Moe M.,   Izzard R.~G.,  2023,
  \mn@doi [Monthly Notices of the Royal Astronomical Society]
  {10.1093/mnras/stad2048}, p. stad2048

\bibitem[\protect\citeauthoryear{Moe \& Stefano}{Moe \&
  Stefano}{2017}]{moeMindYourPs2017}
Moe M.,  Stefano R.~D.,  2017, \mn@doi [The Astrophysical Journal Supplement
  Series] {10.3847/1538-4365/aa6fb6}, 230, 15

\bibitem[\protect\citeauthoryear{Moe, Kratter  \& Badenes}{Moe
  et~al.}{2019}]{moeCloseBinaryFraction2019}
Moe M.,  Kratter K.~M.,   Badenes C.,  2019, \mn@doi [The Astrophysical
  Journal] {10.3847/1538-4357/ab0d88}, 875, 61

\bibitem[\protect\citeauthoryear{Moulton}{Moulton}{1914}]{moulton1914}
Moulton F.~R.,  1914, An Introduction to Celestial Mechanics.
{Macmillan}, \url {https://ui.adsabs.harvard.edu/abs/1914icm..book.....M}

\bibitem[\protect\citeauthoryear{Nomoto}{Nomoto}{1986}]{nomotoFateAccretingWhite1986}
Nomoto K.,  1986, \mn@doi [Progress in Particle and Nuclear Physics]
  {10.1016/0146-6410(86)90020-7}, 17, 249

\bibitem[\protect\citeauthoryear{Ogilvie}{Ogilvie}{2014}]{ogilvieTidalDissipationStars2014}
Ogilvie G.~I.,  2014, \mn@doi [Annual Review of Astronomy and Astrophysics]
  {10.1146/annurev-astro-081913-035941}, 52, 171

\bibitem[\protect\citeauthoryear{Osaki, Hirose  \& Ichikawa}{Osaki
  et~al.}{1993}]{osakiTidalEffectsAccretion1993}
Osaki Y.,  Hirose M.,   Ichikawa S.,  1993, \mn@doi [Accretion Disks in Compact
  Stellar Systems] {10.1142/9789814350976\_0010}, p.~272

\bibitem[\protect\citeauthoryear{Ovenden \& Roy}{Ovenden \&
  Roy}{1961}]{ovendenUseJacobiIntegral1961}
Ovenden M.~W.,  Roy A.~E.,  1961, \mn@doi [Monthly Notices of the Royal
  Astronomical Society] {10.1093/mnras/123.1.1}, 123, 1

\bibitem[\protect\citeauthoryear{Papaloizou \& Pringle}{Papaloizou \&
  Pringle}{1977}]{papaloizouTidalTorquesAccretion1977}
Papaloizou J.,  Pringle J.~E.,  1977, \mn@doi [Monthly Notices of the Royal
  Astronomical Society] {10.1093/mnras/181.3.441}, 181, 441

\bibitem[\protect\citeauthoryear{Pejcha, Metzger  \& Tomida}{Pejcha
  et~al.}{2016}]{pejchaBinaryStellarMergers2016}
Pejcha O.,  Metzger B.~D.,   Tomida K.,  2016, \mn@doi [Monthly Notices of the
  Royal Astronomical Society] {10.1093/mnras/stw1481}, 461, 2527

\bibitem[\protect\citeauthoryear{Phillips \& Podsiadlowski}{Phillips \&
  Podsiadlowski}{2002}]{phillipsIrradiationPressureEffects2002}
Phillips S.~N.,  Podsiadlowski {\relax Ph}.,  2002, \mn@doi [Monthly Notices of
  the Royal Astronomical Society] {10.1046/j.1365-8711.2002.05886.x}, 337, 431

\bibitem[\protect\citeauthoryear{Plavec}{Plavec}{1958}]{plavec49DynamicalInstability1958}
Plavec M.,  1958, in Liege International Astrophysical Colloquia. pp 411--420

\bibitem[\protect\citeauthoryear{Podsiadlowski \& Rees}{Podsiadlowski \&
  Rees}{1994}]{podsiadlowskiSupermassiveBlackHoles1994}
Podsiadlowski P.,  Rees M.~J.,  1994, \mn@doi [AIP Conference Proceedings]
  {10.1063/1.45957}, 308, 71

\bibitem[\protect\citeauthoryear{Pols, Schr{\"o}der, Hurley, Tout  \&
  Eggleton}{Pols et~al.}{1998}]{polsStellarEvolutionModels1998}
Pols O.~R.,  Schr{\"o}der K.-P.,  Hurley J.~R.,  Tout C.~A.,   Eggleton P.~P.,
  1998, \mn@doi [Monthly Notices of the Royal Astronomical Society]
  {10.1046/j.1365-8711.1998.01658.x}, 298, 525

\bibitem[\protect\citeauthoryear{Postnov, Mironov, Lutovinov, Shakura,
  Kochetkova  \& Tsygankov}{Postnov
  et~al.}{2015}]{postnovSpinupSpindownNeutron2015}
Postnov K.~A.,  Mironov A.~I.,  Lutovinov A.~A.,  Shakura N.~I.,  Kochetkova
  A.~Y.,   Tsygankov S.~S.,  2015, \mn@doi [Monthly Notices of the Royal
  Astronomical Society] {10.1093/mnras/stu2155}, 446, 1013

\bibitem[\protect\citeauthoryear{Preece, Hamers, Neunteufel, Schafer  \&
  Tout}{Preece et~al.}{2022}]{preeceEquilibriumTideUpdated2022}
Preece H.~P.,  Hamers A.~S.,  Neunteufel P.~G.,  Schafer A.~L.,   Tout C.~A.,
  2022, \mn@doi [The Astrophysical Journal] {10.3847/1538-4357/ac6fe3}, 933, 25

\bibitem[\protect\citeauthoryear{Pringle \& Rees}{Pringle \&
  Rees}{1972}]{pringleAccretionDiscModels1972}
Pringle J.~E.,  Rees M.~J.,  1972, Astronomy and Astrophysics, \href
  {https://ui.adsabs.harvard.edu/abs/1972A\&A....21....1P} {21, 1}

\bibitem[\protect\citeauthoryear{Raymer}{Raymer}{2012}]{raymerThreedimensionalHydrodynamicSimulations2012}
Raymer E.,  2012, \mn@doi [Monthly Notices of the Royal Astronomical Society]
  {10.1111/j.1365-2966.2012.22090.x}, 427, 1702

\bibitem[\protect\citeauthoryear{Renzo et~al.,}{Renzo
  et~al.}{2019}]{renzoMassiveRunawayWalkaway2019}
Renzo M.,  et~al., 2019, \mn@doi [Astronomy \& Astrophysics]
  {10.1051/0004-6361/201833297}, 624, A66

\bibitem[\protect\citeauthoryear{Riley et~al.,}{Riley
  et~al.}{2022}]{rileyRapidStellarBinary2022}
Riley J.,  et~al., 2022, \mn@doi [The Astrophysical Journal Supplement Series]
  {10.3847/1538-4365/ac416c}, 258, 34

\bibitem[\protect\citeauthoryear{Ritter}{Ritter}{1988}]{ritterTurningMassTransfer1988}
Ritter H.,  1988, Astronomy and Astrophysics, \href
  {https://ui.adsabs.harvard.edu/abs/1988A\&A...202...93R/abstract} {202, 93}

\bibitem[\protect\citeauthoryear{Ruiter, Belczynski, Sim, Hillebrandt, Fink  \&
  Kromer}{Ruiter et~al.}{2010}]{ruiterTypeIaSupernovae2010}
Ruiter A.~J.,  Belczynski K.,  Sim S.~A.,  Hillebrandt W.,  Fink M.,   Kromer
  M.,  2010, in Kalogera V.,  {van der Sluys} M.,  eds,  American Institute of
  Physics Conference Series Vol. 1314, International Conference on Binaries: In
  Celebration of Ron Webbink's 65th Birthday. pp 233--238 (\mn@eprint {arxiv}
  {1009.3661}), \mn@doi{10.1063/1.3536375}

\bibitem[\protect\citeauthoryear{Savonije}{Savonije}{1978}]{savonijeRochelobeOverflowXray1978}
Savonije G.~J.,  1978, Astronomy and Astrophysics, \href
  {https://ui.adsabs.harvard.edu/abs/1978A\&A....62..317S/abstract} {62, 317}

\bibitem[\protect\citeauthoryear{Sepinsky, Willems  \& Kalogera}{Sepinsky
  et~al.}{2007}]{sepinskyEquipotentialSurfacesLagrangian2007}
Sepinsky J.~F.,  Willems B.,   Kalogera V.,  2007, \mn@doi [The Astrophysical
  Journal] {10.1086/513736}, 660, 1624

\bibitem[\protect\citeauthoryear{Sepinsky, Willems, Kalogera  \&
  Rasio}{Sepinsky et~al.}{2010}]{sepinskyInteractingBinariesEccentric2010}
Sepinsky J.~F.,  Willems B.,  Kalogera V.,   Rasio F.~A.,  2010, \mn@doi [The
  Astrophysical Journal] {10.1088/0004-637X/724/1/546}, 724, 546

\bibitem[\protect\citeauthoryear{Shakura \& Sunyaev}{Shakura \&
  Sunyaev}{1973}]{shakuraBlackHolesBinary1973}
Shakura N.~I.,  Sunyaev R.~A.,  1973, \href
  {https://ui.adsabs.harvard.edu/abs/1973A\&A....24..337S} {24, 337}

\bibitem[\protect\citeauthoryear{Shao \& Li}{Shao \&
  Li}{2014}]{shaoFormationBeStars2014}
Shao Y.,  Li X.-D.,  2014, \mn@doi [The Astrophysical Journal]
  {10.1088/0004-637X/796/1/37}, 796, 37

\bibitem[\protect\citeauthoryear{Tauris \& Takens}{Tauris \&
  Takens}{1998}]{taurisRunawayVelocitiesStellar1998}
Tauris T.~M.,  Takens R.~J.,  1998, Astronomy and Astrophysics, \href
  {http://adsabs.harvard.edu/abs/1998A\%26A...330.1047T} {330, 1047}

\bibitem[\protect\citeauthoryear{Tout, Aarseth, Pols  \& Eggleton}{Tout
  et~al.}{1997}]{toutRapidBinaryStar1997}
Tout C.~A.,  Aarseth S.~J.,  Pols O.~R.,   Eggleton P.~P.,  1997, \mn@doi
  [Monthly Notices of the Royal Astronomical Society]
  {10.1093/mnras/291.4.732}, 291, 732

\bibitem[\protect\citeauthoryear{Tsantilas \& {Rovithis-Livaniou}}{Tsantilas \&
  {Rovithis-Livaniou}}{2006}]{tsantilasInfluenceRadiationPressure2006}
Tsantilas S.,  {Rovithis-Livaniou} H.,  2006, \href
  {https://ui.adsabs.harvard.edu/abs/2006ASPC..349..355T} {349, 355}

\bibitem[\protect\citeauthoryear{Ulrich \& Burger}{Ulrich \&
  Burger}{1976}]{ulrichAccretingComponentMassexchange1976}
Ulrich R.~K.,  Burger H.~L.,  1976, \mn@doi [The Astrophysical Journal]
  {10.1086/154406}, 206, 509

\bibitem[\protect\citeauthoryear{Vanbeveren}{Vanbeveren}{1977}]{vanbeverenInfluenceCriticalSurface1977}
Vanbeveren D.,  1977, Astronomy and Astrophysics, \href
  {https://ui.adsabs.harvard.edu/abs/1977A\&A....54..877V} {54, 877}

\bibitem[\protect\citeauthoryear{Virtanen et~al.,}{Virtanen
  et~al.}{2020}]{virtanenSciPyFundamentalAlgorithms2020}
Virtanen P.,  et~al., 2020, \mn@doi [Nature Methods]
  {10.1038/s41592-019-0686-2}, \href {https://rdcu.be/b08Wh} {17, 261}

\bibitem[\protect\citeauthoryear{Warner \& Peters}{Warner \&
  Peters}{1972}]{warnerLocationSizeHot1972}
Warner B.,  Peters W.~L.,  1972, \mn@doi [Monthly Notices of the Royal
  Astronomical Society] {10.1093/mnras/160.1.15}, 160, 15

\bibitem[\protect\citeauthoryear{Zahn}{Zahn}{1975}]{zahnDynamicalTideClose1975}
Zahn J.~P.,  1975, Astronomy and Astrophysics, \href
  {https://ui.adsabs.harvard.edu/abs/1975A\&A....41..329Z} {41, 329}

\bibitem[\protect\citeauthoryear{Zahn}{Zahn}{1977a}]{zahnTidalFrictionClose1977}
Zahn J.~P.,  1977a, Astronomy and Astrophysics, \href
  {https://ui.adsabs.harvard.edu/abs/1977A\&A....57..383Z} {57, 383}

\bibitem[\protect\citeauthoryear{Zahn}{Zahn}{1977b}]{zahnReprint1977AAmp1977}
Zahn J.-P.,  1977b, Astronomy and Astrophysics, \href
  {https://ui.adsabs.harvard.edu/abs/1977A\&A....57..383Z/abstract} {500, 121}

\bibitem[\protect\citeauthoryear{Zahn}{Zahn}{2008}]{zahnTidalDissipationBinary2008}
Zahn J.~P.,  2008, in {{EAS Publications Series}}. {eprint: arXiv:0807.4870},
  pp 67--90, \mn@doi{10.1051/eas:0829002}, \url
  {https://ui.adsabs.harvard.edu/abs/2008EAS....29...67Z}

\makeatother
\end{thebibliography}


\appendix

\section{Description of output datasets}
\label{sec:fiduc-source-distr}
The ballistics stream trajectory summary datasets contain the
parameters described in \Tabref{tab:description_table}, along with
meta-data regarding indices and global configurations. These datasets
can be interpolated on and implemented in other binary stellar
evolution codes to include the effect explored in our paper and the
subsequent changes in the mass transfer properties like the torque on
the orbit, the fraction of self accretion.

\begin{table*}
  \begin{center}
    \begin{tabular}{p{3.5cm}p{6cm}p{3.5cm}p{3.5cm}}
      \toprule
      Name & Description & Reference & Input/Output \\
      \midrule
      Initial thermal velocity $\mathrm{log}_{10}\left(\vthermal\right)$ & Initial thermal velocity of particles at L1. & \Secref{sec:therm-veloc-stre-1} and \Eqref{eq:sigma_thermal} & Input\\
      Synchronicity factor $\fsync$ & Synchronicity factor of the donor. & \Secref{sec:roche-potent-reduc} and \Eqref{eq:synchronicity_factor} & Input\\
      Mass ratio $\qacc$ & Mass ratio of the accretor, $\qacc = M_{\mathrm{accretor}}/M_{\mathrm{donor}}$. &  & Input\\
      \hline
      Total weight successful $w_{\mathrm{succes}}$ & Total weight of the successful trajectory calculations. & \Secref{sec:classifying-and-averaging} and \Eqref{eq:all_succesfull} & Output\\
      Total weighted fraction self-intersecting trajectories $\selfintersect$  & Total weighted fraction of the valid trajectories that self-intersect. & \Secref{sec:intersecting-orbits} & Output\\
      Total weighted fraction other-intersecting trajectories $\otherintersect$ & Total weighted fraction of the valid trajectories that intersect with other orbits at angles above the threshold $\thetathreshold$. & \Secref{sec:intersecting-orbits} & Output\\
      Accretor accretion fraction $\beta_{\mathrm{acc}}$ & Total weighted fraction of trajectories that accrete onto the accretor. & \Secref{sec:classifying-and-averaging} and \Eqref{eq:fraction_accretion_accretor} & Output\\
      Self accretion fraction $\beta_{\mathrm{don}}$ & Total weighted fraction of trajectories that accrete back onto the donor. & \Secref{sec:classifying-and-averaging} & Output\\
      Initial specific angular momentum w.r.t donor $h_{\mathrm{i,\ don}}$ & Average weighted initial specific angular momentum of the mass stream with respect to the donor. & \Secref{sec:self-accr-torq} and \Eqref{eq:angmom_wrt_don} & Output\\
      Self-accretion specific angular momentum factor $h_{\mathrm{f,\ don}} / h_{\mathrm{i,\ don}}$ & Average weighted ratio of final and initial specific angular momentum of the mass that accretes back onto the donor. & \Secref{sec:self-accr-torq} and \Eqref{eq:angmom_wrt_don} & Output\\
      Radius of closest approach $r_{\mathrm{min}}$ & Average weighted radius of closest approach to the accretor of the mass transfer stream for accretion onto accretor. & \Secref{sec:onto-accretor-stream-properties} & Output\\
      Circularisation radius $r_{\mathrm{circ}}$ & Average weighted radius of circularisation based on the angular momentum content of the stream at the radius of closest approach for accretion onto accretor.  & \Secref{sec:onto-accretor-stream-properties} and \Eqref{eq:circularisation radius} & Output\\
      Stream orientation & Average weighted pro- or retrograde orientation of the stream. & & Output\\
      Specific angular momentum along the stream at i-th radius $h_{\mathrm{stream\, i}}$ & Averaged weighted specific angular momentum of the stream at distance $d_{\mathrm{stream\, i}}$. For radii that are smaller than the radius of closest approach $r_{\mathrm{min}}$, this quantity is set to 0. For donor-accretion this quantity is filled with a dummy value. The distances $d_{\mathrm{stream\, i}}$ are expressed in units of the Roche-lobe radius of the accretor. & \Secref{sec:onto-accretor-stream-properties} and \Eqref{eq:distance_for_stream_interpolation} & Output\\
      \bottomrule
    \end{tabular}
  \end{center}
  \caption{Parameters included in the output datasets. The
    Input/Output column indicates whether this parameter is an input
    parameter to the interpolation table, or an output quantity.}
  \label{tab:description_table}
\end{table*}

\section{Lagrange points as a function of synchronicity}
\label{sec:lagrange-point-plot}
\begin{figure*}
  \includegraphics[width=0.75\textwidth]{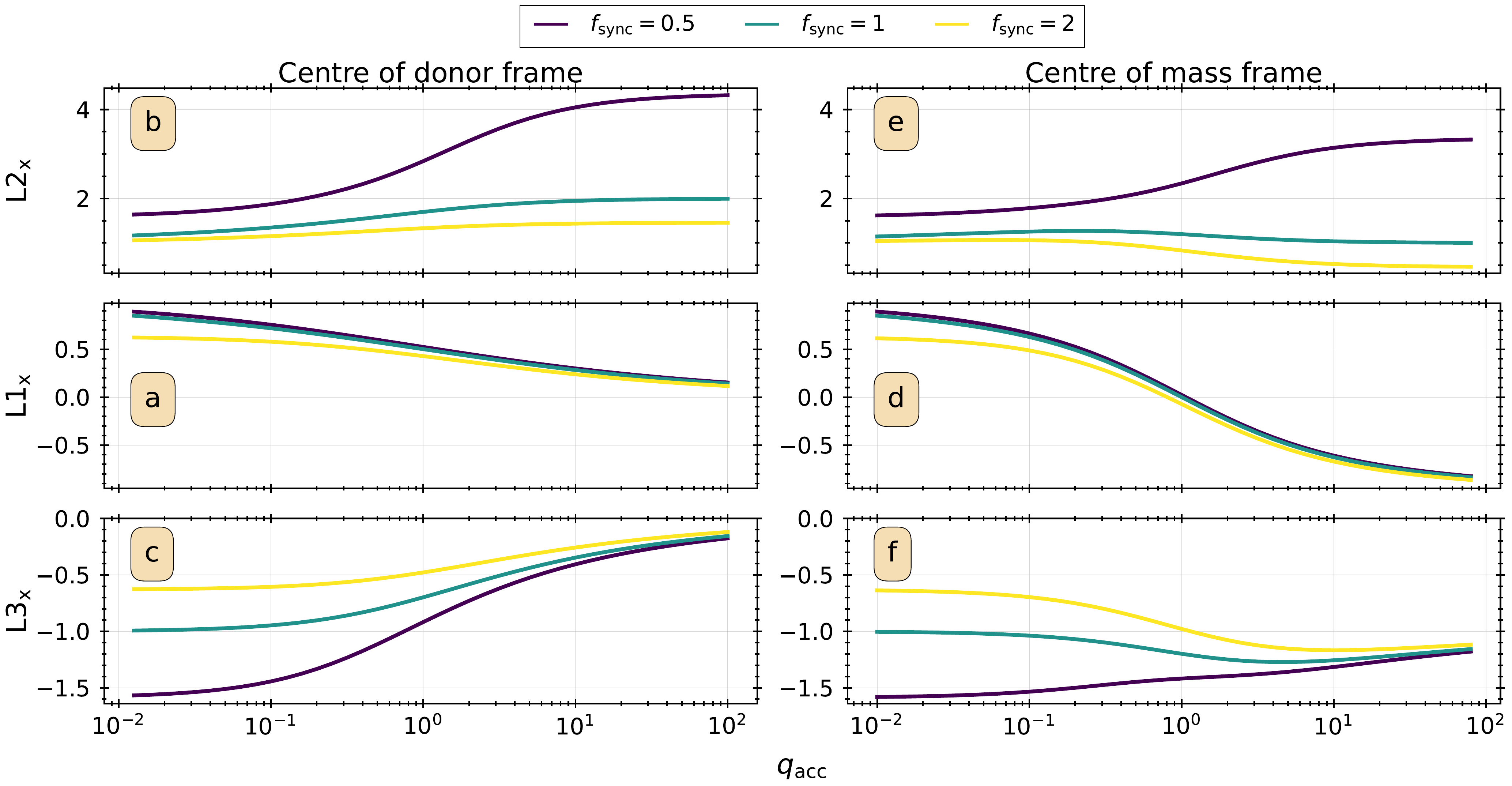}
  \caption{$x$-coordinates of the first three Lagrange points of the
    donor ($\mathrm{L1}_{x}$, $\mathrm{L2}_{x}$, $\mathrm{L3}_{x}$) in
    the reference frame centred on the donor (b, a and c respectively)
    and the reference frame centred on the centre of mass of the
    binary system (e, d and f respectively)
    (\Secref{sec:roche-potent-reduc}).}
  \label{fig:lagrange_point_plot}
\end{figure*}

With \Eqref{eq:critical_surface} we calculate the first three Lagrange
points, the donor for the synchronicity factors $\fsync$ and mass
ratios $\qacc$. We calculate these in the non-inertial reference frame
centred on the donor and transform this to the non-inertial reference
frame centred on the centre of mass of the system
(\Secref{sec:roche-potent-reduc}). In \Figref{fig:lagrange_point_plot}
we show the $x$-coordinate of the first three Lagrange points for both
of these frames.

\bsp    
\label{lastpage}
\end{document}